\documentclass[11pt]{article}

\usepackage{geometry}
\usepackage{graphicx,hyperref,setspace}
\usepackage[boxed,linesnumbered]{algorithm2e}

\usepackage{amsmath, amsthm, amssymb}

\theoremstyle{theorem}
\newtheorem{theorem}{Theorem}
\newtheorem{lemma}[theorem]{Lemma}
\newtheorem{corollary}[theorem]{Corollary}

\theoremstyle{definition}
\newtheorem{definition}[theorem]{Definition}
\theoremstyle{plain}
\newtheorem{remark}{Remark}
\newtheorem{note}{Note}
\title{A technique for solving the polygon inclusion problems}
\author{Kai Jin, Taikun Zhu, Ruixi Luo}
\date{}

\newcommand{\area}{\mathsf{area}}
\newcommand{\peri}{\mathsf{perimeter}}
\newcommand{\Kill}{\mathsf{Kill}}
\newcommand{\List}{\mathsf{List}}
\newcommand{\OPT}{\mathsf{a}}
\newcommand{\D}{\mathsf{D}}
\newcommand{\U}{\mathsf{U}}
\newcommand{\SSS}{\mathcal{S}}
\newcommand{\hyperbola}{h}
\newcommand{\Opt}{\mathsf{opt}}

\begin{document}

\maketitle

\begin{abstract}
We propose a technique called Rotate-and-Kill for solving the polygon inclusion and circumscribing problems. By applying this technique, we obtain $O(n)$ time algorithms for computing (1) the maximum area triangle in a given $n$-sided convex polygon $P$, (2) the minimum area triangle enclosing $P$, (3) the minimum area triangle enclosing $P$ touching edge-to-edge, i.e. the minimum area triangle that is the intersection of three half-planes out of the $n$ half-planes defining $P$, and (4) the minimum perimeter triangle enclosing $P$ touching edge-to-edge.

Our algorithm for computing the maximum area triangle is simpler than the alternatives given in [Chandran and Mount, IJCGA'92] and [Kallus, arXiv'17]. Our algorithms for computing the minimum area or perimeter triangle enclosing $P$ touching edge-to-edge improve the $O(n\log n)$ or $O(n\log^2n)$ time algorithms given in [Boyce \emph{et al.}, STOC'82], [Aggarwal \emph{et al.}, Algorithmica'87], [Aggarwal and J. Park., FOCS'88], [Aggarwal \emph{et al.}, DCG'94], and [Schieber, SODA'95].

\medskip\textbf{keywords.} Polygon inclusion problem, Convex Polygon, Geometric Optimization.
\end{abstract}

\theoremstyle{plain}
\newtheorem{observation}[theorem]{Observation}

\section{Introduction}\label{sect:introduction}

The problems of computing extremal figures inside or outside a polygon region introduced in \cite{Shamos-CG-dissertation,kgon82, kgonE-area-FOCS84, linear-wrong-DS} have been studied extensively over the past four decades due to their enormous applications in operation research, shape recognition, and shape approximation \cite{3d-Suri-JC02,app-shape-approx-1,app-shape-approx-2,app-shape-class-1,app-shape-class-2,app-bio}.
The most typical and fundamental problems in this category include: computing
(1) the maximum (area / perimeter) $k$-gon inside a given convex polygon $P$;
(2) the minimum (area / perimeter) $k$-gon enclosing $P$; and
(3) the minimum (area / perimeter) $k$-gon that is the intersection of $k$ half-planes out of the half-planes defining $P$,
which is also known as the minimum \emph{all-flush} $k$-gon enclosing $P$, where all-flush means that
each edge of the $k$-gon contains an edge of $P$ as a subregion.

Boyce \emph{et al.} \cite{kgon82} showed how to find the maximum $k$-gon in two phases:
In phase~1, choose a vertex $v$ of $P$ and find the maximum $k$-gon $Q$ in $P$ that is rooted at $v$
(which means that it admits $v$ as one of its corner).
Assume $v_1,\ldots,v_n$ are the $n$ vertices of $P$ in clockwise order, and assume $Q=(v_{i_1}=v,v_{i_2},\ldots,v_{i_k})$.
Let $Q_{v'}=(v_{j_1}=v',v_{j_2},\ldots,v_{j_k})$ denote the maximum $k$-gon rooted at $v'$ for vertex $v'$ between $v_{i_1}$ and $v_{i_2}$.
It is proved in \cite{kgon82} that $Q_{v'}$ \emph{interleaves} $Q_v$, namely, $v_{j_x}$ lies between $v_{i_x}$ and $v_{i_{x+1}}$ for every $x$.
In Phase~2, based on the interleaving property and by a binary search,
compute $Q_{v'}$ for each $v'$ between $v_{i_1}$ and $v_{i_2}$;
the largest among these rooted $k$-gons is the maximum $k$-gon in $P$.
This method also works in computing the minimum all-flush $k$-gon \cite{kgon82}.

The two phases were solved in $O(kn\log n)$ and $O(n\log^2 n)$ time respectively in \cite{kgon82},
and were later solved in $O(kn)$ and $O(n\log n)$ time respectively by Aggarwal \emph{et al.} \cite{kgon87}, as applications of their \emph{Matrix-Searching technique}.
Moreover, Aggarwal \emph{et al.} \cite{kgon94} solved phase~1 in $O(n\sqrt{k\log n})$ time using their technique for computing the \emph{minimum weight $k$-link path}, and
this bound was further improved by Schieber \cite{kgon95soda} who optimized the $k$-link path technique.
These bounds on phase~1 are better than $O(kn)$ when $k=\Omega(\log n)$, but the $O(kn)$ time algorithm is optimal for $k=O(1)$.
To our best, there is no improvement on phase~2 over the $O(n\log n)$ bound even for $k=3$.

Other approaches were proposed for solving the maximum area $k$-gon problem, and the problem was solved in $O(n)$ time for $k\leq 4$.
Observe that each diagonal of the maximum area $4$-gon connects an antipodal pair (pair $v_i,v_j$ is an \emph{antipodal pair} if it admits parallel lines of support).
It is not difficult to compute the maximum area $4$-gon in $O(n)$ time after preprocessing the antipodal pairs (which takes $O(n)$ time \cite{Shamos-CG-dissertation,rotatingcaliper}).
Chandran and Mount \cite{linear-correct} gave the first (correct) linear time algorithm (\emph{Alg-CM}) for finding the maximum area triangle, which
actually computes the maximum triangle in $P$ and the minimum triangle enclosing $P$ simultaneously.
Kallus \cite{Kallus17b} recently posted another linear time algorithm (\emph{Alg-K}) for the maximum triangle problem.
Both Alg-CM and Alg-K are clever applications of the well-known \emph{Rotating-Caliper technique} \cite{rotatingcaliper}.
However, neither of them extend to the perimeter case, nor to the case of finding the minimum area all-flush triangle.
As a side note, Alg-K turns out to be essentially equivalent to Alg-CM; we discuss this in section~\ref{sect:compare}.

Before Chandran and Mount \cite{linear-correct}, Dobkin and Snyder \cite{linear-wrong-DS} gave an algorithm (\emph{Alg-DS}) for computing the maximum area triangle, which has been found incorrect by Hoog \emph{et al.} \cite{linear-wrong-IPL}.

O'Rourke \emph{et al.} \cite{Tri-Enclose-Area} gave the first linear time algorithm for computing the minimum area triangle enclosing $P$,
which improves over an $O(n\log^2n)$ time algorithm of \cite{Tri-Enclose-Area-nlogn2}.
Alg-CM is an extension of this linear algorithm.
The minimum perimeter enclosing triangle is also solved in $O(n)$ time \cite{Tri-Enclose-Peri}.
For general $k$, \cite{kgon-FOCS88} computed the minimum area $k$-gon in $O(kn+n\log n)$ time,
whereas \cite{kgonE-peri-IPL08} computed the minimum perimeter $k$-gon in much larger polynomial time.

\smallskip In this paper, we propose a unified technique (which we call Rotate-and-Kill) for solving all the three problems (1) -- (3) for the case $k=3$.
As its prototype applications, we design linear time algorithms for finding the maximum area triangle in $P$,
the minimum area triangle enclosing $P$,
the minimum area all-flush triangle, and
the minimum perimeter all-flush triangle.
We remark that the previously best known algorithms for finding the minimum area / perimeter all-flush triangle
take nearly linear time \cite{kgon82,kgon87,kgon-FOCS88,kgon94,kgon95soda}, that is, $O(n\log n)$ or $O(n\log^2n)$ time.

Subsection~\ref{subsect:techover} presents a technique overview of Rotate-and-Kill,
from which we can see this technique is different from the Rotating-Caliper technique (see discussions in subsection~\ref{subsect:difference}).
One major difference is that the Rotating-Caliper uses only one parameter (e.g. some angle $\theta$), whereas
the Rotate-and-Kill uses two (which are not functions of some hidden $\theta$).

In fact, the Rotate-and-Kill technique is powerful in solving other polygon inclusion / circumscribing problems,
but the running time is sometimes more than $O(n)$. The reason is as follows.
The inclusion / circumscribing problems usually admit the property that the set of locally optimal solutions are pairwise interleaving \cite{kgon82}. Once this property is admitted and $k=3$, we show that
an iteration process (also referred to as \emph{Rotate-and-Kill}) can be applied for searching all the locally optimal solutions, and thus find the optimum solution.
The mentioned iteration process will be conducted by a specialized function $\Kill$ (which we call the \emph{killing function}), and the running time of the iteration process is $O(n\cdot t_\Kill)$, where $t_\Kill$ denotes the expected running time of the function $\Kill$. It is usually not too difficult to give a killing function that runs in $O(\log n)$ time.
Toward a linear time algorithm, however, we need a killing function with $t_\Kill=O(1)$, which becomes challenging.

There seems to be no unified way for designing such a constant time killing function.

\subparagraph{Other related work.}
Melissaratos and Souvaine \cite{shortestpathforGO} computed the maximum area triangle in a simple polygon.
Zhou and Suri~\cite{3d-Suri-JC02} and Vivien and Wicker~\cite{3d-Vivien-CGTA04} computed extremal polytopes in three dimensional space.
Brass and Na \cite{BRASS10} computes the maximum intersection of $k$ half-planes out $n$ half-planes in arbitrary position.
Computing extremal equilateral triangles, squares, rectangles, parallelograms, disks, and ellipses in or outside a given region received attention from many researchers; see references in \cite{Jin15}.


\medskip The extremal triangle problems have found applications in
collision detection \cite{kgon82,3d-Suri-JC02},
shape approximation and convexity measuring, \cite{3d-Suri-JC02,app-shape-approx-1,app-shape-approx-2},
and shape classification \cite{app-shape-class-1,app-shape-class-2}.
In particular, computing the maximum area triangle serves as a subroutine in \cite{app-shape-approx-1,app-shape-class-1,app-shape-class-2}.

These problems have also found applications in operation design (VLSI design / robot motion plan),
Earth science \cite{3d-Suri-JC02}, and bioinformatics \cite{app-bio}.

It is shown in \cite{Jin15} that the simplest case of the Heilbronn triangle problem
reduces to finding the maximum area triangle and the maximum area parallelogram inside a polygon $P$.

\subsection{Preliminary: 3-stable, G-3-stable, and F-3-stable triangles}\label{subsect:preliminary}

Let $v_1,\ldots,v_n$ be a \emph{clockwise} enumeration of the \emph{vertices} of $P$.
Assume that any three vertices are not collinear.
Denote by $e_1,\ldots,e_n$ the $n$ \emph{edges} of $P$, such that $e_i$ is the line segment connecting $v_i$ and $v_{i+1}$ (where $v_{n+1}=v_1$).
Throughout, regard $P$ as a polygon region which contains its interior and boundary.
Denote by $\partial P$ the boundary of $P$.
Denote by $p_i$ the (closed) half-plane delimited by the extended line $\ell_i$ of $e_i$ and contains $P$.
Denote by $\area(\phi)$ the area of any region $\phi$.
For distinct points $X$ and $Y$, denote by $\overleftrightarrow{XY}$ and $\overrightarrow{XY}$
the \emph{line} connecting $X$ and $Y$ and the \emph{directed line segment} from $X$ to $Y$, respectively.
Given points $X,X'$ on the boundary of $P$,
denote by $[X\circlearrowright X']$ the boundary portion of $P$ that starts from $X$ and clockwise to $X'$ which contains its two endpoints $X$ and $X'$
($[X\circlearrowright X]=X$).

Consider any triangle $T=\triangle X_1X_2X_3$ with $X_1,X_2,X_3$ lying in $\partial P$ and lying in clockwise order.
For $i\in \{1,2,3\}$, corner $X_i$ is \emph{stable} in $T$ if $X_i$ has the largest distance to $\overleftrightarrow{X_{i+1}X_{i-1}}$
among all points in $P$ that lie on the right of $\overrightarrow{X_{i+1}X_{i-1}}$ (subscripts taken modulo 3).
Throughout, when we write $\triangle X_1X_2X_3$, we assume that \emph{$X_1,X_2,X_3$ lie in clockwise order}.
\begin{description}
	\item[3-stable.] Consider three vertices $v_i,v_j,v_k$ of $P$ in clockwise order, which forms a triangle $T=\triangle v_iv_jv_k$.
	This triangle is \emph{3-stable} if all its three corners $v_i,v_j,v_k$ are stable in $T$.
	\item[G-3-stable.] Consider any triangle $T=\triangle X_1X_2X_3$ with $X_1,X_2,X_3$ lying in $\partial P$ and lying in clockwise order.
	It is \emph{G-3-stable} if all its three corners $X_1,X_2,X_3$ are stable in $T$.
	
	\item[All-flush.] For distinct edges $e_i,e_j,e_k$ that are in clockwise order, the intersecting region $p_i\cap p_j\cap p_k$ is denoted by $\triangle e_ie_je_k$ and is called an \emph{all-flush triangle}.
	Throughout, when we write $\triangle e_ie_je_k$, we assume that \emph{$e_i,e_j,e_k$ are distinct and lie in clockwise order}.
	
	\item[F-3-stable.] For triangle $T=\triangle e_ie_je_k$ with a finite area,
	$e_i$ is \emph{stable} in $T$ if no all-flush triangle $\triangle e_{i'}e_je_k$ has a smaller area than $T$;
	$e_j$ is \emph{stable} in $T$ if no all-flush triangle $\triangle e_ie_{j'}e_k$ has a smaller area than $T$; and
	$e_k$ is \emph{stable} in $T$ if no all-flush triangle $\triangle e_ie_je_{k'}$ has a smaller area than $T$.
	Moreover, $T$ is \emph{F-3-stable} if $e_i,e_j,e_k$ are all stable in $T$.
\end{description}

\begin{definition}[\cite{kgon82}]\label{def:interleaving}
	Triangles $\triangle A_1A_2A_3$ and $\triangle B_1B_2B_3$ with all corners lying in $\partial P$
	are \emph{interleaving} if
	we can permutate $A_1,A_2,A_3,B_1,B_2,B_3$ to $Z_1,\ldots,Z_6$ so that $Z_1,\ldots,Z_6$ lie in clockwise order (in a non-strict manner; so neighbors may be identical), in which $\{Z_1,Z_3,Z_5\}=\{A_1,A_2,A_3\}$ and $\{Z_2,Z_4,Z_6\}=\{B_1,B_2,B_3\}$.
	Two all-flush triangles $\triangle e_re_se_t$ and $\triangle e_ie_je_k$ are \emph{interleaving} if
	we can permutate $r,s,t,i,j,k$ to $z_1,\ldots,z_6$ so that $e_{z_1},\ldots,e_{z_6}$ lie in clockwise order (in a non-strict manner; so neighbors may be identical), in which $\{z_1,z_3,z_5\}=\{r,s,t\}$ and $\{z_2,z_4,z_6\}=\{i,j,k\}$.
\end{definition}

\begin{lemma}\label{lemma:interleaving}
	1. \cite{linear-wrong-IPL} The 3-stable triangles are pairwise interleaving.
	
	2. The F-3-stable triangles are pairwise interleaving.
	
	3. The G-3-stable triangles are pairwise interleaving.
\end{lemma}

\begin{corollary}\label{corol:number_n}
	The number of 3-stable or F-3-stable triangles is $O(n)$.
\end{corollary}

Easy proofs of Lemma~\ref{lemma:interleaving} and Corollary~\ref{corol:number_n} can be found in appendix~\ref{sect:misc}.


\subsection{Technique overview}\label{subsect:techover}

It is easy to compute one 3-stable triangle in $O(n)$ time; we show how to do this in section~\ref{sect:one}\footnote{Alg-DS fails to find one 3-stable triangle and so we introduce the algorithm in section~\ref{sect:one}. This algorithm in section~\ref{sect:one} is not the same as and does not originate from Alg-DS (see appendix~\ref{subsect:diff-Alg-DS}).}.
Denote the computed 3-stable triangle by $\triangle v_rv_sv_t$ and assume $r,s,t$ are given in the following.

Due to the interleaving property of 3-stable triangles (Lemma~\ref{lemma:interleaving}),
these triangles can be listed as $\triangle v_{a_1}v_{b_1}v_{c_1},\ldots,\triangle v_{a_m}v_{b_m}v_{c_m}$, where
$v_{a_1}$, \ldots, $v_{a_m}$, $v_{b_1}$, \ldots, $v_{b_m}$, $v_{c_1}$, \ldots, $v_{c_m}$ lie in clockwise order (neighbors may be identical). Assume $(a_1,b_1,c_1)=(r,s,t)$ without loss of generality.
Starting from the 3-stable triangle $\triangle v_rv_sv_t$, we search all the 3-stable triangles one by one by iterations.
During this iteration process, maintain variables $a,b,c$, which are initially assigned to $r,s,t$ respectively.
At each iteration, select one of $a,b,c$ and increase it (increasing $x$ means that $x\leftarrow x \mod n + 1$).
By carefully selecting the variable to increase at each iteration,
which will be elaborated right below,
we make sure that $\triangle v_av_bv_c$ arrives at each 3-stable triangle $\triangle v_{a_1}v_{b_1}v_{c_1}$, $\ldots$, $\triangle v_{a_m}v_{b_m}v_{c_m}$ one after another,
and that it takes  $O(a_i-a_{i-1} + b_i-b_{i-1}+c_i-c_{i-1})$ iterations
to arrive $\triangle v_{a_i}v_{b_i}v_{c_i}$ from $\triangle v_{a_{i-1}}v_{b_{i-1}}v_{c_{i-1}}$ for each $i>1$.
Applying the fact that $v_{a_1}$, \ldots, $v_{a_m}$, $v_{b_1}$, \ldots, $v_{b_m}$, $v_{c_1}$, \ldots, $v_{c_m}$ lie in clockwise order, we see the searching process takes only $O(n)$ iterations.
Furthermore, it takes $O(n)$ time if the selection at each iteration took amortized $O(1)$ time.
By finding all the 3-stable triangles, we immediately obtain the maximum area triangle in $P$.

How to select among $a,b,c$ to increase is essential to the algorithm.
We must make decision rapidly yet make sure that no 3-stable triangle is missing.
This is difficult. Two previous algorithms, Alg-DS (FOCS'79) and an algorithm given in \cite{kgon82} (STOC'82),
were based on the above approach and they presented two simple methods for selection.
Yet their selection methods would miss 3-stable triangles and even worse,
sometimes they miss the largest 3-stable triangle, as shown by \cite{linear-wrong-IPL}.

We come up with a different selection method which is sketched in the following.
Above all, we state that a vertex pair $(v_j,v_k)$ is \emph{dead} if there is no $v_i$ such that $\triangle v_iv_jv_k$ is 3-stable.

Assume $b\in \{s,\ldots,t\}$ and $c\in \{t,\ldots,r\}$.
The interleaving property of 3-stable triangles implies that
(I) $(v_b,v_{c+1}),\ldots,(v_b,v_r)$ are dead \textbf{or}
(II) $(v_{b+1},v_c),\ldots,(v_t,v_c)$ are dead.
(Otherwise, there exist two 3-stable triangles that are not interleaving;
see an easy proof below Observation~\ref{obs:dead-I-II}.)

Suppose for the time being there is a function $\Kill(b,c)$ defined for each $(b,c)$ such that $b\in \{s,\ldots,t\}$ and $c\in \{t,\ldots,r\}$,
which returns either `b' or `c' and which owns the following properties:
\textcircled{1} \emph{it returns `b' only if (I) hold and returns `c' only if (II) hold}.
\textcircled{2} \emph{it costs amortized $O(1)$ time to compute}.

By utilizing this function, we can select among $a,b,c$ to increase at each iteration as follows.
If $\Kill(b,c)=$`b', select $b$ to increase; if $\Kill(b,c)=$`c', select $c$ to increase.
In the subsequent iterations, select $a$ to increase until $v_a$ has the largest distance to $\overleftrightarrow{v_bv_c}$.
The running time is clearly $O(n)$ due to \textcircled{2},
and the correctness easily follows from property \textcircled{1}, as we will analyze in section~\ref{sect:3-stable}.

The process for searching the 3-stable triangles as above is referred to as a \emph{Rotate-and-Kill process}.
Obviously, its kernel lies in the function $\Kill$, which we call the \emph{killing function}.

\begin{figure}[h]
	\centering \includegraphics[width=\textwidth]{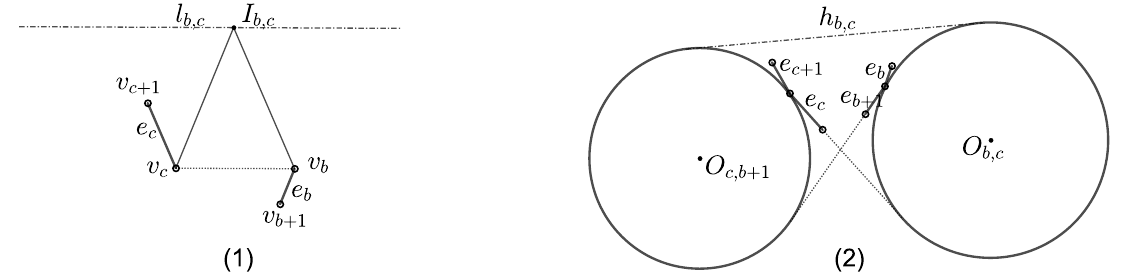}
	\caption{Illustration of the killing functions.}\label{fig:sketch-KF}
\end{figure}

\subparagraph{Design of killing function $\Kill$.}
There exists an surprisingly simple design of $\Kill(b,c)$.
Assume the line parallel to $e_{c}$ at $v_{b}$ and the line parallel to $e_{b}$ at $v_{c}$ intersect at $I_{b,c}$.
Make a parallel line $\ell_{b,c}$ of $v_{b}v_{c}$ at $I_{b,c}$, as shown in Figure~\ref{fig:sketch-KF}~(1).
Define $\Kill(b,c)=$`b' if $v_a$ is below $\ell_{b,c}$ and $\Kill(b,c)=$`c' otherwise, where $v_a$ denotes the vertex of $P$
on the right of $\overrightarrow{v_bv_c}$ furthest to $\overleftrightarrow{v_bv_c}$.
The proof that this definition of $\Kill(b,c)$ does satisfy \textcircled{1} and \textcircled{2} will be presented in subsection~\ref{subsect:kill}.

\subsubsection{Minimum perimeter all-flush triangle (technique overview)}

In the following, we sketch how to apply the Rotate-and-Kill process to solve a polygon circumscribing problem, that is,
finding the all-flush triangle with the minimum perimeter.
The details are given in section~\ref{sect:MPFT}.

We say $e_i$ is \emph{stable} in $\triangle e_ie_je_k$ if $\peri (\triangle e_ie_je_k)$ cannot be reduced by changing $e_i$ to other edges.
We state that $\triangle e_ie_je_k$ is \emph{3-stable}, if $e_i,e_j,e_k$ are all stable in $\triangle e_ie_je_k$.
Pair $(e_i,e_j)$ is \emph{dead} if there is no $e_k$ such that $\triangle e_ie_je_k$ is 3-stable.
A triangle refers to an all-flush triangle in the following.
We first compute a 3-stable triangle $\triangle e_re_se_t$. This can be done using the ideas given in section~\ref{sect:one}.

All the 3-stable triangles are pairwise interleaving (see Lemma~\ref{lemma:MFPT-interleaving}, which is similar to Lemma~\ref{lemma:interleaving}).
As a corollary, for $b\in \{s,\ldots,t\}$ and $c\in \{t,\ldots,r\}$,
at least one of the following holds: (I) $(e_b,e_{c+1}),\ldots,(e_b,e_r)$ are dead \textbf{or}
(II) $(e_{b+1},e_c),\ldots,(e_t,e_c)$ are dead.

To apply Rotate-and-Kill, the challenging part lies in designing a killing function $\Kill(b,c)$ that satisfies \textcircled{1} and \textcircled{2}.
Unfortunately, we do not know how to design such a function.
Instead, we provide a function $\Kill_p(b,c)$ satisfying some weaker properties to be elaborated below.
This function is insufficient for finding all 3-stable triangles, yet sufficient for finding the minimum perimeter (all-flush) triangle.

\subparagraph{Design of killing function $\Kill_p$.}
See Figure~\ref{fig:sketch-KF}~(2). Make a circle $O_{c,b+1}$ tangent to $\ell_c,\ell_{c+1},\ell_{b+1}$ on the right of $e_{b+1}$ and left of $e_c,e_{c+1}$, and a circle $O_{b,c}$ tangent to $\ell_b,\ell_{b+1},\ell_c$ on the right of $e_c$ and left of $e_b,e_{b+1}$.
Moreover, find their common tangent $h_{b,c}$ (that intersects the ray $\overrightarrow{v_cv_{c+1}}$) as shown in the picture.
We define $\Kill_p(b,c)=$`b' if $e_a$ is (entirely) below $h_{b,c}$, and $\Kill_p(b,c)=$`c' otherwise,
where $a$ denotes the index in $\{a\mid e_a\prec e_b,e_c\prec e_a\}$ that minimizes $\peri(\triangle e_ae_be_c)$.

We say $(e_i,e_j)$ is \emph{DEAD} if there is no $e_k$ such that $\triangle e_ie_je_k$ is the minimum perimeter triangle
(among all all-flush triangles). Note that``dead'' implies ``DEAD'' but the reverse is not true.
We prove (in Lemma~\ref{lemma:MFPT-kill-function}) that
if $\Kill_p(b,c)$ returns `c', (II) holds; and if $\Kill_p(b,c)$ returns `b',
$(e_b,e_{c+1}),\ldots,(e_b,e_r)$ are DEAD (this is weaker than (I)).
Then, during the Rotate-and-Kill process, the pair $(e_b,e_c)$ will meet all pairs that are not DEAD, which implies that the algorithm finds the minimum perimeter (all-flush) triangle. However, it is not guaranteed that all the pairs that are not dead will be meet,
so the algorithm possibly misses some 3-stable triangles.

\subsubsection{More applications of Rotate-and-Kill.}
In section~\ref{sect:F-3-stable}, we compute the F-3-stable triangles in linear time by a Rotate-and-Kill process,
and thus find the minimum area all-flush triangle.
In section~\ref{sect:G-3-stable}, we compute the G-3-stable triangles in linear time by a Rotate-and-Kill process,
and thus find the minimum area triangle enclosing $P$ (see an explanation at the end of section~\ref{sect:G-3-stable}).

As we will see, the killing functions in these problems are also simple. Nevertheless, it requires painstaking effort and creativity to derive the killing functions.

It remains to be investigated how to design killing functions for other related problems.

\subsubsection{Rotate-and-Kill technique v.s. Rotating-Caliper technique}\label{subsect:difference}

An application of Toussaint's Rotating-Caliper (RC) \cite{rotatingcaliper} technique usually adopts the following framework:
choose a variable $\theta$ (e.g., a direction over a discrete space),
define some objects (e.g. $A_\theta$, $B_\theta$, and $C_\theta$) as functions of $\theta$, and
prove that (i) each of $A_\theta,B_\theta,C_\theta$ has a monotonicity property with respect to $\theta$,
and then design an algorithm for updating $A_\theta,B_\theta,C_\theta$ after each change of $\theta$.
Updating $A_\theta,B_\theta,C_\theta$ should be taken care in amortized $O(1)$ time.
Sometimes, however, this is not that easy, even if condition~(i) holds. For example, Klee and Laskowski \cite{Tri-Enclose-Area-nlogn2} spent $O(\log^2n)$ time for updating $B_\theta$ or $C_\theta$ in finding the minimum area enclosing triangle based on RC.
By using a nontrivial speedup technique,
O'Rourke \emph{et al.} \cite{Tri-Enclose-Area} further showed that for the minimum area enclosing triangle problem,
$B_\theta,C_\theta$ effect each other and that $(B_\theta,C_\theta)$ as a unity can be updated together in amortized $O(1)$ time.
Nevertheless, O'Rourke \emph{et al.} did not break the framework of RC.

In Rotate-and-Kill, a pair of variables $(b,c)$ are maintained, which as a unity serves as a \emph{2-dimensional variable}.
Unlike RC, $b,c$ here are not functions of another variable $\theta$.
On the contrary, we may need to use a function $\theta$ of variable $(b,c)$; see the description of $\Kill_F$ in subsection~\ref{subsect:Kill-F-MFTA} for an example.
As such, the flow of Rotate-and-Kill is different from RC.


\subsubsection{Comparison to Alg-K and Alg-CM}\label{subsect:compareKKK}

We compare our algorithm (\emph{Alg-A}) for finding the maximum area triangle
with its two alternatives Alg-CM and Alg-K in section~\ref{sect:compare}.
We think Alg-A is better in almost every aspect. This is because it is essentially simpler.
Among other merits, Alg-A is much \textbf{faster}, because it has a smaller constant behind the asymptotic complexity $O(n)$ than the others:
\begin{description}
	\item[1. Less candidates.] Alg-A computes at most $n$ candidate triangles (proof is trivial and omitted) whereas Alg-CM computes at most $5n$ triangles (proved in \cite{linear-correct}) and so as Alg-K.
	(by experiment, Alg-CM and Alg-K have to compute roughly $4.66 n$ candidate triangles.)
	\item[2. Simpler primitives.]
	Alg-A has simpler primitives because (1) the candidate triangles considered in it have all corners lying on $P$'s vertices and (2) searching the next candidate from a given one is much easier -- the code length for this is \textbf{1:7} in Alg-A and in Alg-CM.
\end{description}

More importantly, Alg-A is \textbf{more stable} and the brief reason is as follows.
Alg-A has no accumulation of inaccuracy; it does not compute but reads the constant coordinates of the corners (which are on vertices) every iteration.
See a detailed explanation in section~\ref{sect:compare}.

We implement all algorithms by C++ programs and test them on random convex polygons \cite{implementation}.
Our experiment result shows that Alg-A is much faster and more stable than the others;
see section~\ref{subsect:experiment}.
In particular, Alg-CM suffers from float-issue as it accumulates inaccuracy. If Alg-CM is implemented based on the original paper, it encounters problems when $n$ arrives at 300. With optimizations such as avoiding trigonometric functions, it can merely handle $n=10000$. The implementation of Alg-A easily handles $n$ as large as 10000000.

\section{Find all 3-stable triangles}\label{sect:3-stable}

In this section, assuming that we are given indices $r,s,t$ so that $\triangle v_rv_sv_t$ is 3-stable,
we show how to compute all the 3-stable triangles by a Rotate-and-Kill process.

Recall that a vertex pair $(v_j,v_k)$ is \emph{dead} if there is no 3-stable triangle $\triangle v_iv_jv_k$.

We start with an observation which is an immediate corollary of Lemma~\ref{lemma:interleaving}~part~1.

\begin{observation}\label{obs:dead-I-II}
	Given $(b,c)$ where $b \in \{s,\ldots, t\}$, $c \in \{t, \ldots, r\}$,
	at least one of the following holds.
	
	(I) $(v_b,v_{c+1}),\ldots,(v_b,v_r)$ are dead. \qquad (II) $(v_{b+1},v_c),\ldots,(v_t,v_c)$ are dead.
\end{observation}

\begin{figure}[h]
	\begin{minipage}[b]{.37\textwidth}
		\includegraphics[width=.9\textwidth]{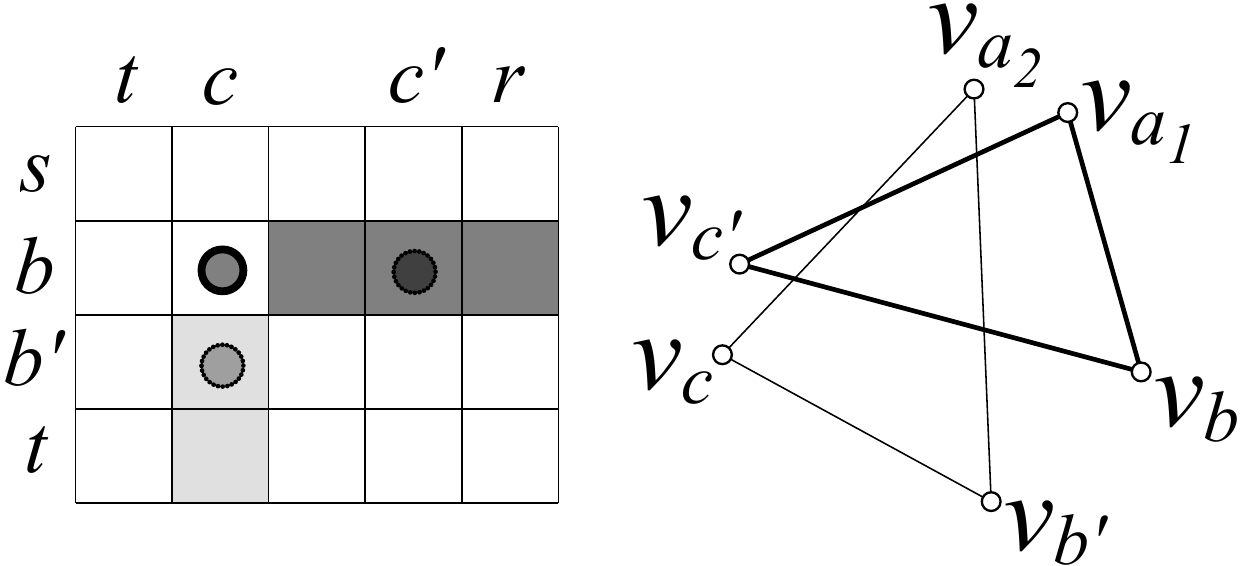}
		\caption{Proof of Observation~\ref{obs:dead-I-II}.}\label{fig:dead-I-II}
	\end{minipage}
	\begin{minipage}[b]{.61\textwidth}
		\includegraphics[width=\textwidth]{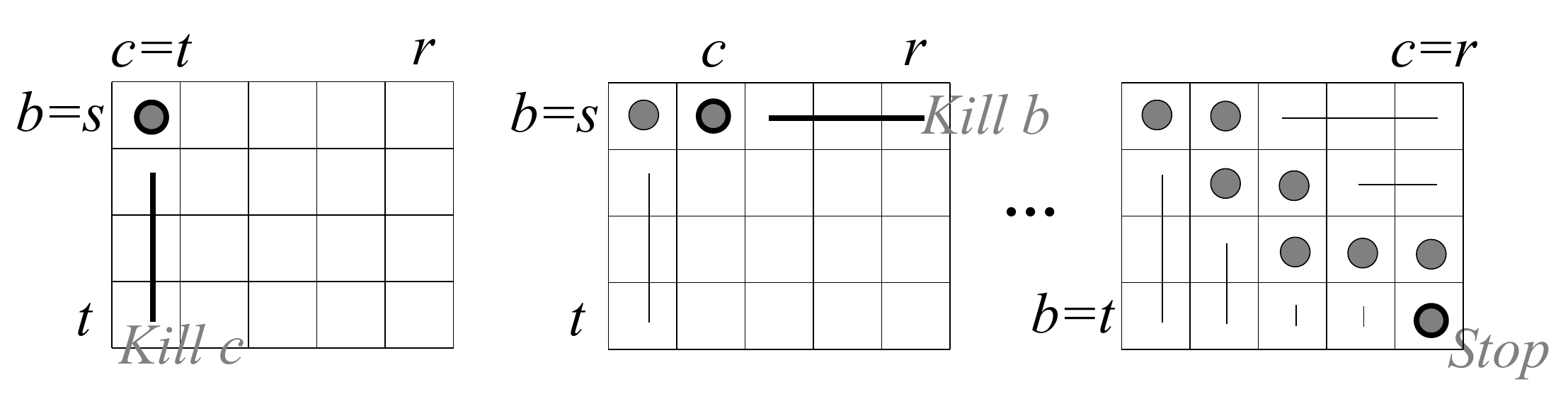}
		\caption{Illustration of the Rotate-and-Kill process.}\label{fig:example-RK}
	\end{minipage}
\end{figure}

\begin{proof}
	Suppose both (I) and (II) fail. This implies $(v_b,v_{c'})$ and $(v_{b'},v_c)$ which are not dead, where $c'\in \{c+1,\ldots, r\}$
	and $b'\in \{b+1,\ldots,t\}$, which implies
	3-stable triangles $\triangle v_{a_1}v_bv_{c'}$ and $\triangle v_{a_2}v_{b'}v_c$ that are not interleaving (see Figure~\ref{fig:dead-I-II}), contradicting Lemma~\ref{lemma:interleaving}~part~1.
\end{proof}

Assume for the time being that we have a function (or oracle) $\Kill$, which inputs parameters $(b,c)$, where $b\in \{s,\ldots,t\}$ and $c\in \{t,\ldots,r\}$, and which has the following two properties:

\textcircled{1} \emph{It equals `b' or `c', and it equals b' only if (I) holds and equals `c' only if (II) holds.}

\textcircled{2} \emph{It can be computed in amortized $O(1)$ time.}

\paragraph*{The Rotate-and-Kill process (sketch).}
Initially, set $(b,c)=(s,t)$.
In each iteration, output all 3-stable triangles $\triangle XYZ$ with $Y=v_b$ and $Z=v_c$ (to be elaborated right below),
and call $\Kill(b,c)$.
If this function returns `b', set $b\leftarrow b+1$.
If it returns `c', set $c\leftarrow c+1$.
If $(b,c)\neq (t,r)$, go to the next iteration; otherwise, terminate the process.
See Figure~\ref{fig:example-RK} for an illustration.
(In this description of Rotate-and-Kill, the aforementioned variable $a$ is hidden. But we will see it later.)

\begin{description}
	\item[Correctness.]
	Since $\triangle v_sv_tv_r$ is 3-stable, $(v_t,v_r)$ is not dead.
	Therefore,  $\Kill(b,c)$ cannot return `b' when $b=t$ and $c\neq r$, and
	cannot return `c' when $c=r$ and $b\neq t$, due to property \textcircled{1}.
	This means $(b,c)$ eventually reaches $(t,r)$. So the process terminates.
	Moreover, let $\U=\{(v_j,v_k) \mid j\in \{s,\ldots,t\},k\in \{t,\ldots,r\}\}$.
	As another corollary of property \textcircled{1},
	\emph{for each pair $(v_j,v_k)$ in $\U$ that is not dead, it holds that $(b,c)=(j,k)$ in some iteration}.
	Consider an arbitrary 3-stable triangle $\triangle v_iv_jv_k$. By Lemma~\ref{lemma:interleaving}, it interleaves $\triangle v_rv_sv_t$.
	Therefore, without loss of generality we can assume $(v_j,v_k)\in \U$.
	Since $\triangle v_iv_jv_k$ is 3-stable, $(v_j,v_k)$ is not dead.
	In the iteration where $(b,c)=(j,k)$, the 3-stable triangle $\triangle v_iv_jv_k$ will be output.
\end{description}

It remains to be shown how do we compute 3-stable triangles $\triangle XYZ$ with $Y=v_b$ and $Z=v_c$ (for those $(b,c)$ that will be visited by the Rotate-and-Kill process).
This reduces to finding the vertex with the largest distance to $\overleftrightarrow{v_bv_c}$ on the right of $\overrightarrow{v_bv_c}$,
which costs amortized $O(1)$ time since the direction of $\overrightarrow{v_bv_c}$ monotonously rotates in clockwise throughout the Rotate-and-Kill process.

Further according to property \textcircled{2}, the Rotate-and-Kill process runs in $O(n)$ time.

\subsection{Description of the $\Kill$ function}\label{subsect:kill}

The core of the above process lies in function $\Kill$, which is yet to be described.

Above all, it is easy to find a function $\Kill$ that satisfies \textcircled{1}. Here are two candidates:

\quad Candidate 1: determine whether (I) holds, and returns `b' if (I) holds and `c' otherwise.

\quad Candidate 2: determine whether (II) holds, and returns `c' if (II) holds and `b' otherwise.

Because (I) or (II) holds, \textcircled{1} is satisfied by both candidates.
Unfortunately, we do not know how to determine (I) (or (II)) efficiently in amortized $O(1)$ time.
(In fact, this can be done in $O(\log n)$ time using a binary search, yet we cannot afford $O(\log n)$ time.)

\newcommand{\R}{\mathcal{R}}

To introduce our killing function $\Kill$ that satisfies both \textcircled{1} and \textcircled{2}, we need some notations.

Regard each edge $e_i$ of $P$ as \emph{directed}. The direction of $e_i$ is from $v_i$ to $v_{i+1}$ ($v_{n+1}=v_1$).

\begin{definition}\label{def:R}
	Assume $v_j,v_k$ are distinct vertices of $P$. See Figure~\ref{fig:QIJKDef}.
	Let $r_1$ and $r_2$ respectively denote the ray originating from $v_j$ with the same direction as $e_{k-1}$ and $e_k$.
	Let $r'_1$ and $r'_2$ respectively denote the ray originating from $v_k$ with the opposite direction to $e_{j-1}$ and $e_j$.
	The intersection of the wedge defined by $r_1,r_2$ and the wedge defined by $r'_1,r'_2$ defines a region, denoted by $\R_{j,k}$.
	
	Denote by $H_{j,k},I_{j,k},J_{j,k},K_{j,k}$ the intersecting points between $r_1,r'_1$,
	between $r_2,r'_2$, between $r_1,r'_2$, and between $r_2,r'_1$, respectively.
	When two rays do not intersect, their intersecting point is defined as $\infty$.
	Therefore, $H_{j,k},I_{j,k},J_{j,k},K_{j,k}$ are always defined for $v_j\neq v_k$.
	
	\begin{figure}[h]
		\centering \includegraphics[width=\textwidth]{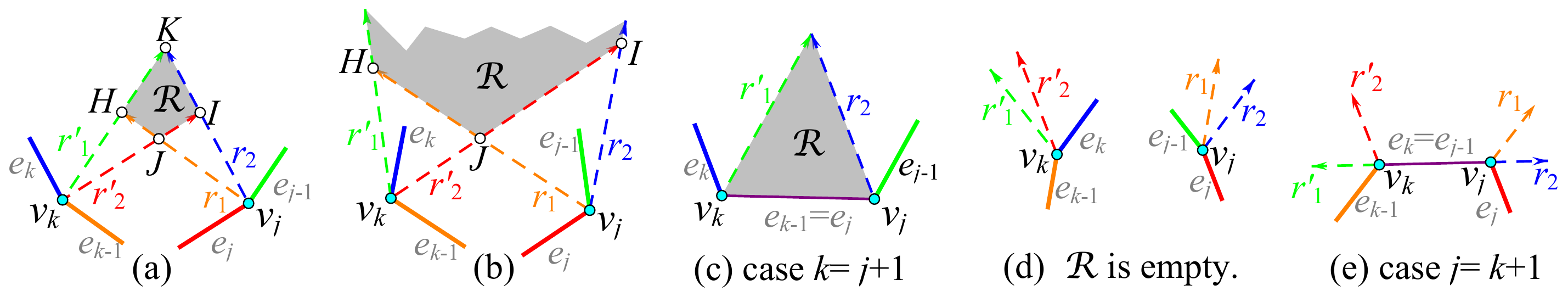}
		\caption{Illustration of Definition~\ref{def:R}. Subscripts of $\R,H,I,J,K$ are omitted for conciseness.}\label{fig:QIJKDef}
	\end{figure}
\end{definition}

\begin{algorithm}[b]
	\caption{The Rotate-and-Kill process for finding all 3-stable triangles.}\label{alg:all-r-k}
	$(a,b,c)\leftarrow (r,s,t)$; \qquad // $\triangle v_rv_sv_t$ is 3-stable.\\
	\Repeat{$(b,c)=(t,r)$}
	{
		\lWhile{$v_{a+1}>v_a$ in the distance to $\overleftrightarrow{v_bv_c}$}
		{$a\leftarrow a+1$ \label{code:A++}}
		Output $\triangle v_av_bv_c$ if it is 3-stable and output $\triangle v_{a+1}v_bv_c$ if it is 3-stable \label{code:output}\;
		\lIf{$v_a\leq I_{b,c}$ in the distance to $\overleftrightarrow{v_bv_c}$}
		{$b\leftarrow b+1$ \label{code:B++Key}}
		\lElse{$c\leftarrow c+1$ \label{code:C++Key}}
	}
\end{algorithm}

Our function $\Kill$ is as follows. Let $v_a$ be the vertex on the right of $\overrightarrow{v_bv_c}$ furthest to $\overleftrightarrow{v_bv_c}$.
\begin{equation}\label{eqn:def-Kill}
	\text{$\Kill(b,c)=$`b' if $v_a \leq I_{b,c}$ in the distance to $\overleftrightarrow{v_bv_c}$,
		and $\Kill(b,c)=$`c' otherwise.}
\end{equation}

We present a pseudo code of the entire Rotate-and-Kill process in Algorithm~\ref{alg:all-r-k}.

\begin{note}\label{note:rigorous}
	Since $I_{b,c}$ is undefined for $(b,c)=(t,t)$, the value of $\Kill(t,t)$ is unspecified. For convenience, we can define $\Kill(t,t)=$`c'.
	However, this is unnecessary, because we can prove that $(b,c)\neq (t,t)$ throughout the process.
	Suppose in some iteration $(b,c)=(t,t)$. Then, in the previous iteration, $(b,c)=(t-1,t)$ and $\Kill(b,c)=$`b'.
	For $(b,c)=(t-1,t)$, however, we have $c=b+1$ and so $I_{b,c}=v_{b}$, hence $v_a>I_{b,c}$ in the distance to $\overleftrightarrow{v_bv_c}$,
	which means $\Kill(b,c)=$ `c'. Contradiction.
\end{note}

\subsection{Analysis of the $\Kill$ function}

Function $\Kill$ defined in (\ref{eqn:def-Kill}) clearly satisfies \textcircled{2}.
Lemma~\ref{lemma:correct} states that it also satisfies \textcircled{1}.

\begin{lemma}\label{lemma:correct} Assume $b\in\{s,\ldots,t\},c\in \{t,\ldots, r\}$, $(b,c)\neq (t,t)$, and $v_{a}$ is defined as in (\ref{eqn:def-Kill}).
	\begin{enumerate}
		\item If $v_{a}\leq I_{b,c}$ in the distance to $\overleftrightarrow{v_bv_c}$, then $(v_b,v_{c+1}),\ldots,(v_b,v_r)$ are dead.
		\item If $v_{a}>   I_{b,c}$ in the distance to $\overleftrightarrow{v_bv_c}$, then $(v_{b+1},v_c),\ldots,(v_t,v_c)$ are dead.
	\end{enumerate}
\end{lemma}

We first give some observations and then use them to prove Lemma~\ref{lemma:correct}.

\begin{definition}
	For edges $e_i\neq e_j$, we state that $e_i$ is \emph{chasing} $e_j$, denoted by $e_i\prec e_j$, if $\ell_i$ intersects $\ell_j$ and the intersecting point lies between $e_i$ and $e_j$ clockwise.
	For example, in Figure~\ref{fig:QIJKDef}~(b), $e_{j-1}$ is chasing $e_j$ and $e_{k-1}$ but not $e_k$,
	whereas $e_k$ is chasing $e_{j-1}$ but not $e_j$.
\end{definition}

\begin{observation}\label{obs:R}
	\begin{enumerate}
		\item Fix $v_j$ and $v_k$~$(j\neq k)$. If there is a vertex $v_i$ on the right of $\overrightarrow{v_jv_k}$ such that $v_j,v_k$ are stable in $\triangle v_iv_jv_k$, then $v_i\in \R_{j,k}$. (Throughout, assume $\R_{j,k}$ contains its boundary.)
		\item A pair $(v_j,v_k)$ is dead if $\R_{j,k}$ does not intersect $P$.
		\item A pair $(v_j,v_k)$ is dead if $e_k\prec e_j$.
		\item A pair $(v_j,v_k)$ is dead if there exists a point $A$ in $P$ which lies on the right of $\overrightarrow{v_jv_k}$ and is further than $K_{j,k}$ in the distance to $\overleftrightarrow{v_jv_k}$. (If $K_{j,k}=\infty$, there could be no such $A$.)
	\end{enumerate}
\end{observation}

\begin{proof}
	Recall the four rays $r_1,r_2,r'_1,r'_2$ used in Definition~\ref{def:R}. See Figure~\ref{fig:QIJKDef}.
	
	1. Since $v_j$ needs to be stable, $v_i$ must lie in the wedge bounded by $r'_1$ and $r'_2$.
	Since $v_k$ needs to be stable, $v_i$ must lie in the wedge bounded by $r_1$ and $r_2$.
	Together, $v_i$ must lie in $\R_{j,k}$.
	
	\smallskip
	2. Assume $\R_{j,k}$ does not intersect $P$.
	Then, due to part~1 of this observation,
	there is no vertex $v_i$ on the right of $\overrightarrow{v_jv_k}$ such that $v_j,v_k$ are stable in $\triangle v_iv_jv_k$
	(otherwise $\R_{j,k}$ contains $v_i$ and hence intersects $P$).
	The nonexistence of such $v_i$ implies that $(v_j,v_k)$ is dead by the definition of dead.
	
	\smallskip
	3. Assume $e_k\prec e_j$. We first argue that $\R_{j,k}$ does not intersect $P$.
	This holds obviously when $j=k+1$; see Figure~\ref{fig:QIJKDef}~(e). We assume $j\neq k+1$ in the following.
	
	First, observe that (i) \emph{$\R_{j,k}$ does not contain $v_k$}. This is because
	the wedge bounded by $r_1,r_2$ cannot contain $v_k$.
	(This wedge contains $v_k$ only when $k=j+1$. Yet $e_k\prec e_j$ implies that $k\neq j+1$.)
	
	Second, observe that (ii) \emph{$\R_{j,k}$ does not intersect $P\setminus \{v_k\}$}.
	Since $e_k\prec e_j$ and $j\neq k+1$, rays $r'_1,r'_2$ are on the left of $e_k$ (except for the originate $v_k$ of $r'_1,r'_2$).
	So, the wedge bounded by $r'_1,r'_2$ is on the left of $e_k$ (except for $v_k$).
	Therefore, this wedge, including $\R_{j,k}$, does not intersect $P\setminus \{v_k\}$.
	
	Together, $\R_{j,k}$ does not intersect $P$. Hence $(v_j,v_k)$ is dead according to part~2.
	
	\smallskip
	4. Suppose to the opposite that $(v_j,v_k)$ is not dead but $\triangle v_iv_jv_k$ is 3-stable for some vertex $v_i$.
	By part~1 of this observation, $v_i$ is contained in $\R_{j,k}$.
	By the definition of $\R_{j,k}$ (see Figure~\ref{fig:QIJKDef}~(a)),
	this means $v_i$ is not further than $K_{j,k}$ in the distance to $\overleftrightarrow{v_jv_k}$.
	However, by the assumption, we know $A$ is further than $K_{j,k}$ in the distance to $\overleftrightarrow{v_jv_k}$.
	Therefore, $A$ is further than $v_i$ in the distance to $\overleftrightarrow{v_jv_k}$.
	This means that $v_i$ is not stable in $\triangle v_iv_jv_k$. Contradictory.
\end{proof}

\begin{proof}[Proof of Lemma~\ref{lemma:correct}]
	
	\textbf{1.} If $e_b$ is not chasing $e_c$, we get $e_{c+1}\prec e_b,\ldots,e_r\prec e_b$, which implies that $(v_b,v_{c+1}),...,(v_b,v_r)$ are dead (by Observation~\ref{obs:R}~part~3).
	Therefore, we only need to consider the case that \emph{$e_b\prec e_c$}, as shown in Figure~\ref{fig:march-VV}~(a).
	Assume $v_{a}\leq I_{b,c}$ in the distance to $\overleftrightarrow{v_bv_c}$.
	
	Let $v_{c^*}$ denote the last vertex in the sequence $v_c,v_{c+1},\ldots,v_r$ so that $e_b \prec e_{c^*}$ or $e_b\parallel e_{c^*}$.
	We know that $e_{c^*+1}\prec e_b,\ldots,e_r\prec e_b$, which implies that $(v_b,v_{c^*+1}),...,(v_b,v_r)$ are dead (by Observation~\ref{obs:R}~part~3).
	It remains to prove that $(v_b,v_{c+1}),...,(v_b,v_{c^*})$ are also dead.
	
	Let $l$ denote the unique line at $I_{b,c}$ that is parallel to $\overleftrightarrow{v_bv_c}$.
	We state two facts about $l$.
	
	(i) Polygon $P$ lies in the (closed) half-plane delimited by $l$ and containing $v_b,v_c$.
	
	(ii) $\R_{b,c+1},\ldots,\R_{b,c^*}$ lie in the (open) half-plane delimited by $l$ and not containing $v_b,v_c$.
	
	Combining the two facts, no one in $\R_{b,c+1},\ldots,\R_{b,c^*}$ intersects $P$,
	which implies that $(v_b,v_{c+1})$, $\ldots$, $(v_b,v_{c^*})$ are dead (by Observation~\ref{obs:R}~part~2).
	We prove facts (i) and (ii) in the following.
	
	\noindent \emph{Proof of (i)}.
	By assumption, $v_{a}\leq I_{b,c}$ in the distance to $\overleftrightarrow{v_bv_c}$.
	However, $v_{a}$ has the largest distance among all vertices of $P$ that lie on the right of $\overrightarrow{v_bv_c}$.
	Together, we get fact (i).
	
	\newcommand{\Q}{\mathcal{Q}}
	
	\noindent \emph{Proof of (ii)}.
	For convenience, we shall introduce another region $\mathcal{Q}_{j,k}$ for each pair $(v_j,v_k)$ such that $j\neq k,j+1\neq k$, and $e_j\prec e_k$.
	Recall $H_{j,k},I_{j,k},J_{j,k}$ in Definition~\ref{def:R}. See Figure~\ref{fig:QIJKDef}~(a) and (b).
	Region $\Q_{j,k}$ is defined as the wedge bounded by $\overrightarrow{J_{j,k}I_{j,k}},\overrightarrow{J_{j,k}H_{j,k}}$ and containing $\R_{j,k}$.
	
	As illustrated by Figure~\ref{fig:march-VV}~(b), we have $\Q_{b,c}\supseteq \Q_{b,c+1}$.
	Similarly, we have $\Q_{b,c+1}\supseteq \Q_{b,c+2}\supseteq \ldots \supseteq \Q_{b,c^*},$
	which implies that $\R_{b,c+1},\ldots,\R_{b,c^*}$ lie in $\Q_{b,c+1}$.
	So to prove (ii) reduces to show $\Q_{b,c+1}\subseteq \phi$, where $\phi$ denotes the (open) half-plane delimited by $l$ and not containing $v_b,v_c$.
	
	As $v_b,v_c,v_{c+1}$ lie in clockwise order, $v_{c+1}$ is on the right of $\overrightarrow{v_bv_c}$.
	Further since $\overrightarrow{I_{b,c}J_{b,c+1}}$ is a translate of $\overrightarrow{v_cv_{c+1}}$ and $l\parallel v_bv_c$,
	we get $J_{b,c+1}\in\phi$.
	Thus, $\Q_{b,c+1}\subseteq \phi$, as $J_{b,c+1}$ is the apex of $\Q_{b,c+1}$.
	
	\begin{figure}[h]
		\begin{minipage}[b]{.42\textwidth}
			\centering \includegraphics[width=.97\textwidth]{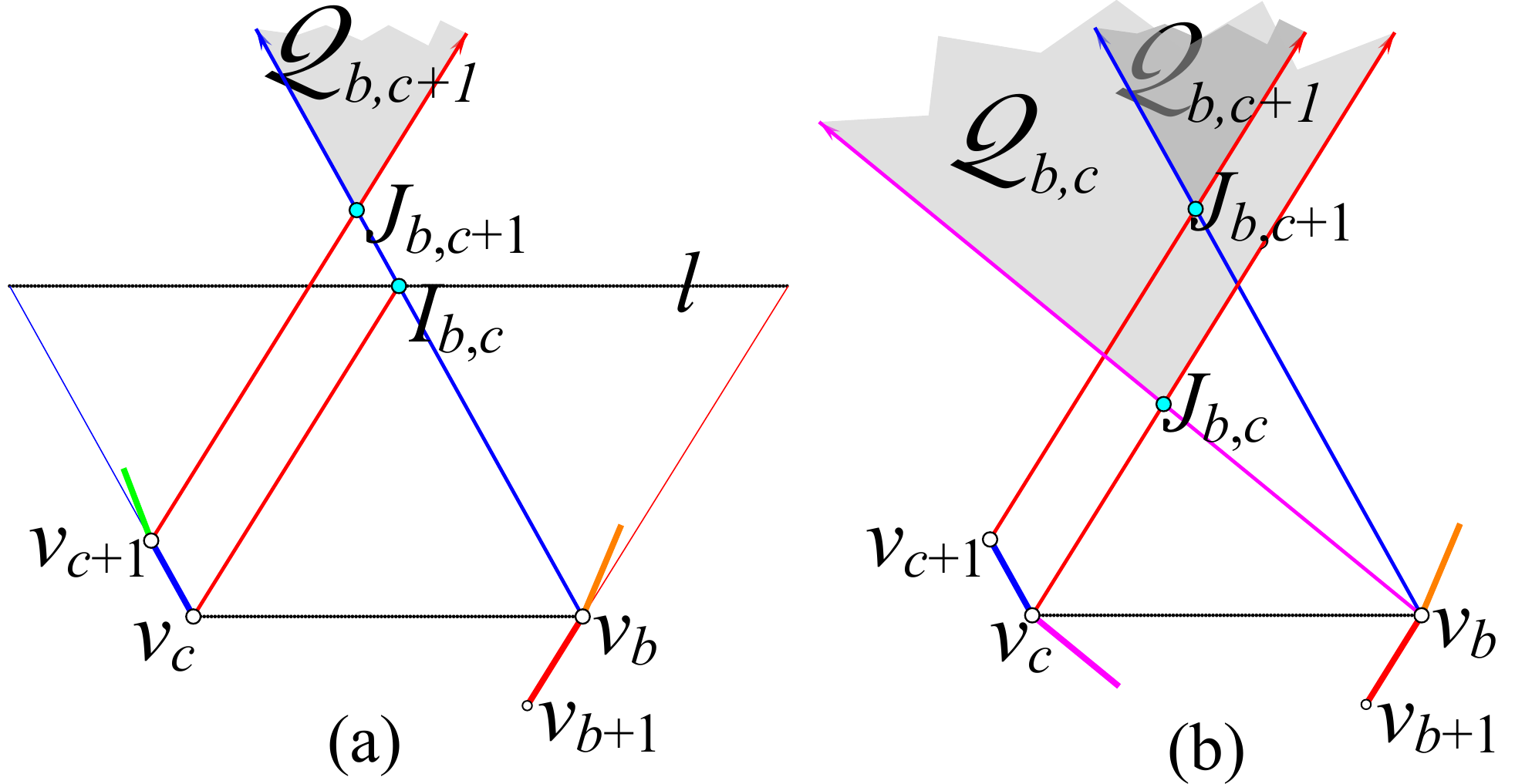}
		\end{minipage}
		\begin{minipage}[b]{.02\textwidth}
			~
		\end{minipage}
		\begin{minipage}[b]{.55\textwidth}
			\centering \includegraphics[width=\textwidth]{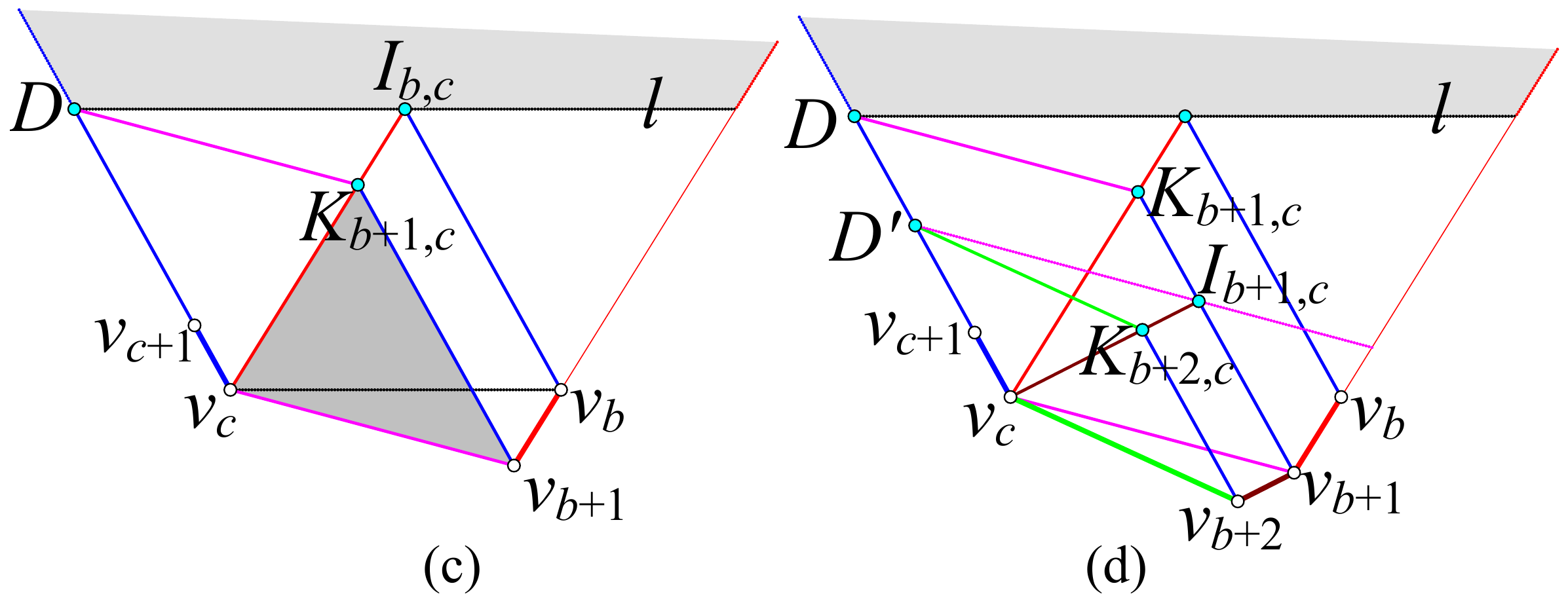}
		\end{minipage}
		\caption{Illustration of the proof of Lemma~\ref{lemma:correct}.}\label{fig:march-VV}
	\end{figure}
	
	\noindent \textbf{2.} Assume $v_{a}>I_{b,c}$ in the distance to $\overleftrightarrow{v_bv_c}$.
	This assumption implies that $I_{b,c}\neq \infty$. Hence $e_b\prec e_c$, as shown in Figure~\ref{fig:march-VV}~(c).
	Let $l$ and $\phi$ be the same as above. Let $D$ be the intersecting point between $\overleftrightarrow{v_cv_{c+1}}$ and $l$.
	Let $d_{j,k}(X)$ denote the distance from $X$ to $\overleftrightarrow{v_jv_k}$ for any vertex pair $(v_j,v_k)$ such that $j\neq k$.
	We claim two inequalities: $\text{(i) } d_{b+1,c}(v_{a})>d_{b+1,c}(D) \text{ and (ii) } d_{b+1,c}(D)=d_{b+1,c}(K_{b+1,c}).$
	
	Since $v_{a}>I_{b,c}$ in the distance to $\overleftrightarrow{v_bv_c}$, we get $v_{a}\in \phi$.
	Since $P$ is convex, $v_{a}$ lies in the (closed) half-plane delimited by $\overleftrightarrow{v_cv_{c+1}}$ and containing $P$. Together, they imply (i).
	Since both $v_{b+1}K_{b+1,c}$ and $v_cD$ are translates of $v_bI_{b,c}$, region $v_{b+1}K_{b+1,c}Dv_c$ is a parallelogram, which implies (ii).
	
	\medskip Combining inequalities (i) and (ii), $d_{b+1,c}(v_{a})>d_{b+1,c}(K_{b+1,c})$. This implies that $(v_{b+1},v_c)$ is dead (by Observation~\ref{obs:R}~part~4).
	Next, we argue that $(v_{b+2},v_c)$ is also dead.
	Applying Observation~\ref{obs:R}~part~4, it is enough to prove the following inequality (iii): $d_{b+2,c}(v_{a})>d_{b+2,c}(K_{b+2,c})$.
	The proof of this inequality is illustrated in Figure~\ref{fig:march-VV}~(d).
	Briefly, we can prove that $d_{b+2,c}(v_a)>d_{b+2,c}(D)>d_{b+2,c}(D')=d_{b+2,c}(K_{b+2,c})$.
	The proof is almost a copy of the proofs of (i) and (ii) and hence is omitted.
	Furthermore, by induction, $(v_{b+1},v_c),(v_{b+2},v_c),\ldots (v_t,v_c)$ are all dead.
\end{proof}


\section{Find all F-3-stable triangles in $O(n)$ time}\label{sect:F-3-stable}

This section finds \textbf{all} F-3-stable triangles based on Rotate-and-Kill,
assuming we are given $r,s,t$ such that $\triangle e_re_se_t$ is F-3-stable.
We state some basic observations and lemmas first.
\newcommand{\hG}{\hyperbola_{c+1,b}}
\newcommand{\hg}{\hyperbola_{b+1,c+1}}
\newcommand{\hH}{\hyperbola_{c+1,b+1}}
\newcommand{\hh}{\hyperbola_{b+1,c}}

An edge pair $(e_b,e_c)$ is \emph{dead} if there is no edge $e_a$ such that $\triangle e_ae_be_c$ is F-3-stable.

\begin{observation}\label{obs:F-dead-I-II}
	Given $b\in \{s,\ldots,t\}$ and $c\in \{t,\ldots,r\}$, at least one of the following holds.
	
	\quad (I) $(e_b,e_{c+1}),\ldots,(e_b,e_r)$ are dead. \quad (II) $(e_{b+1},e_c),\ldots,(e_t,e_c)$ are dead.
\end{observation}

Observation~\ref{obs:F-dead-I-II} is a counterpart of Observation~\ref{obs:dead-I-II}. It directly follows from Lemma~\ref{lemma:interleaving}~part~2.

\begin{observation}\label{Obs:area-finite}
	$\area(\triangle e_ie_je_k)$ is finite if and only if: $e_i\prec e_j,e_j\prec e_k$ and $e_k\prec e_i$.
\end{observation}

For each edge $e_i$, denote by $\D_i$ the clockwise first vertex with the furthest distance to $\ell_i$,
and denote by $\D^*_i$ the clockwise last vertex with the furthest distance to $\ell_i$ ($\D_i=\D^*_i$ in most cases).

\begin{definition}\label{def:OPT}
	For $(e_b,e_c)$ such that $e_b \prec e_c$, denote by $\OPT_{b,c}$ the edge $e_x$ in $e_{c+1},\ldots,e_{b-1}$ such that
	$\area(\triangle e_xe_be_c)$ is minimized.
	To be specific, if there are multiple edges that minimize $\area(\triangle e_xe_be_c)$, the first edge in $e_{c+1},\ldots,e_{b-1}$
	that minimizes $\area(\triangle e_xe_be_c)$ is denoted by $\OPT_{b,c}$.
	If $\area(\triangle e_ae_be_c)$ are infinite for all $a\in \{c+1,\ldots,b-1\}$
	(this occurs if $\D^*_b=\D_c$), let $\OPT_{b,c}$ be the previous edge of $\D_c$.
\end{definition}

\begin{lemma}[A unimodality of $\area(\triangle e_ae_be_c)$]\label{lemma:area-unimodal}
	Given $(e_b,e_c)$ such that $e_b\prec e_c$ and $\D^*_b\neq \D_c$, $\area(\triangle e_ae_be_c)$ is unimodal for $e_a\in [\D^*_b \circlearrowright \D_c]$.
	Specifically, let $e_y$ denote the next edge of $\D^*_b$, $e_z$ denote the previous edge of $\D_c$, and assume $\OPT_{b,c}=e_x$ (observe that $x\in \{y,\ldots,z\}$). Then,
	
	1. $\area(\triangle e_ae_be_c)$ strictly decreases when $a$ goes from $y$ to $x$; and
	
	2. $\area(\triangle e_ae_be_c)$ strictly increases when $a$ goes from $x+1$ to $z$.
\end{lemma}

\begin{lemma}[Bi-monotonicity of $\OPT_{b,c}$]\label{lemma:OPTmono} Denote by $E=\{e_{c+1},\ldots,e_{b-1}\}$ in the following claims.
	1. Assume $e_b\prec e_c,e_b\prec e_{c+1}$.
	Notice that  $\OPT_{b,c}$ and $\OPT_{b,c+1}$ lie in $E$ according to Definition~\ref{def:OPT}.
	We claim that $\OPT_{b,c},\OPT_{b,c+1}$ lie in clockwise order non-strictly in $E$.
	2. Assume $e_b\prec e_c,e_{b+1}\prec e_c$.
	Notice that $\OPT_{b,c}$ and $\OPT_{b+1,c}$ lie in $E$ according to Definition~\ref{def:OPT}.
	We claim that $\OPT_{b,c},\OPT_{b+1,c}$ lie in clockwise order non-strictly in $E$.
\end{lemma}

We are ready to describe our algorithm. See Algorithm~\ref{alg:RK-F}.
As in the case of finding all 3-stable triangles, we need a function $\Kill_F$ in the algorithm.
This function must satisfy properties \textcircled{1} and \textcircled{2}
(note that (I) and (II) in the description of \textcircled{1} now refer to (I) and (II) in Observation~\ref{obs:F-dead-I-II}).
The design of this function is crucial and is shown right below.

\begin{algorithm}[h]
	\caption{The Rotate-and-Kill process for finding all F-3-stable triangles.}\label{alg:RK-F}
	$(a,b,c)\leftarrow (r,s,t)$\;
	\Repeat{$(b,c)=(t,r)$}
	{
		Compute $e_a=\OPT_{b,c}$ \quad (recall $\OPT_{b,c}$ in Definition~\ref{def:OPT})\;
		Output $\triangle e_ae_bv_c$ if it is F-3-stable and output $\triangle e_{a+1}e_be_c$ if it is F-3-stable\;
		\lIf{$\Kill_F(b,c)$=`b'} {$b\leftarrow b+1$}
		\lElse{$c\leftarrow c+1$}
	}
\end{algorithm}

\begin{theorem}\label{thm:F-3-stable}
	If $\Kill_F$ satisfies \textcircled{1} and \textcircled{2}, Algorithm~\ref{alg:RK-F} finds all F-3-stable triangles in $O(n)$ time.
\end{theorem}

The easy proofs of Lemmas~\ref{lemma:area-unimodal}, \ref{lemma:OPTmono} and Theorem~\ref{thm:F-3-stable} are deferred to subsection~\ref{subsect:F-omitted}.
(Hint: Lemmas~\ref{lemma:area-unimodal} and \ref{lemma:OPTmono} together imply that $\OPT_{b,c}$ can be computed in amortized $O(1)$ time.)

\subsection{Description of the killing function $\Kill_F$}\label{subsect:Kill-F-MFTA}

\begin{figure}[h]
	\begin{minipage}[b]{.5\textwidth}
		\centering \includegraphics[width=.25\textwidth]{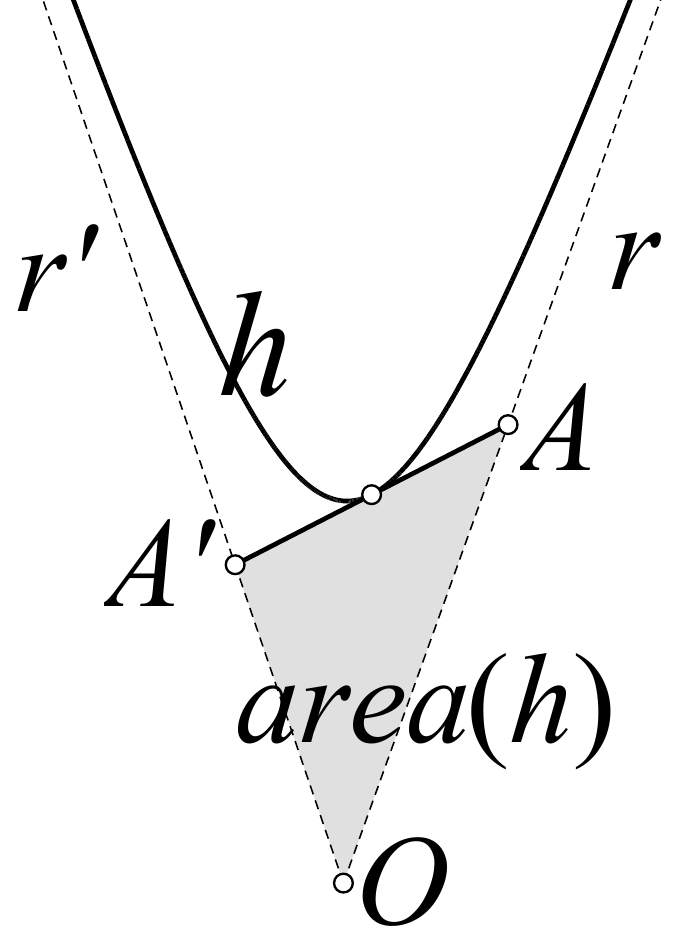}
		\caption{$\area(h)$.}\label{fig:def-h-area}
	\end{minipage}
	\begin{minipage}[b]{.5\textwidth}
		\centering \includegraphics[width=.5\textwidth]{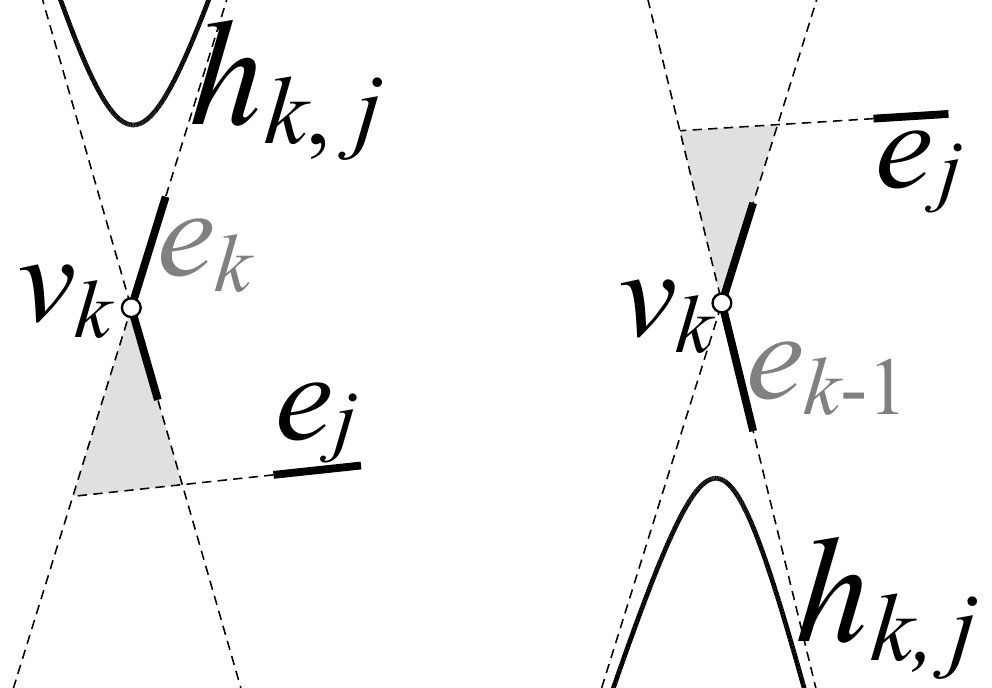}
		\caption{$\hyperbola_{k,j}$.}\label{fig:def-h}
	\end{minipage}
\end{figure}

We need several notations in order to introduce our killing function $\Kill_F$.
See Figure~\ref{fig:def-h-area}. Fix two rays $r,r'$ originating at some point $O$ and a hyperbola branch $h$ asymptotic to $r,r'$.
Take an arbitrary tangent line of $h$ and assume it intersects $r,r'$ at $A,A'$ respectively.
Then, $\area(\triangle OA'A)$ is a constant. This area is called the \emph{triangle-area} of $h$, denoted by $\area(h)$.


Consider $(v_k,e_j)$ where $e_j\notin \{e_{k-1},e_k\}$. See Figure~\ref{fig:def-h}.
If $e_j\prec e_k$, denote by $\hyperbola_{k,j}$ the hyperbola branch asymptotic to $\ell_{k-1},\ell_k$ in $p_{k-1}$ with triangle-area equal to $\area(p^C_{k-1}\cap p_k\cap p_j)$.
If $e_{k-1}\prec e_j$, denote by $\hyperbola_{k,j}$ the hyperbola branch asymptotic to $\ell_{k-1},\ell_k$ in $p_k$ with triangle-area equal to $\area(p_{k-1}\cap p^C_k\cap p_j)$.
(Note that $h_{k,j}$ is undefined sometimes, e.g., when $e_k\prec e_j$ and $e_j\prec e_{k-1}$.)

%
%
%

\begin{figure}[h]
	\begin{minipage}[b]{.5\textwidth}
		\centering \includegraphics[width=.75\textwidth]{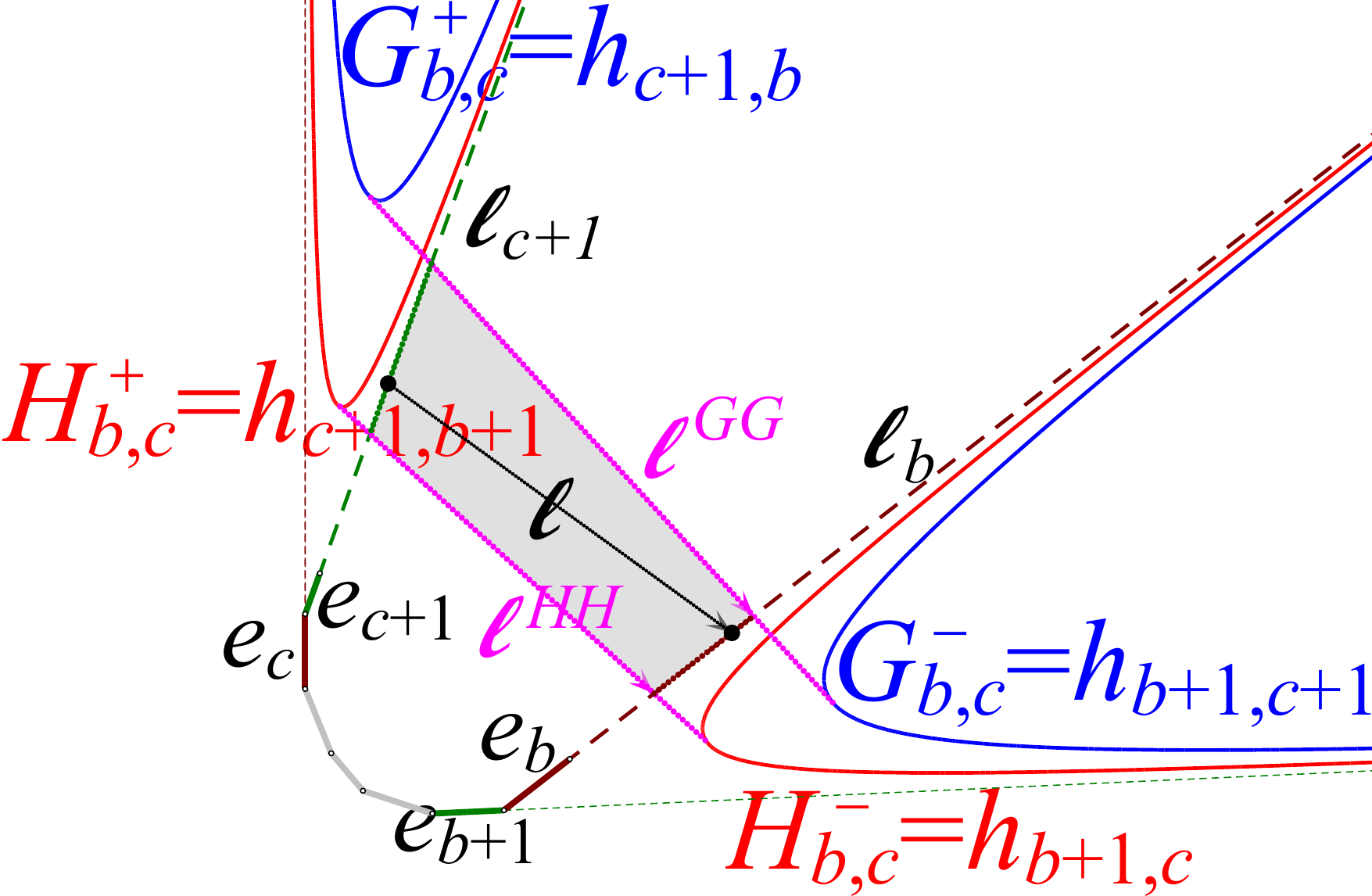}
		\caption{$G^+,G^-,H^+,H^-$.}\label{fig:kill-l}
		
	\end{minipage}
	\begin{minipage}[b]{.5\textwidth}
		\centering \includegraphics[width=.55\textwidth]{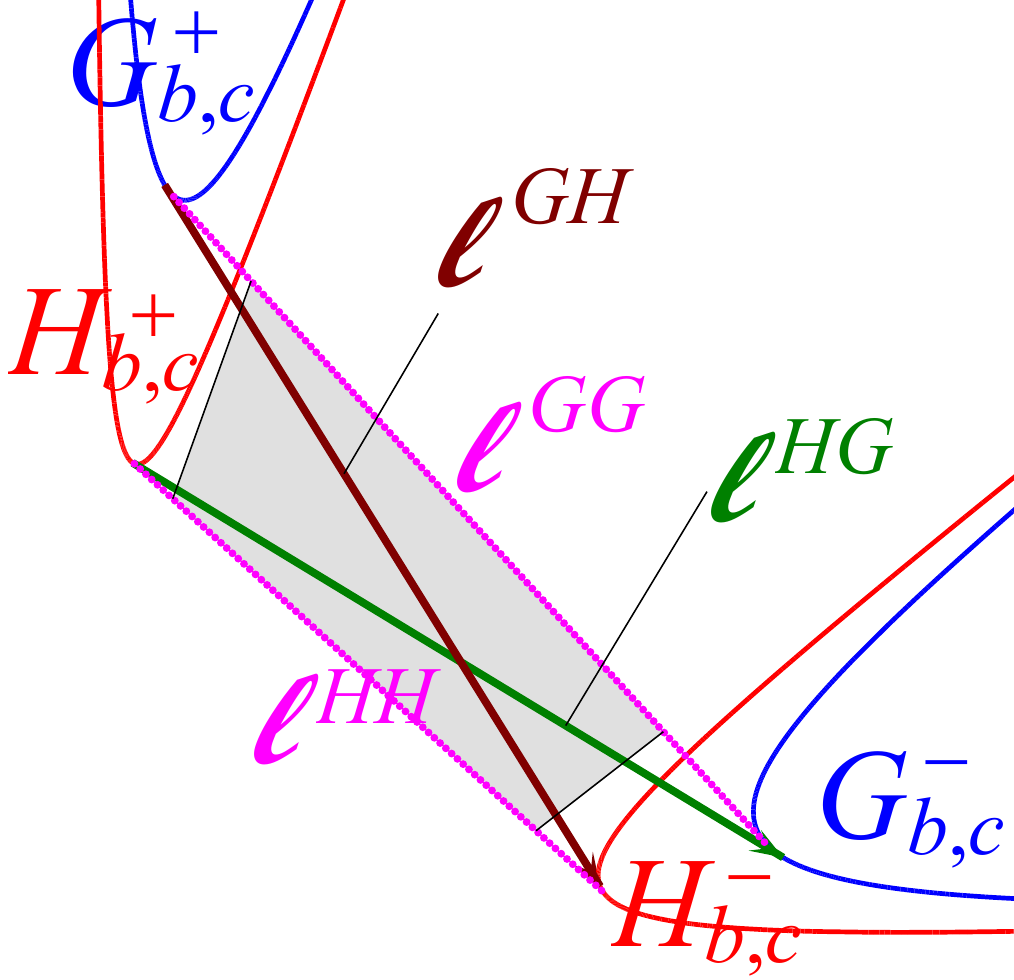}
		\caption{$L^{GH},L^{HG}$.}\label{fig:kill-l-2}
	\end{minipage}
\end{figure}

Denote $\U=\{(b,c) \mid b\in \{s,\ldots,t\}, c\in \{t,\ldots,r\}\}$.
Unless otherwise stated, assume hereafter in this section.
\begin{equation}\label{eqn:condition}
	(b,c)\in \U, e_b,e_{b+1},e_c,e_{c+1}\hbox{ are distinct, and }e_b\prec e_{c+1}.
\end{equation}
See Figure~\ref{fig:kill-l}. Let $G^+_{b,c},H^+_{b,c},G^-_{b,c},H^-_{b,c}$ respectively denote $\hG,\hH,\hg,\hh$.
(It is easy to check that $\hG,\hH,\hg,\hh$ are defined when \eqref{eqn:condition} holds.)

Moreover, denote by $\ell_{b,c}^{GH}, \ell_{b,c}^{HG},\ell_{b,c}^{GG},\ell_{b,c}^{HH}$ the common tangent of $G^+_{b,c},H^-_{b,c}$,
the common tangent of $H^+_{b,c},G^-_{b,c}$, the common tangent of $G^+_{b,c},G^-_{b,c}$, and the common tangent of $H^+_{b,c},H^-_{b,c}$, respectively.
See Figure~\ref{fig:kill-l-2} for examples ($b,c$ are omitted when they are clear in context).
Regard these common tangents as \emph{directed}; the direction of a common tangent $\ell$ is from $\ell\cap \ell_{c+1}$ to $\ell\cap \ell_b$.

For any directed line $\ell$, let $\theta[\ell]$ denote its direction, which is an angle in $[0,2\pi)$,
adopting the convention that \textbf{$\theta[\overrightarrow{OA}]$ increases when $A$ rotates clockwise around $O$}.

\paragraph*{Definition of $\Kill_F$.} Take $(b,c)\in \U$.
If $b=c$ or $b+1=c$, define $\Kill_F(b,c)=$ `c'.
If $e_b\prec e_{c+1}$ is not true, define $\Kill_F(b,c)=$ `b'.
Otherwise, \eqref{eqn:condition} holds.
Choose some direction $\theta\in [\theta[\ell^{HG}], \theta[\ell^{GH}]]$ (specified below),
and find any line $\ell_{b,c}=\overrightarrow{XY}$ with $\theta[\overrightarrow{XY}] = \theta$ and with the following property:

(P1) \emph{$X$ lies strictly between $\ell^{GG}\cap \ell_{c+1}$ and $\ell^{HH}\cap \ell_{c+1}$, and $Y$ lies strictly between $\ell^{GG}\cap \ell_b$ and $\ell^{HH}\cap \ell_b$.}
(A point $A$ lies strictly between $B,C$ if $A$ is in segment $BC$ and $A$ neither equals $B$ nor equals $C$.)
(The existence of such a line is obvious and is formally stated in the next lemma.)
Define
\begin{equation}\label{eqn:Kill-F}
	\Kill_F(b,c)= \left\{
	\begin{array}{ll}
		\text{`b'}, & \hbox{if $P$ lies on the right of $\ell_{b,c}$ ;} \\
		\text{`c'}, & \hbox{Otherwise.}
	\end{array}
	\right.
\end{equation}

Direction $\theta$ is a variable of our algorithm. Its value is updated in each iteration.
Initially, $\theta\leftarrow \theta[\ell^{HG}_{s,t}]$. Subsequently,
$\theta\leftarrow \left\{                              \begin{array}{ll}
	\theta, &  \hbox{if }\theta \in  [\theta[\ell_{b,c}^{HG}], \theta[\ell_{b,c}^{GH}]];\\
	\theta[\ell_{b,c}^{HG}], & \hbox{otherwise.}
\end{array}
\right.
$.


\begin{lemma}\label{lemma:kill-l}
	For any $\theta\in [\theta[\ell^{HG}], \theta[\ell^{GH}]]$, we can compute in constant time a directed line $\ell=\overrightarrow{XY}$
	with $\theta[\ell]=\theta$ and with the aforementioned property (P1).
	Moreover, for any line $\ell$ satisfying (P1), the following hold:
	If $P$ lies on the right of $\ell$, condition (I) holds. Otherwise, condition (II) holds.
\end{lemma}

\begin{lemma}\label{lemma:d-mono}
	Let $\theta_1,\theta_2$ be the opposite directions of $e_{s+1},e_r$. Assume $[\theta_1,\theta_2]\subset [0,2\pi)$ without loss of generality.
	The variable $\theta$ defined above increases (non-strictly) within the range $[\theta_1,\theta_2]$ during the algorithm.
\end{lemma}

\begin{theorem}\label{thm:Kill-F}
	The function $\Kill_F$ satisfies \textcircled{1} and \textcircled{2}.
\end{theorem}

Theorem~\ref{thm:Kill-F} is easily built on the Lemma~\ref{lemma:kill-l} and Lemma~\ref{lemma:d-mono}.
Lemma~\ref{lemma:d-mono} is very challenging to prove and it is crucial for proving property \textcircled{2}.
Due to space limit, we show the proofs of Theorem~\ref{thm:Kill-F}, Lemma~\ref{lemma:kill-l}, and Lemma~\ref{lemma:d-mono} in subsection~\ref{subsect:F-kill-omitted}.

Combining Theorem~\ref{thm:Kill-F} and \ref{thm:F-3-stable}, we find all F-3-stable triangles in $O(n)$ time.

\subsection{Analysis of the Kill function $\Kill_F$ introduced in subsection~\ref{subsect:Kill-F-MFTA}}\label{subsect:F-kill-omitted}

We first prove Theorem~\ref{thm:Kill-F} and then prove Lemma~\ref{lemma:kill-l} and Lemma~\ref{lemma:d-mono}.

\begin{proof}[Proof of Theorem~\ref{thm:Kill-F}]
	Property \textcircled{1} states that for any $(b,c)\in \U$, $\Kill_F(b,c)$ returns `b' only if (I) $(e_b,e_{c+1})$, $\ldots$, $(e_b,e_r)$ are dead, and `c' only if (II) $(e_{b+1},e_c),\ldots,(e_t,e_c)$ are dead.
	
	When $b=c$ or $b+1=c$,  $\Kill_F(b,c)$ returns `c' by definition.
	Moreover, we have $(b,c)\in \{(t,t),(t-1,t),(t,t+1)\}$, hence condition~(II) holds trivially.
	When $e_b\prec e_{c+1}$ is not true, $\Kill_F(b,c)$ returns `b' by definition.
	In this case, $e_{c+1},\ldots,e_r$ are chasing or parallel to $e_b$,
	which implies that $(e_b,e_{c+1}),\ldots,(e_b,e_{r})$ are dead (by Observation~\ref{Obs:area-finite}).
	Assume now $b\neq c,b+1\neq c$, and $e_b\prec e_{c+1}$.
	In other words, condition~\eqref{eqn:condition} holds.
	There are two cases: $\Kill_F(b,c)$=`b', or $\Kill_F(b,c)$=`c'.
	When $\Kill_F(b,c)$=`b', $P$ is on the right of $\ell_{b,c}$, which implies condition~(I) by Lemma~\ref{lemma:kill-l}.
	When $\Kill_F(b,c)$=`c', $P$ is not on the right of $\ell_{b,c}$, which implies condition~(II) by Lemma~\ref{lemma:kill-l}.
	
	\smallskip Property \textcircled{2} states that $\Kill_F(b,c)$ can be computed in amortized $O(1)$ time.
	Denote by $\ell_\theta$ the tangent of $P$ with direction $\theta$ and with $P$ on its right.
	As $\theta$ increases monotonously (within $[0,2\pi)$) during the algorithm (due to Lemma~\ref{lemma:d-mono}),
	$\ell_\theta$ can be computed in amortized $O(1)$ time.
	Moreover, $\ell_{b,c}$ can be computed in $O(1)$ time (by Lemma~\ref{lemma:kill-l}).
	Therefore, it takes amortized $O(1)$ time to check whether $\ell_\theta$ is on the right of $\ell_{b,c}$.
	Further since polygon $P$ is on the right of $\ell_{b,c}$ if and only if $\ell_\theta$ is on the right of $\ell_{b,c}$,
	function $\Kill_F(b,c)$ in \eqref{eqn:Kill-F} can be computed in amortized $O(1)$ time.
\end{proof}

The following observation follows from elementary knowledge of hyperbolas; proof omitted.
\begin{description}
	\item[Intersects \& avoids]
	Assume $h$ is a hyperbola branch (see Figure~\ref{fig:def-h-area}).
	If $h\cap p_a\neq \varnothing$, we state that $p_a$ \emph{intersects} $h$.
	If $h\cap (p_a\setminus \ell_a)=\varnothing$, we state that $p_a$ \emph{avoids} $h$.
\end{description}
\begin{observation}\label{obs:area-to-intersection}
	See Figure~\ref{fig:def-h-area}. Let $\phi$ be the wedge bounded by $r,r'$ and containing $h$.
	If $\area(p_a \cap \phi) \geq \area(h)$, then $p_a$ intersects $h$. If $\area(p_a \cap \phi) \leq \area(h)$, then $p_a$ avoids $h$.
\end{observation}

\begin{proof}[Proof of Lemma~\ref{lemma:kill-l}]
	The first part of this lemma is rather obvious. See Figure~\ref{fig:kill-l-2} for an illustration.
	We can compute tangents $\ell^{GG}$ and $\ell^{HH}$ in constant time and then compute the intersecting points between these two tangents and the two extended lines $\ell_b,\ell_{c+1}$.
	It is then easy to find two points $X,Y$ with $\theta(\overrightarrow{XY})=\theta$ and with $X,Y$ satisfying property (P1).
	Further details are omitted.
	
	The ``moreover'' part is less obvious. Before proving it, we first prove the following facts.
	
	\noindent 1. \emph{If $p_a\in \{p_{c+2},\ldots,p_{b-1}\}$ intersects both $G^+_{b,c}$ and $G^-_{b,c}$,
		$v_a$ or $v_{a+1}$ is not on the right of $\ell^{GG}$.}
	
	\noindent 2. \emph{If $p_a\in \{p_{c+2},\ldots,p_{b-1}\}$ avoids both $H^+_{b,c}$ and $H^-_{b,c}$,
		$P$ is on the right of $\ell^{HH}$.}
	
	\noindent 3. \emph{Assume $\triangle e_ae_be_{c'}$ is F-3-stable for some $e_{c'}\in \{e_{c+1},\ldots,e_r\}$.
		Then,}
	
	\emph{\quad (a) $p_a$ intersects $G^+_{b,c}=\hG$; (b) $p_a$ intersects $G^-_{b,c}=\hg$; and (c) $a\in \{c+2,\ldots,b-1\}$.}
	
	\noindent 4. \emph{Assume $\triangle e_ae_{b'}e_c$ is F-3-stable for some $e_{b'}\in \{e_{b+1},\ldots,e_t\}$.
		Then,}
	
	\emph{\quad (a) $p_a$ avoids $H^+_{b,c}=\hH$; (b) $p_a$ avoids $H^-_{b,c}=\hh$; and (c) $a\in \{c+2,\ldots,b-1\}$.}

	\begin{figure}[t]
		\centering
		\includegraphics[width=.28\textwidth]{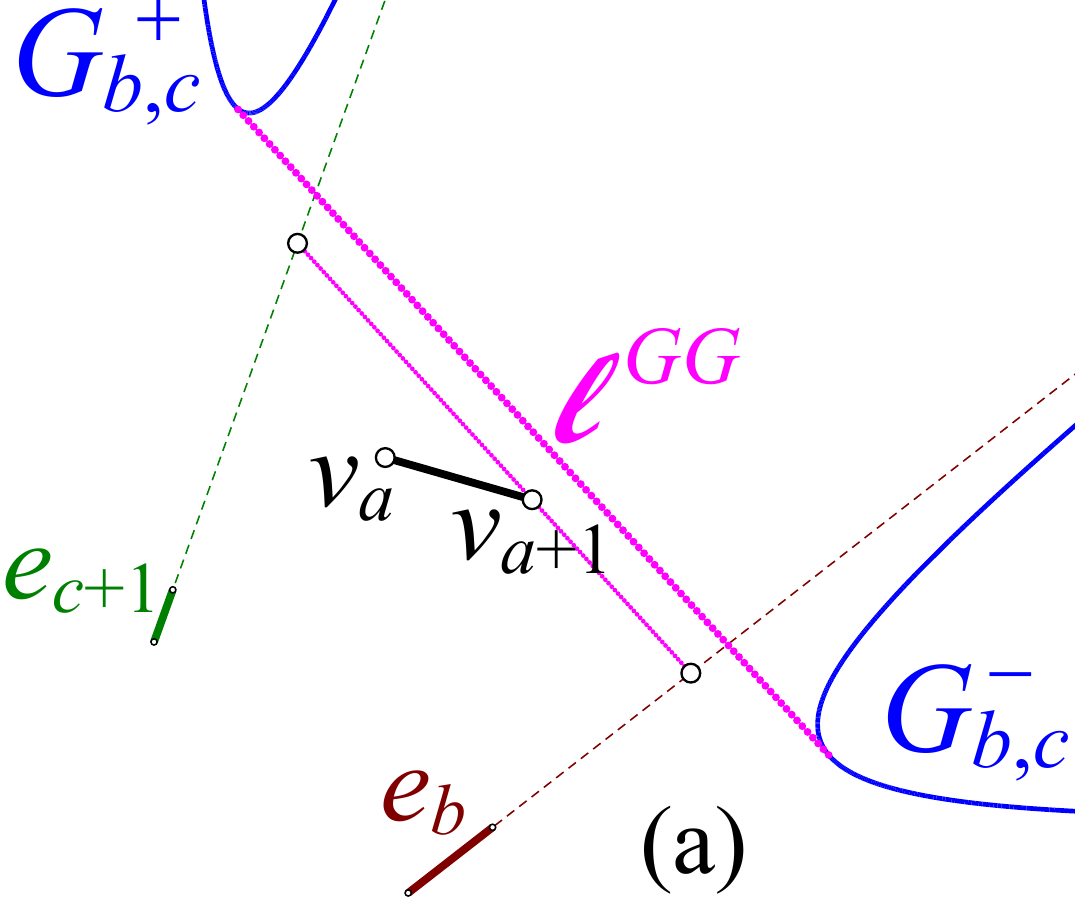}
		\includegraphics[width=.22\textwidth]{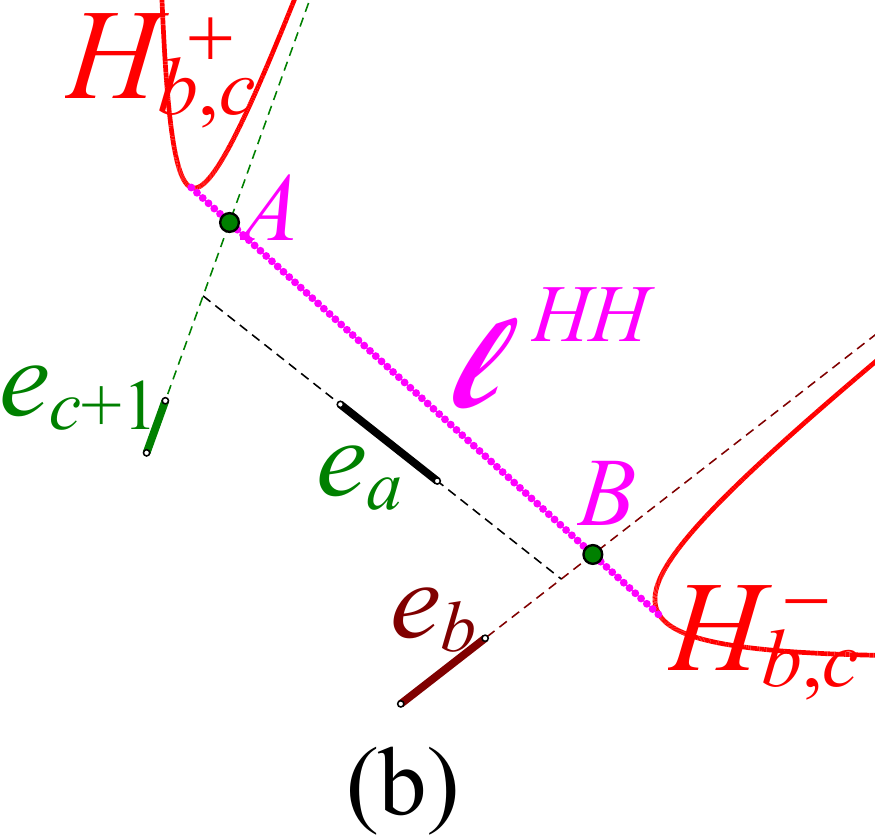}
		\includegraphics[width=.22\textwidth]{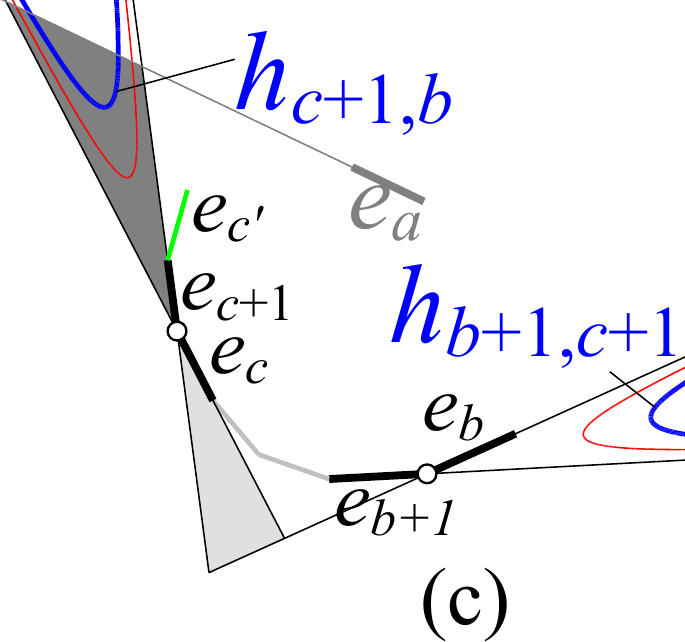}
		\includegraphics[width=.22\textwidth]{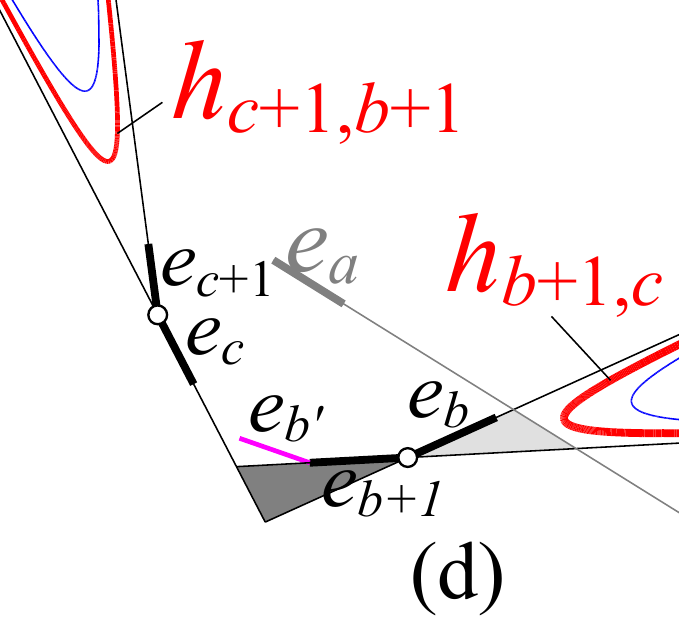}
		\caption{Illustration of the proof of Lemma~\ref{lemma:kill-l}.}\label{fig:kill-l-proof}
	\end{figure}
	
	\smallskip \noindent \emph{Proof of fact~1.} See Figure~\ref{fig:kill-l-proof}~(a).
	We prove it by contradiction.
	Suppose $v_a,v_{a+1}$ are both on the right of $\ell^{GG}$.
	We discuss two cases distinguished by whether $v_a$ or $v_{a+1}$ is closer to $\ell^{GG}$.
	If $v_{a+1}$ is closer to $\ell^{GG}$, it is easy to see that $p_a$ cannot intersect $G^+_{b,c}$.
	If $v_a$ is closer to $\ell^{GG}$, it is easy to see that $p_a$ cannot intersect $G^-_{b,c}$.
	Either way, $p_a$ cannot intersect both $G^+_{b,c}$ and $G^-_{b,c}$.
	
	\smallskip \noindent \emph{Proof of fact~2.} See Figure~\ref{fig:kill-l-proof}~(b).
	Assume $p_a$ avoids $H^+_{b,c}$ and $H^-_{b,c}$.
	This means that both $H^+_{b,c}$ and $H^-_{b,c}$ lie on the left of $\ell_{a}$.
	It follows that points $A=\ell^{HH}\cap \ell_{c+1}$ and $B=\ell^{HH}\cap \ell_b$ lie on the left of $\ell_a$.
	Therefore, $e_a,e_b,e_{c+1}$ are all on the right of $\overrightarrow{AB}=\ell^{HH}$.
	Thus, $P$ lies on the right of $\ell^{HH}$.
	
	\smallskip \noindent
	\emph{Proof of fact~3~(a).} See Figure~\ref{fig:kill-l-proof}~(c). Because $e_{c'}\in \{e_{c+1},\ldots, e_r\}$ and is stable in $\triangle e_ae_be_{c'}$,
	following the unimodality stated in Lemma~\ref{lemma:area-unimodal}, $\area(p_a\cap (p_c \cap p^C_{c+1}))\geq \area(p_b\cap (p^C_c \cap p_{c+1}))=\area(\hG)$.
	Applying Observation~\ref{obs:area-to-intersection}, this means $p_a$ intersects $\hG$.
	
	\noindent \emph{Proof of fact~3~(b).}
	Because $e_b$ is stable in $\triangle e_ae_be_{c'}$,
	$\area(\triangle e_ae_{b+1}e_{c'})\geq \area(\triangle e_ae_be_{c'})$.
	In other words, $\area(p_a\cap (p_b^C \cap p_{b+1}) \geq \area(p_{c'}\cap (p_{b}\cap p^C_{b+1}))$.
	It follows that $\area(p_a\cap (p^C_{b}\cap p_{b+1})) \geq \area(p_{c+1}\cap (p_b \cap p^C_{b+1}))=\area(\hg)$.
	Applying Observation~\ref{obs:area-to-intersection}, this means $p_a$ intersects $\hg$.
	
	\noindent
	\emph{Proof of fact~3~(c).} Since $c'\in \{c+1,\ldots,r\}$ and $e_a,e_b,e_{c'}$ are in clockwise order, $a\in \{c+2,\ldots,b-1\}$.
	
	\smallskip \noindent
	\emph{Proof of fact~4~(a).} See Figure~\ref{fig:kill-l-proof}~(d).
	Because $e_{b'}\in \{e_{b+1},\ldots, e_t\}$ and is stable in $\triangle e_ae_{b'}e_c$,
	following the unimodality stated in Lemma~\ref{lemma:area-unimodal},
	$\area(p_a\cap (p^C_b \cap p_{b+1}))\leq \area(p_c\cap (p_b\cap p^C_{b+1})) =\area(\hh)$.
	Applying Observation~\ref{obs:area-to-intersection}, this means $p_a$ avoids $\hh$.
	
	\noindent
	\emph{Proof of fact~4~(b).} Because $e_c$ is stable in $\triangle e_ae_{b'}e_c$,
	$\area(\triangle e_ae_{b'}e_c) \leq \area(\triangle e_ae_{b'}e_{c+1})$.
	In other words, $\area(p_a \cap (p_c\cap p^C_{c+1})) \leq  \area(p_b' \cap (p^C_c\cap p_{c+1}))$.
	It follows that $\area(p_a \cap (p_c\cap p^C_{c+1})) \leq  \area(p_{b+1} \cap (p^C_c\cap p_{c+1}))=\area(\hH)$.
	Applying Observation~\ref{obs:area-to-intersection}, this means $p_a$ avoids $\hH$.
	
	\noindent
	\emph{Proof of fact~4~(c).}
	First, because $e_a,e_{b'},e_c$ lie in clockwise order, $a\in \{c+1,\ldots, b'-1\}$.
	Since $\triangle e_ae_{b'}e_c$ is F-3-stable, $e_c\prec e_a$.
	However, edges $e_{b},\ldots, e_{b'-1}$ are chasing $e_c$, so they do not contain $e_a$.
	Therefore, $a\in \{c+1,\ldots, b-1\}$.
	Because $e_b\prec e_{c+1}$, we have $e_{b'}\prec e_{c+1}$. However, $e_a\prec e_{b'}$ because $\triangle e_ae_{b'}e_c$ is F-3-stable.
	Therefore, $a\neq c+1$. Altogether, $a\in \{c+2,\ldots, b-1\}$.
	
	\medskip We are ready to prove the ``moreover'' part.
	Assume $P$ is on the right of $\ell$. Then, $P$ must be on the right of $\ell^{GG}$.
	According to fact~1, this implies that \emph{no one in $\{p_{c+2},\ldots,p_{b-1}\}$ intersects both $G^+_{b,c}$ and $G^-_{b,c}$}.
	Further by applying fact~3, $(e_b,e_{c+1}),\ldots,(e_b,e_r)$ are dead, i.e., condition (I) holds.
	
	Assume in the following $P$ is not on the right of $\ell$. Then, $P$ is not on the right of $\ell^{HH}$.
	According to fact~2, \emph{no one in $\{p_{c+2},\ldots,p_{b-1}\}$ avoids both $H^+_{b,c}$ and $H^-_{b,c}$}.
	Further by applying fact~4, $(e_{b+1},e_c),\ldots,(e_t,e_r)$ are dead, i.e., condition (II) holds.
\end{proof}

\paragraph*{Proof of Lemma~\ref{lemma:d-mono}}\label{subsect:Kill-F-part-II}

\begin{figure}[h]
	\begin{minipage}[b]{.65\textwidth}
		\centering \includegraphics[width=\textwidth]{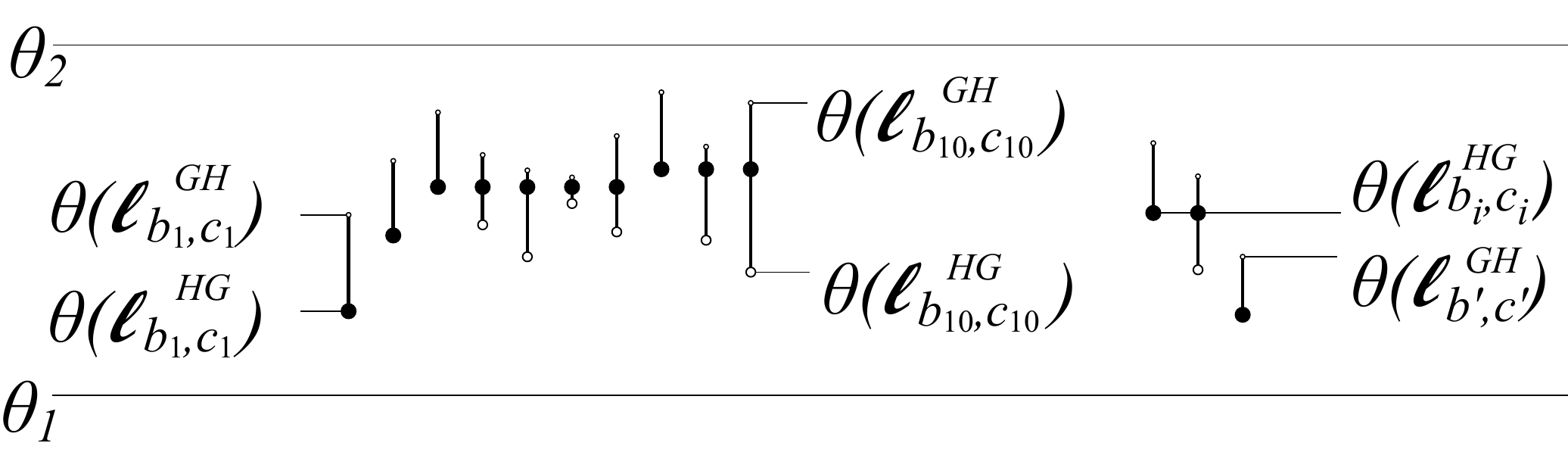}
		\caption{Illustration of the proof of Lemma~\ref{lemma:d-mono}.}\label{fig:Obs-mono-direction}
	\end{minipage}
	\begin{minipage}[b]{.34\textwidth}
		\centering\includegraphics[width=.85\textwidth]{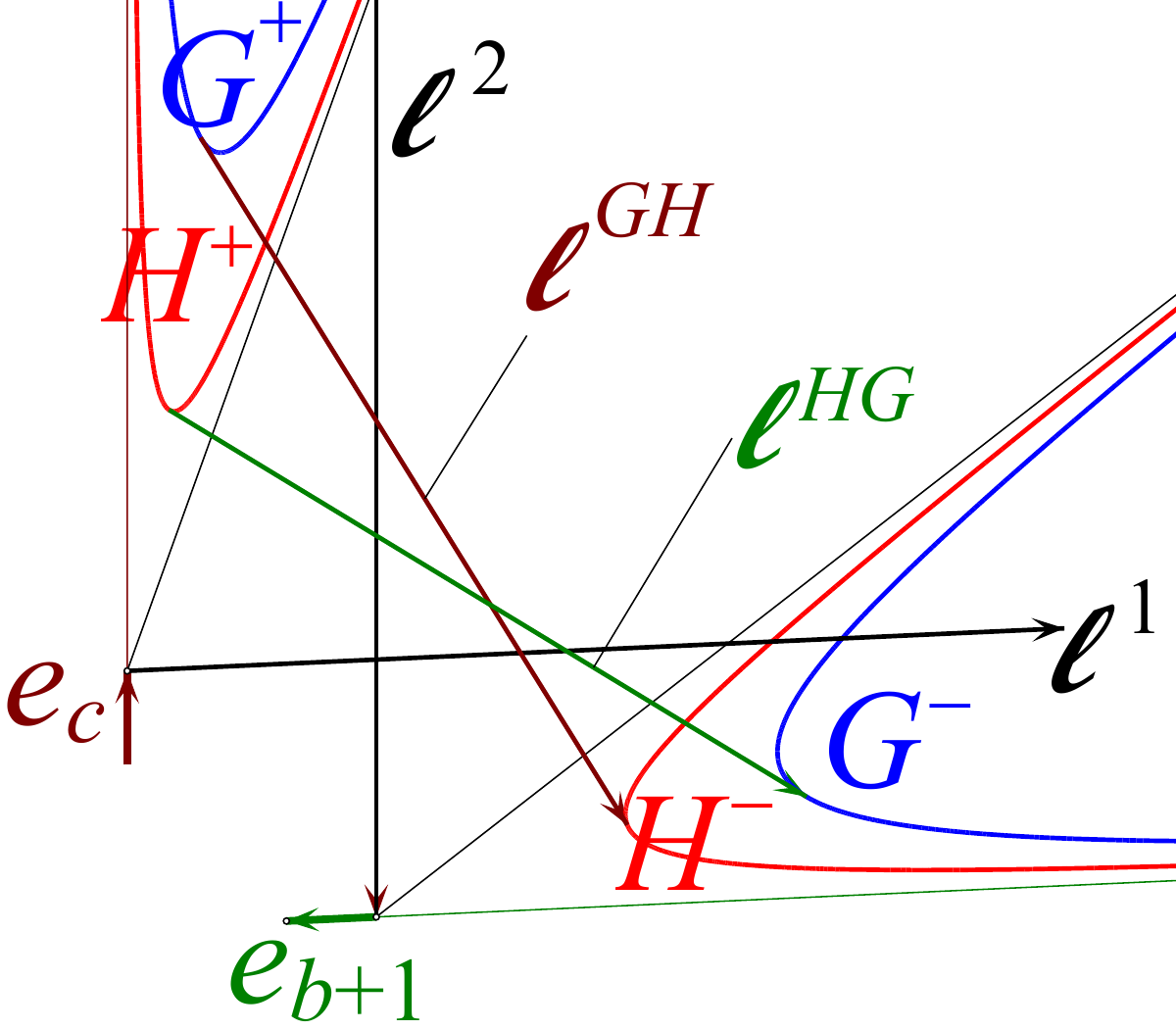}
		\caption{Observation~\ref{obs:d-mono-1}}\label{fig:Obs-mono-range}
	\end{minipage}
\end{figure}

Recall $\theta_1,\theta_2$ in Lemma~\ref{lemma:d-mono}.
Our proof of Lemma~\ref{lemma:d-mono} applies the following two observations.

\begin{observation}\label{obs:d-mono-1}
	When \eqref{eqn:condition} holds, $[\theta[\ell^{HG}_{b,c}], \theta[\ell^{GH}_{b,c}]] \subset [\theta_1,\theta_2]$.
\end{observation}

\begin{observation}\label{obs:d-mono-2}
	Assume we were given $(b,c)$ in some iteration and $(b',c')$ in a later iteration,
	both $(b,c)$ and $(b',c')$ satisfying condition~\eqref{eqn:condition}, then $\theta[\ell^{GH}_{b',c'}]\geq \theta[\ell^{HG}_{b,c}]$.
\end{observation}

\begin{proof}[Proof of Lemma~\ref{lemma:d-mono}]
	See Figure~\ref{fig:Obs-mono-direction} for an illustration.
	Assume we have visited $m$ pairs $(b_1,c_1)$, \ldots, $(b_m,c_m)$ satisfying condition~\eqref{eqn:condition} thus far
	in the Rotate-and-Kill process,
	and we are at the beginning of another iteration $(b',c')$ satisfying \eqref{eqn:condition}.
	Currently, $\theta$ equals to $\theta[\ell^{HG}_{b_i,c_i}]$ for some $i\leq m$.
	Therefore, $\theta\leq \theta[\ell^{GH}_{b',c'}]$ according to Observation~\ref{obs:d-mono-2}.
	After visiting $(b',c')$,
	according to the formula $\theta\leftarrow \left\{                              \begin{array}{ll}
		\theta, &  \hbox{if }\theta \in  [\theta[\ell_{b',c'}^{HG}], \theta[\ell_{b',c'}^{GH}]];\\
		\theta[\ell_{b',c'}^{HG}], & \hbox{otherwise.}
	\end{array}
	\right.
	$,
	$\theta$ either increases (if $\theta<\theta[\ell^{HG}_{b',c'}]$) or remains unchanged (if $\theta\in [\theta[\ell^{HG}_{b',c'}],\theta[\ell^{HH}_{b',c'}]]$). Further by Observation~\ref{obs:d-mono-1},
	$\theta$ increases within $[\theta_1,\theta_2]$ during the algorithm.
	
	(The lemma fails if Observation~\ref{obs:d-mono-2} is not true; see the right part of Figure~\ref{fig:Obs-mono-direction} for an example.)
\end{proof}
%

\begin{proof}[Proof of Observation~\ref{obs:d-mono-1}]
	1. See Figure~\ref{fig:Obs-mono-range}.
	Let $\ell^1_{b,c}$ be the line at $v_{c+1}$ with the opposite direction to $e_{b+1}$, and
	$\ell^2_{b,c}$ be the line at $v_{b+1}$ with the opposite direction to $e_c$.
	Clearly, $\ell_{b,c}^1$ intersects $G^-_{b,c}$ but not $H^+_{b,c}$, whereas $\ell_{b,c}^2$ intersect $G^+_{b,c}$ but not $H^-_{b,c}$.
	On the other hand, $\ell^{HG}_{b,c}$ is tangent to $G^-_{b,c}$ and $H^+_{b,c}$, whereas $\ell^{GH}_{b,c}$ is tangent to $G^+_{b,c}$ and $H^-_{b,c}$.
	Altogether, we can deduce that $[\theta[\ell^{HG}_{b,c}], \theta[\ell^{GH}_{b,c}]] \subset [\theta[\ell_{b,c}^1],\theta[\ell_{b,c}^2]]$.
	However, $[\theta[\ell^1_{b,c}],\theta[\ell^2_{b,c}]]\subset [\theta[\ell^1_{s,t}],\theta[\ell^2_{t,r}]]=[\theta_1,\theta_2]$.
	So $[\theta[\ell^{HG}_{b,c}], \theta[\ell^{GH}_{b,c}]]\subset [\theta_1,\theta_2]$.
\end{proof}

\begin{proof}[Proof of Observation~\ref{obs:d-mono-2}]
	Let $A=\ell_{b+1}\cap \ell_{c+1}, B=\ell_{b'}\cap \ell_{c'}$.
	
	\smallskip \textbf{First, consider the case where $b'>b$ and $c'>c$.} See Figure~\ref{fig:lemma-mono1}. Denote
	
	\quad $M^-=\left\{
	\begin{array}{ll}
		v_{b'+1}, & b'=b+1; \\
		\ell_{b+1}\cap \ell_{b'}, & b'\geq b+2
	\end{array}
	\right. \text{and }
	M^+=\left\{
	\begin{array}{ll}
		v_{c'+1}, & c'=c+1; \\
		\ell_{c+1}\cap \ell_{c'}, & c'\geq c+2.
	\end{array}
	\right
	.$
	
	Denote the reflections of $A$ around $M^-,M^+$ by $A^-,A^+$ respectively.
	Denote the reflections of $B$ around $M^-,M^+$ by $B^-,B^+$ respectively.
	Let $\ell=\overrightarrow{A^+A^-}$ and $\ell'=\overrightarrow{B^+B^-}$.
	
	We state three equalities or inequalities which together imply the claimed inequality.
	
	\quad (i) $\theta[\ell_{b,c}^{HG}]\leq \theta[\ell] $. \qquad (ii) $\theta[\ell]=\theta[\ell']$. \qquad  (iii) $\theta[\ell']\leq \theta[\ell_{b',c'}^{GH}]$.
	
	\begin{figure}[h]
		\begin{minipage}[b]{1\textwidth}
			\centering \includegraphics[width=.455\textwidth]{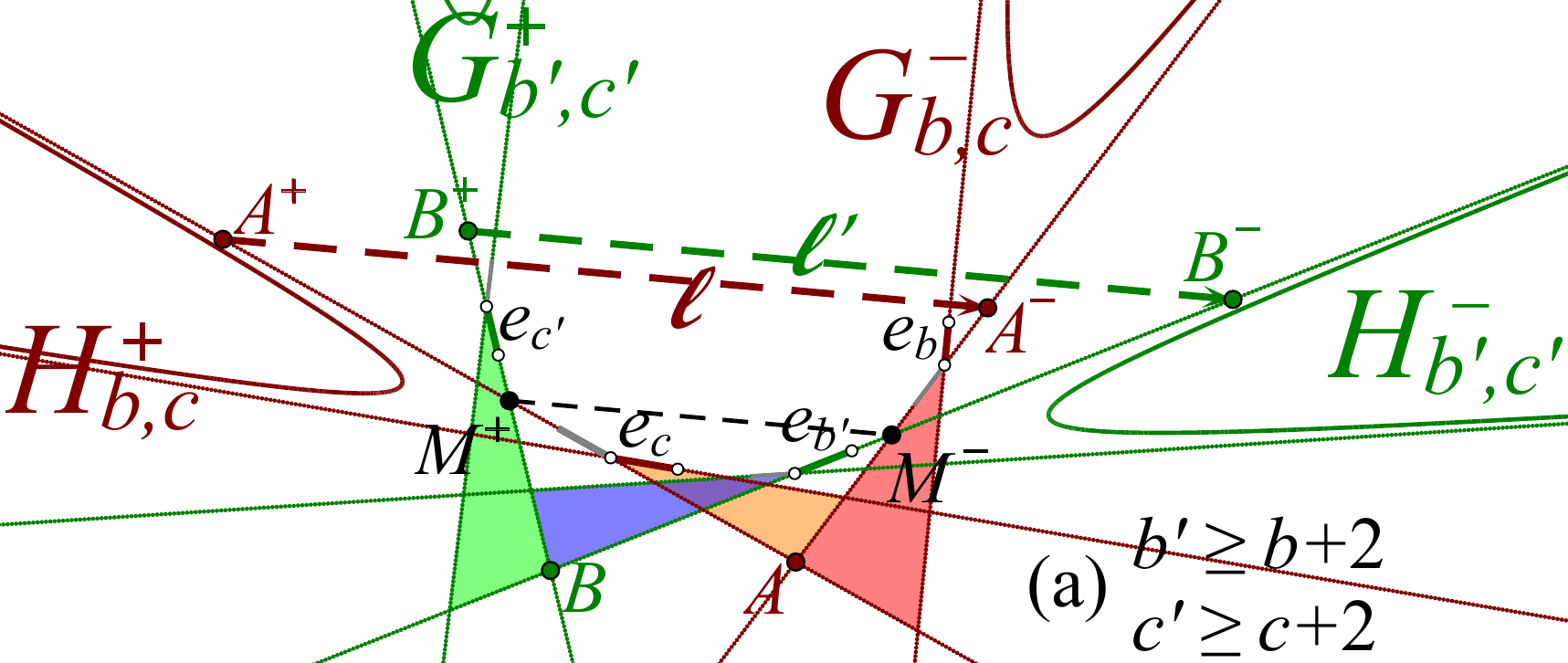} \quad \includegraphics[width=.44\textwidth]{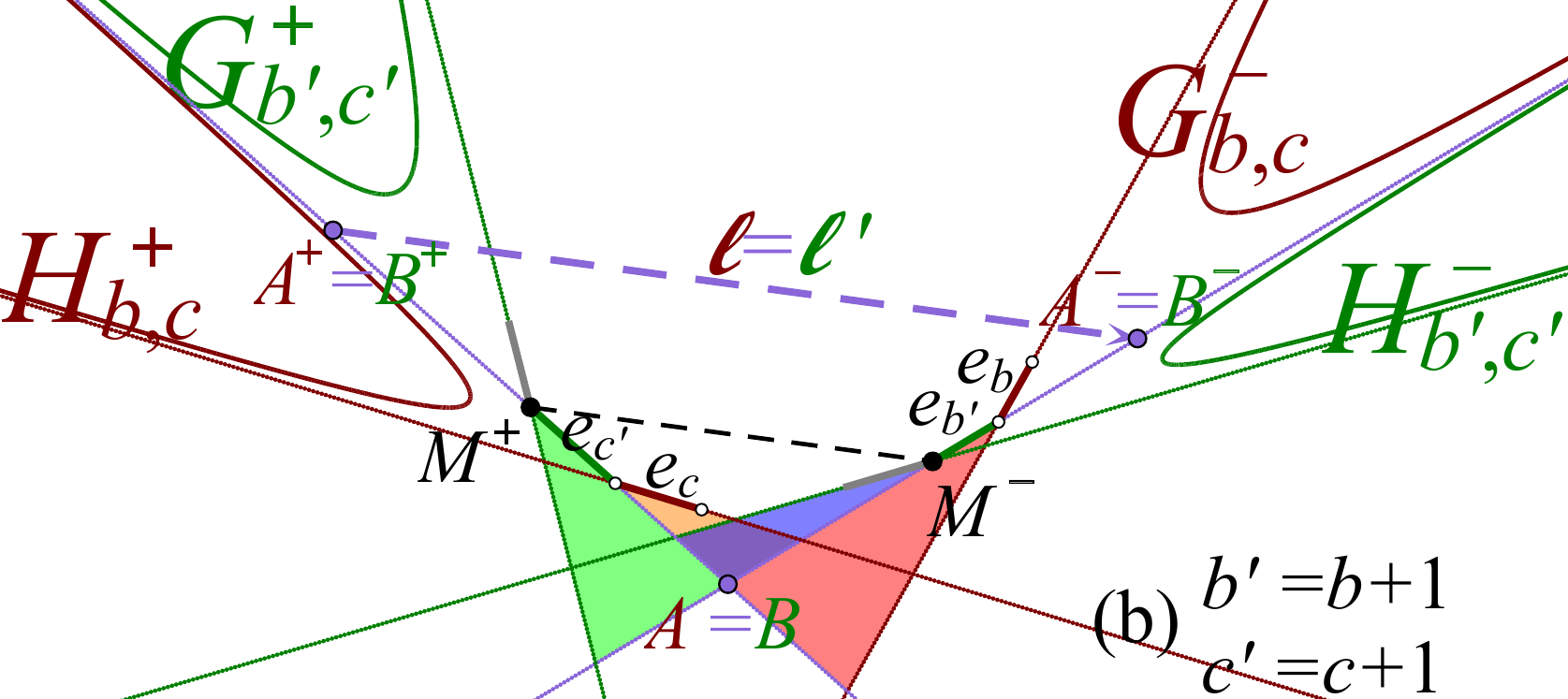} \bigskip
		\end{minipage}
		\begin{minipage}[b]{1\textwidth}
			\vspace{15pt}
			\centering \includegraphics[width=.455\textwidth]{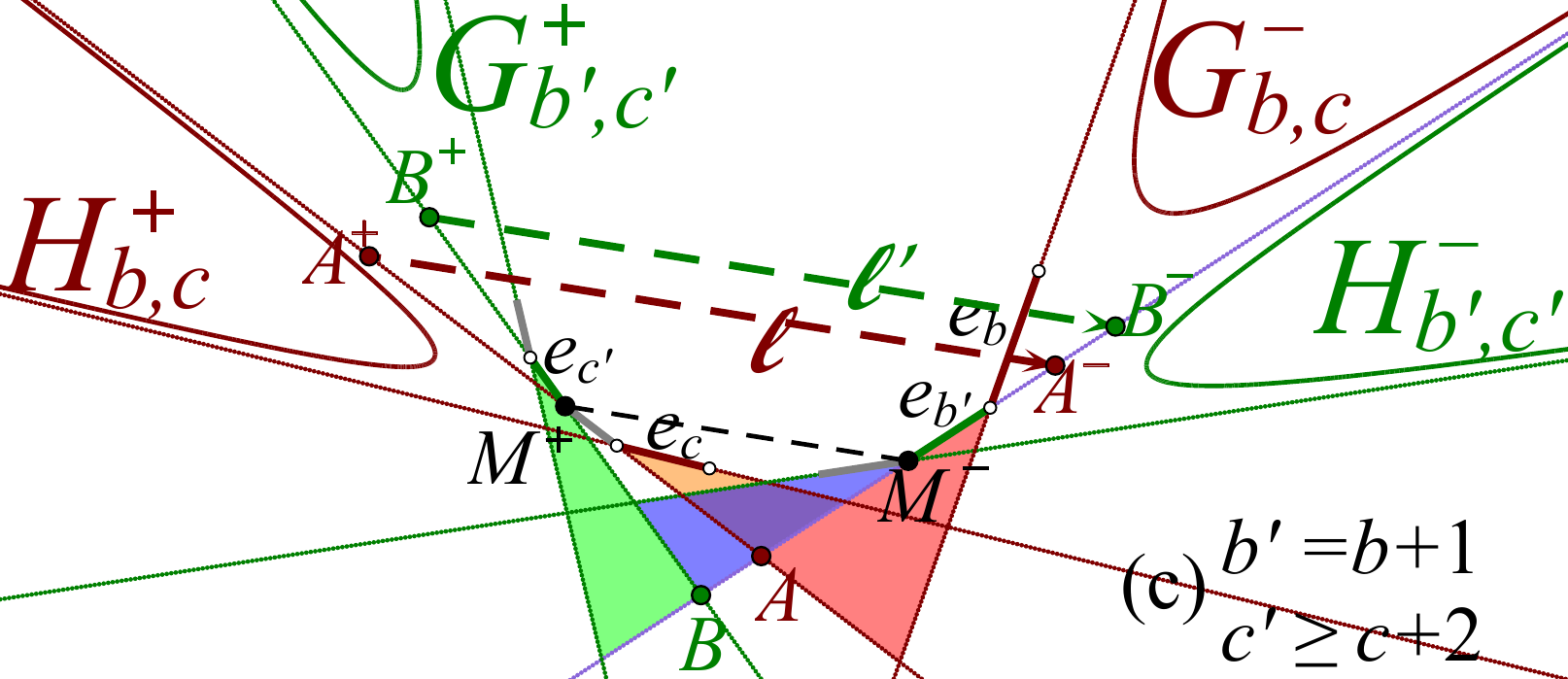} \quad \includegraphics[width=.44\textwidth]{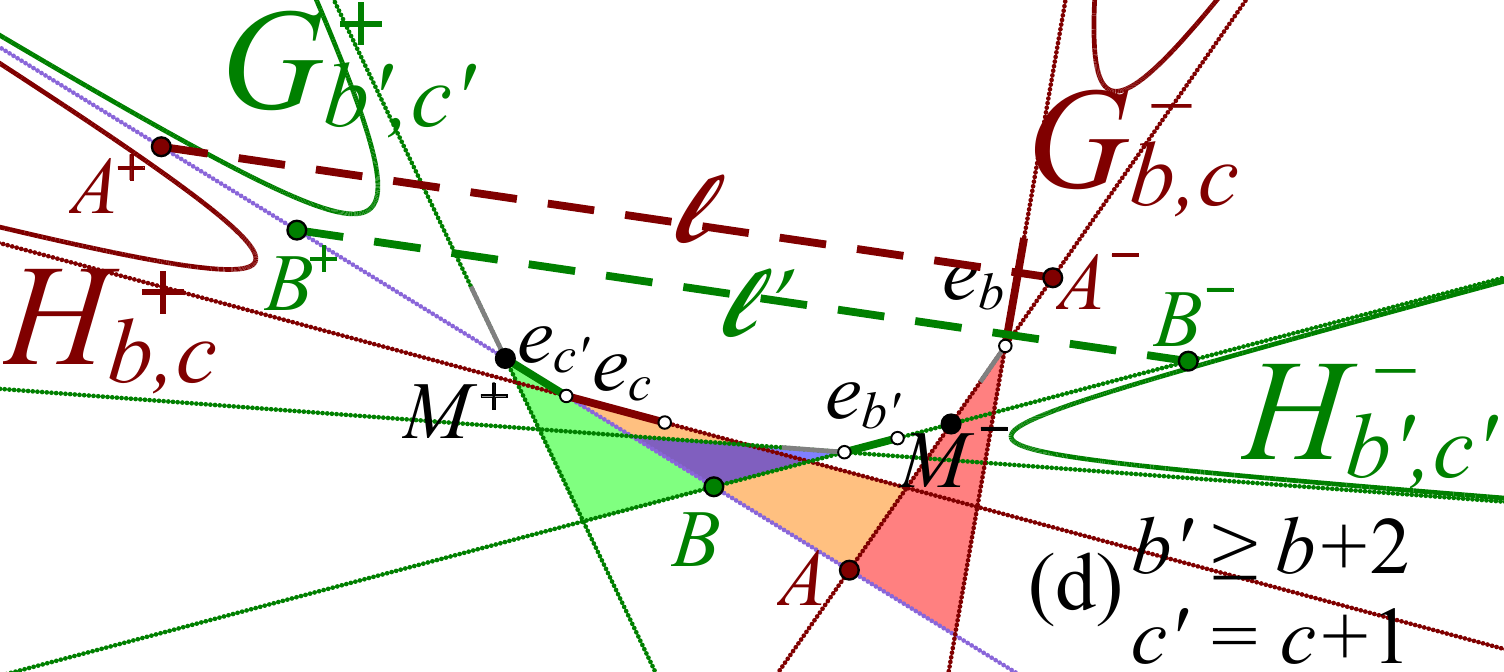} \bigskip
		\end{minipage}
		\begin{minipage}[b]{1\textwidth}
			\vspace{10pt}
			\centering \includegraphics[width=.3\textwidth]{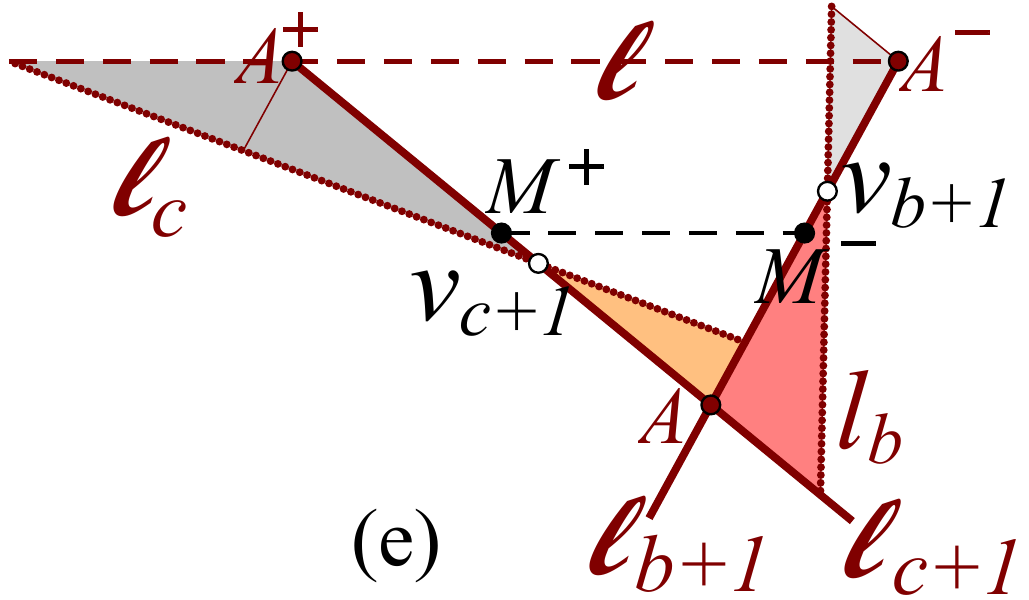} \qquad \qquad \qquad \includegraphics[width=.3\textwidth]{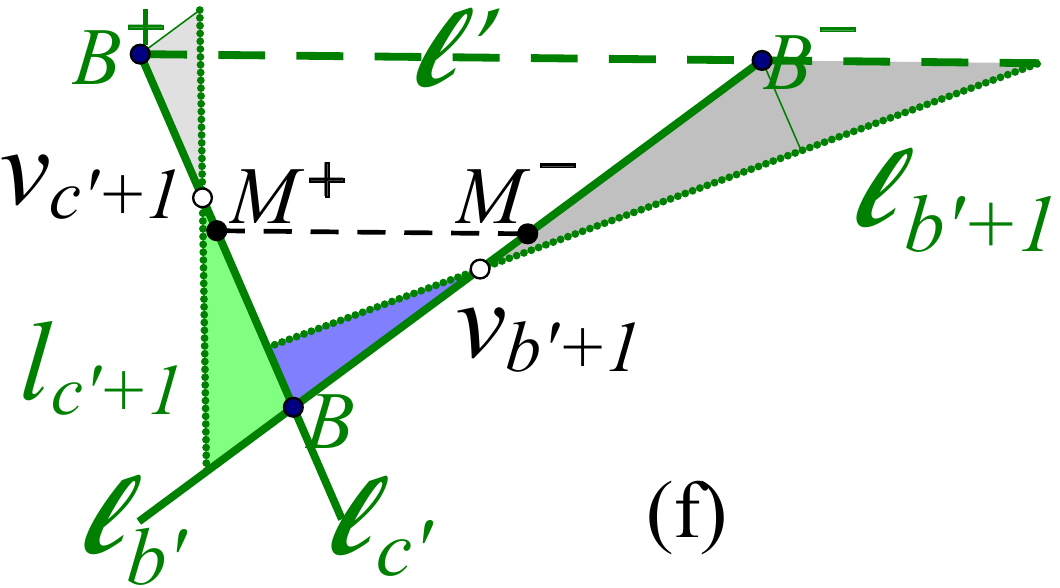}
		\end{minipage}
		\caption{Proof of Observation~\ref{obs:d-mono-2}~part~I.}\label{fig:lemma-mono1}
	\end{figure}
	
	By the definition of $\ell$ and $\ell'$, we have $\theta[\ell]=\theta[\overrightarrow{M^+M^-}]=\theta[\ell']$, thus equality (ii) holds.
	
	\noindent \emph{Proof of (i):}
	This reduces to showing that (i.1) $\ell$ intersects $H^+_{b,c}$ and (i.2) $G^-_{b,c}$ is on the left of $\ell$.
	Now, let us focus on the objects shown in Figure~\ref{fig:lemma-mono1}~(e).
	Notice that $A,v_{c+1},M^+,A^+$ lie in this order in line $\ell_{c+1}$.
	Further since $|AM^+|=|A^+M^+|$, we know $|A^+v_{c+1}|>|Av_{c+1}|$.
	Thus the area of $h\cap (p_c\cap p^C_{c+1})$, where $h$ denotes the half-plane parallel to $\ell_{b+1}$ that admits $A^+$ on its boundary and contains $v_{c+1}$, is larger than the area of the (yellow) triangle $p_{b+1}\cap p^C_c\cap p_{c+1}$, which equals $\area(\hyperbola_{c+1,b+1})$.
	So the triangle area bounded by $\ell_c,\ell_{c+1}$ and $\ell$ is even larger than $\area(\hyperbola_{c+1,b+1})$.
	By Observation~\ref{obs:area-to-intersection}, this means $\ell$ intersects $\hyperbola_{c+1,b+1}$, i.e.\ (i.1) holds.
	
	Assume $A,M^-,v_{b+1},A^-$ lie in this order in $\ell_{b+1}$ (otherwise the order would be $A$, $M^-$, $A^-$, $v_{b+1}$, which is easier).
	Similarly, the area bounded by $\ell_b,\ell_{b+1}$ and $\ell$ is smaller than $\area(\hyperbola_{b+1,c+1})$,
	which by Observation~\ref{obs:area-to-intersection} means that $\hyperbola_{b+1,c+1}$ is on the left of $\ell$, i.e.\ (i.2) holds.
	
	\smallskip \noindent \emph{Proof of (iii):} This reduces to showing that (iii.1) $\ell'$ intersects $H^-_{b',c'}$ and (iii.2) $G^+_{b',c'}$ is on the left of $\ell'$.
	See Figure~\ref{fig:lemma-mono1}~(f); they are symmetric to (i.1) and (i.2) respectively; proof omitted.
	
	\textbf{In the following, assume $b=b'$ or $c=c'$.} We discuss four subcases.
	\begin{itemize}
		\item[Case~1] $b'=b+1,c'=c$. See Figure~\ref{fig:lemma-mono2}~(a). Notice that $H^+_{b,c}=G^+_{b',c'}$.
		Denote by $A^-$ be the reflection of $A$ around $v_{b+1}$ and $B^-$ the reflection of $B$ around $v_{b'+1}$.
		Let $\ell$ be the tangent line of $H^+_{b,c}$ passing through $A^-$,
		and $\ell'$ the tangent line of $G^+_{b',c'}$ passing through $B^-$.
		We argue that (i) $\theta[\ell_{b,c}^{HG}]\leq \theta[\ell] $ and (iii) $\theta[\ell']\leq \theta[\ell_{b',c'}^{GH}]$ still hold in this case.
		They follow from the observations that the triangles with light color in the figure are smaller than their opposite triangles with dark color,
		which is because $|Bv_{b'+1}|=|B^-v_{b'+1}|$ and $|Av_{b+1}|=|A^-v_{b+1}|$.
		Clearly, $A,B,B^-,A^-$ lie in this order in $\ell_{b+1}$, thus $\theta[\ell]<\theta[\ell']$.
		Altogether, $\theta[\ell^{HG}_{b,c}]\leq \theta[\ell^{GH}_{b',c'}]$.
		
		\item[Case~2] $b'=b,c'=c+1$. See Figure~\ref{fig:lemma-mono2}~(b); symmetric to Case~1; proof omitted.
		
		\begin{figure}[h]
			\begin{minipage}[b]{.5\textwidth}
				\centering \includegraphics[width=.72\textwidth]{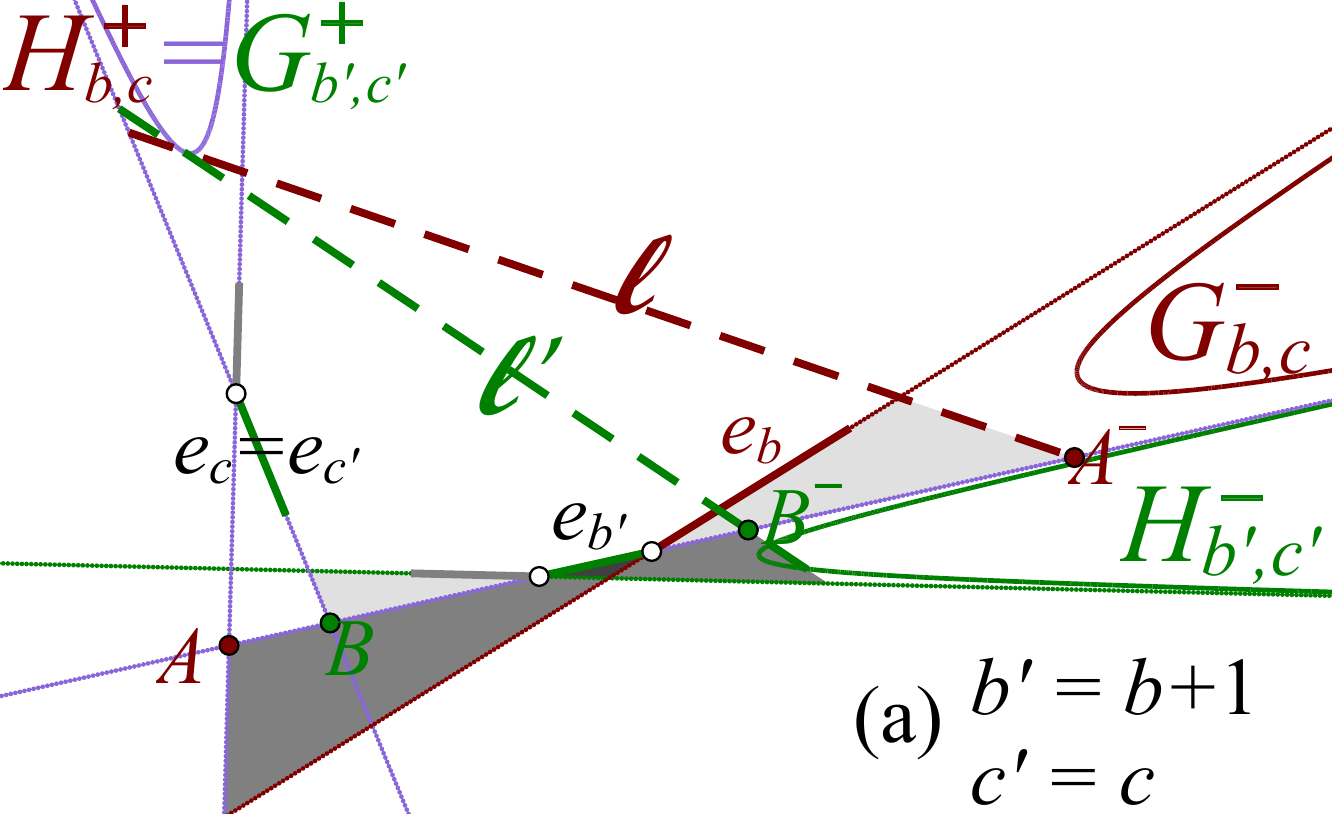}
			\end{minipage}
			\begin{minipage}[b]{.5\textwidth}
				\centering \includegraphics[width=.81\textwidth]{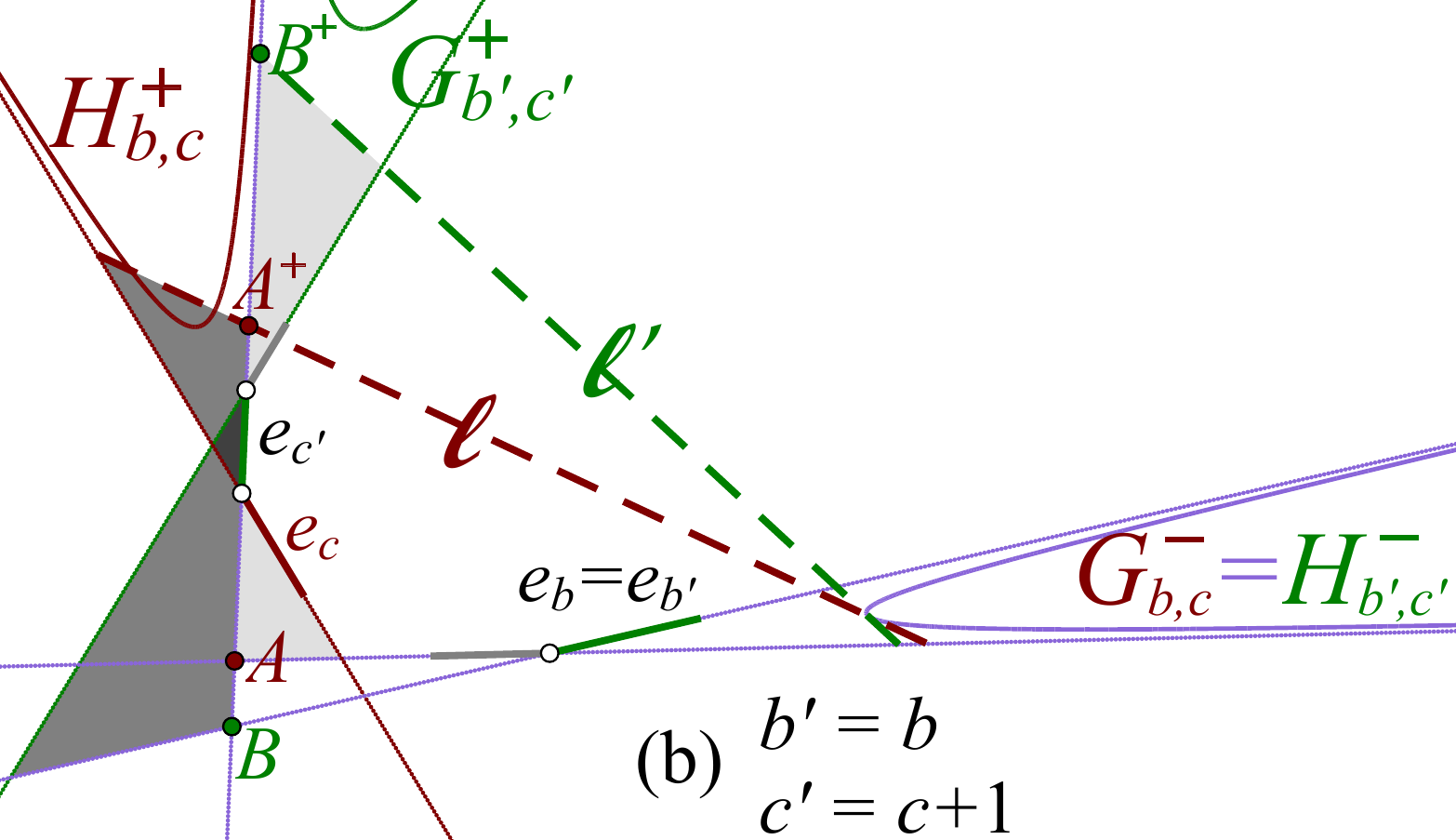}
			\end{minipage}
			\caption{Proof of Observation~\ref{obs:d-mono-2}~part~II.}\label{fig:lemma-mono2}
		\end{figure}
		
		\newpage
		
		\item[Case~3] $b'\geq b+2,c'=c$.
		\textbf{Hint: This case is more difficult because $G^+_{b',c'}$ is now ``below'' $H^+_{b,c}$} as shown in Figure~\ref{fig:lemma-mono3}~(a); \textbf{the following proof contains several more tricks.}
		
		Let $A^-,B^-$ be the reflection of $A,B$ around $M^-$ respectively and $\ell$ be the tangent line of $H^+_{b,c}$ that passes through $A^-$,
		and $\ell'$ the tangent line of $G^+_{b',c'}$ that passes through $B^-$.
		As in the previous cases, (i) and (iii) hold and thus it reduces to showing that $\theta[\ell']>\theta[\ell]$.
		
		See Figure~\ref{fig:lemma-mono3}~(b).
		Make a parallel line $\ell''$ of $\ell$ at $B^-$, which intersects $\ell_c,\ell_{c+1}$ at $B_1,B_2$ respectively.
		It reduces to showing that the area bounded by $\ell'',\ell_c,\ell_{c+1}$, namely $\area(\triangle v_{c+1}B_1B_2)$, is smaller than the area bounded by $\ell',\ell_c,\ell_{c+1}$.
		The latter equals $\area(G^+_{b',c'})=\area(\triangle v_{c+1}BX)$, where $X=\ell_{c+1}\cap \ell_{b'}$.
		Let $X^-$ be the reflection of $X$ around $M^-$. Assume the parallel line of $\ell$ at $X^-$ intersects $\ell_c,\ell_{c+1}$ at $X_1,X_2$ respectively.
		Let $D=\ell_{b+1}\cap \ell_{c}$, and $E$ be the point on $\ell_c$ so that $XE\parallel AD$.
		Clearly, $\area(\triangle v_{c+1}B_1B_2)<\area(\triangle v_{c+1}X_1X_2)$ and $\area(\triangle v_{c+1}BX)>\area(\triangle v_{c+1}EX)$.
		So it further reduces to proving that (I) $\area(\triangle v_{c+1}X_1X_2)<\area(\triangle v_{c+1}EX)$.
		
		Assume $\ell$ intersects $\ell_c,\ell_{c+1}$ at $A_1,A_2$. We know
		(a) $\area(\triangle v_{c+1}A_1A_2)=\area(\triangle v_{c+1}DA)$ since $\ell$ is tangent to $H^+_{b,c}$,
		and so
		(b): $|v_{c+1}A_2|<|v_{c+1}A|$.
		Since segment $A^-X^-$ is a translate of $XA$,
		(c) $|AX|=|A_2X_2|$.
		Combining (b) with (c), $|A_2X_2|/|v_{c+1}A_2|>|AX|/|v_{c+1}A|$, so
		(d) $|v_{c+1}X_2|/|v_{c+1}A_2| <|v_{c+1}X|/|v_{c+1}A|$.
		Further since $X_1X_2\parallel A_1A_2$ whereas $XE\parallel AD$, we have
		(e) $|v_{c+1}X_1|/|v_{c+1}A_1| <|v_{c+1}E|/|v_{c+1}D|$.
		Therefore,
		(f) $\area(\triangle v_{c+1}X_1X_2) / $ $\area(\triangle v_{c+1}A_1A_2) < \area(\triangle  v_{c+1}EX) / \area(\triangle v_{c+1}DA)$.
		Combining (a) and (f) results (I).
		
		\item[Case~4] $b'=b,c'\geq c+2$. Symmetric to Case~3; proof omitted.
		
		\begin{figure}[h]
			\begin{minipage}[b]{.5\textwidth}
				\centering\includegraphics[width=.72\textwidth]{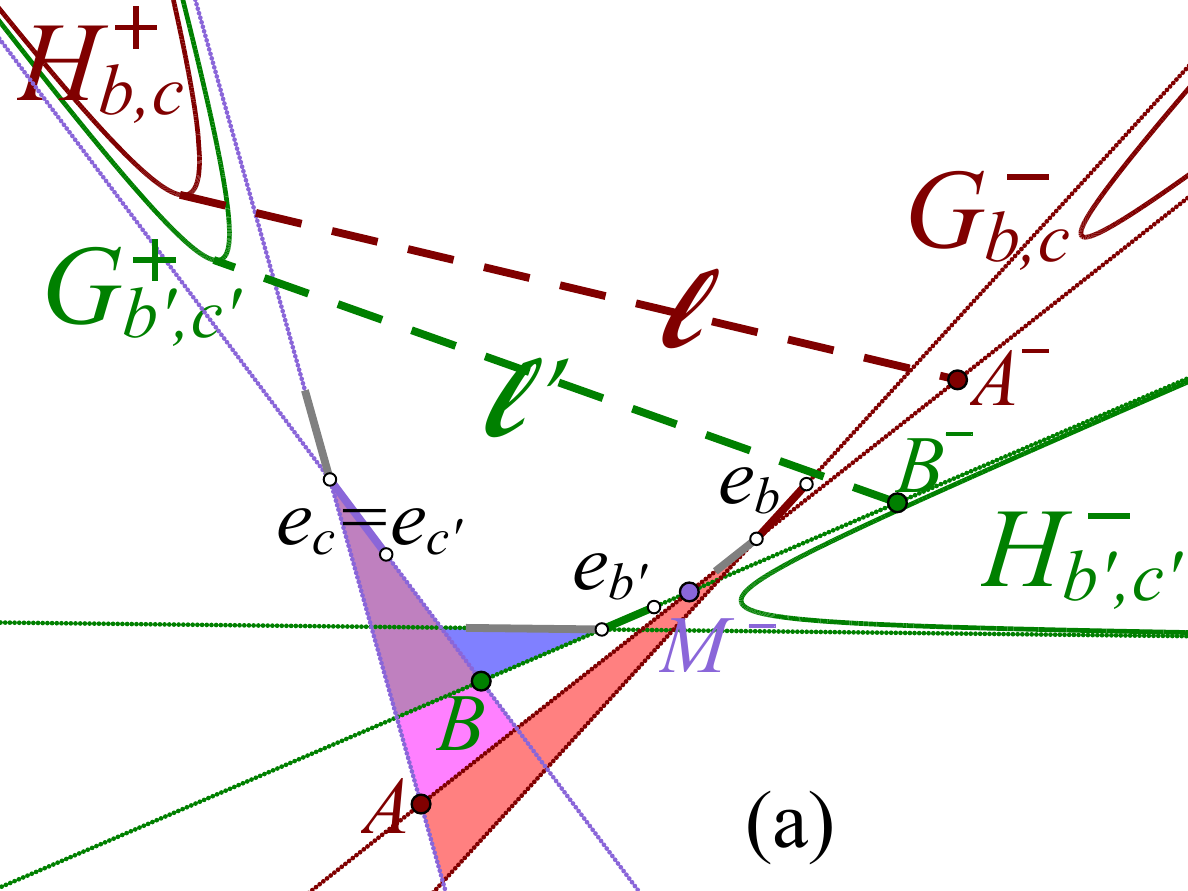}
			\end{minipage}
			\begin{minipage}[b]{.5\textwidth}
				\centering \includegraphics[width=.7\textwidth]{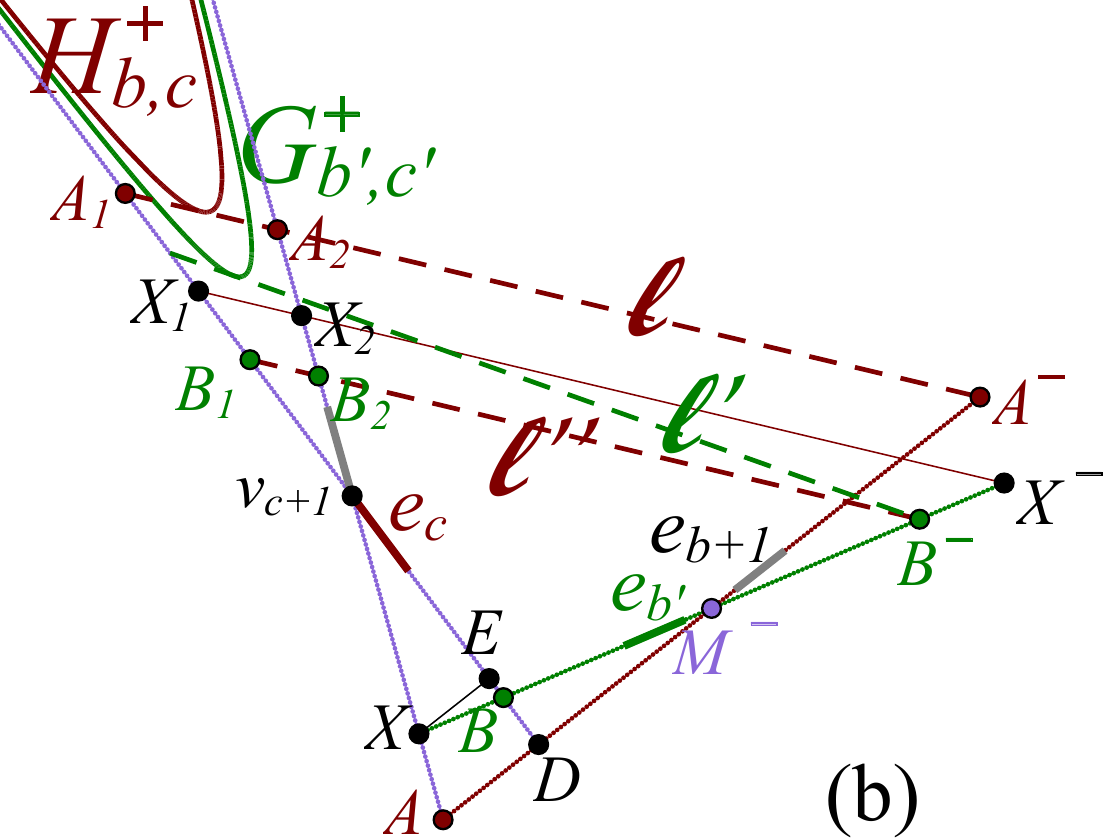}
			\end{minipage}
			\caption{Proof of Observation~\ref{obs:d-mono-2}~part~III.}\label{fig:lemma-mono3}
		\end{figure}
	\end{itemize}
	
\end{proof}

\subsection{Proofs of Lemma~\ref{lemma:area-unimodal}, Lemma~\ref{lemma:OPTmono}, and Theorem~\ref{thm:F-3-stable}}\label{subsect:F-omitted}

\begin{proof}[Proof of Lemma~\ref{lemma:area-unimodal}]
	For each edge pair $(e_i,e_j)$ such that $e_i\nparallel e_j$, let $O_{i,j}$ be the intersecting point between $\ell_i$ and $\ell_j$.
	We classify the edges in $[\D^*_b \circlearrowright \D_c]$ into two categories:
	$e_a$ is \emph{negative} if $|O_{a,c}v_{a+1}|\leq |O_{a,b}v_{a+1}|$ and \emph{positive} otherwise;
	see Figure~\ref{fig:unimodal}.
	We have three observations.
	
	(i) If $e_a$ is positive, its clockwise next edge $e_{a+1}$ must also be positive.
	
	(ii) If $e_a,e_{a+1}$ are both negative, $\area(\triangle e_{a+1}e_be_c)<\area(\triangle e_ae_be_c)$.
	
	(iii) If $e_a,e_{a+1}$ are both positive, $\area(\triangle e_{a+1}e_be_c)>\area(\triangle e_ae_be_c)$.
	
	\begin{figure}[h]
		\centering \includegraphics[width=\textwidth]{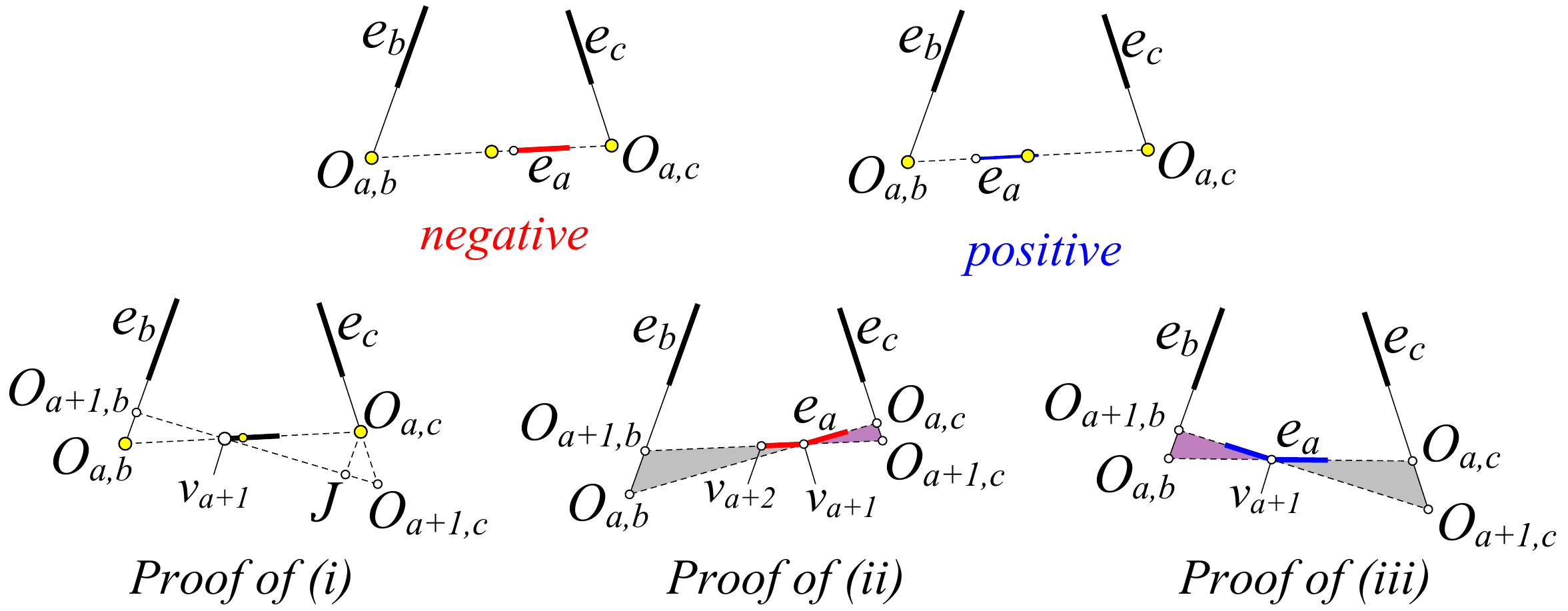}\\
		\caption{Illustration of the proof of Lemma~\ref{lemma:area-unimodal}}\label{fig:unimodal}
	\end{figure}
	
	\noindent \emph{Proof of (i):} Make a line at $O_{a,c}$ parallel to $\ell_b$ and assume it intersects $\ell_{a+1}$ at $J$.
	Since $e_a$ is positive, $|O_{a,c}v_{a+1}|>|O_{a,b}v_{a+1}|$.
	It implies that $|Jv_{a+1}| > |O_{a+1,b}v_{a+1}|$. Therefore, $|O_{a+1,c}v_{a+1}|>|O_{a+1,b}v_{a+1}|$.
	Furthermore, $|O_{a+1,c}v_{a+2}|>|O_{a+1,b}v_{a+2}|$, i.e., $e_{a+1}$ is positive.\smallskip
	
	\noindent \emph{Proof of (ii):} Because $e_a,e_{a+1}$ are negative, $|O_{a,c}v_{a+1}|\leq |O_{a,b}v_{a+1}|$
	and $|O_{a+1,c}v_{a+1}| < |O_{a+1,b}v_{a+1}|$.
	Therefore, $|O_{a,c}v_{a+1}|\cdot |O_{a+1,c}v_{a+1}|<|O_{a,b}v_{a+1}|\cdot |O_{a+1,b}v_{a+1}|$.
	In other words, $\triangle v_{a+1}O_{a,c}O_{a+1,c}$ is smaller than $\triangle v_{a+1}O_{a,b}O_{a+1,b}$, i.e., $\area(\triangle e_{a+1}e_be_c)<\area(\triangle e_ae_be_c)$.
	
	\noindent \emph{Proof of (iii):} Because $e_a$ is positive, $|O_{a,c}v_{a+1}|>|O_{a,b}v_{a+1}|$ and $|O_{a+1,c}v_{a+1}|>|O_{a+1,b}v_{a+1}|$ (use the proof of (i)).
	Therefore, $|O_{a,c}v_{a+1}|\cdot |O_{a+1,c}v_{a+1}|>|O_{a,b}v_{a+1}|\cdot |O_{a+1,b}v_{a+1}|$,
	i.e.\ $\triangle v_{a+1}O_{a,c}O_{a+1,c}$ is larger than $\triangle v_{a+1}O_{a,b}O_{a+1,b}$, i.e., $\area(\triangle e_{a+1}e_be_c)>\area(\triangle e_ae_be_c)$.
	
	\medskip Next, we prove this lemma from observations~(i)--(iii).
	Assume $e_{p-1}$ is negative and $e_{p}$ is positive (the case where all edges are positive or all edges are negative is easier and can be proved similarly). Applying (i), $e_y,\ldots,e_{p-1}$ are negative whereas $e_p,\ldots, e_z$ are positive.
	Denote $s_a=\area(\triangle e_ae_be_c)$.
	
	As $e_y,\ldots,e_{p-1}$ are negative, applying (ii), we get (A): $s_y>s_{y+1}>\ldots>s_{p-1}$.
	
	As $e_p,\ldots, e_z$ are positive, applying (iii), we get (B): $s_p<s_{p+1}<\ldots<s_z$.
	
	Together, $x$ equals either $p-1$ or $p$.
	By definition of $x$, we also have (C): $s_{x-1}>s_x$.
	
	No matter $x$ equals $p-1$ or $p$, we obtain the following from formulas (A), (B), and (C):
	
	1. $s_a$ \textbf{\emph{strictly}} decreases when $a$ goes from $y$ to $x$; and
	
	2. $s_a$ \textbf{\emph{strictly}} increases when $a$ goes from $x+1$ to $z$.
\end{proof}

\begin{figure}[h]
	\begin{minipage}[b]{0.55\textwidth}
		\centering \includegraphics[width=\textwidth]{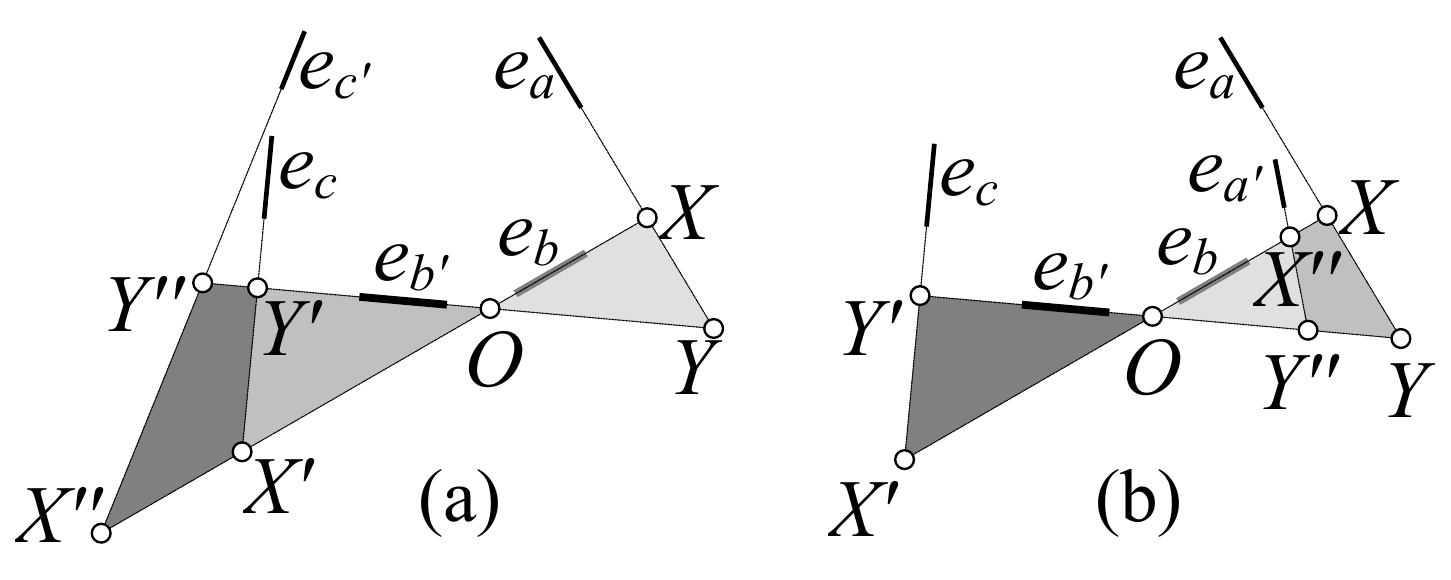}\\
		\caption{Illustration of Observation~\ref{Obs:back-forw-F}.}\label{fig:Obs-back-forw-F}
	\end{minipage}
	\begin{minipage}[b]{0.45\textwidth}
		\centering \includegraphics[width=.85\textwidth]{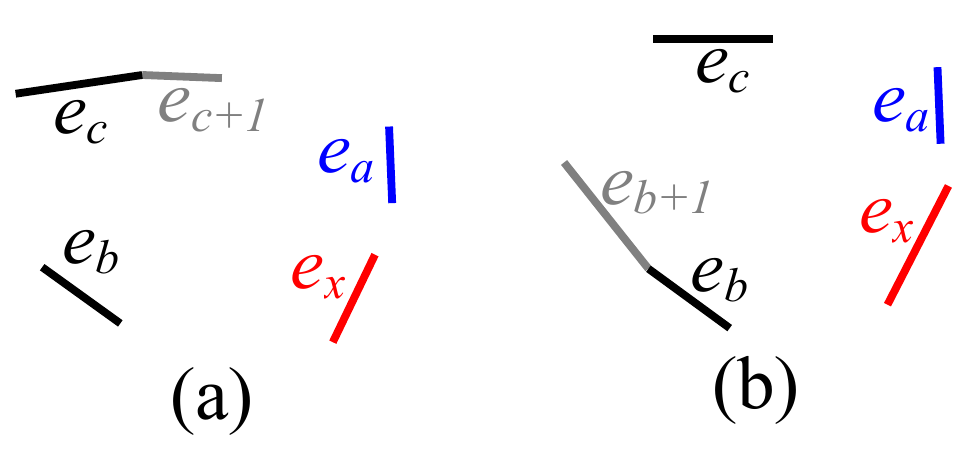}\\
		\caption{Proof of Lemma~\ref{lemma:OPTmono}.}\label{fig:OPTmono}
	\end{minipage}
\end{figure}

Next, we prove an observation, which is preliminary to the proof of Lemma~\ref{lemma:OPTmono} and Lemma~\ref{lemma:interleaving}.
\begin{observation}\label{Obs:back-forw-F}
	1. Assume $e_a,e_b,e_{b'},e_c,e_{c'}$ are distinct and lie in clockwise order (see Figure~\ref{fig:Obs-back-forw-F}~(a)).
	Suppose $\area(\triangle e_ae_{b'}e_c)$ and $\area(\triangle e_ae_{b'}e_{c'})$ are both finite -- this means $e_a\prec e_{b'},e_{b'}\prec e_{c'},e_c\prec e_a$.
	Then, $\area(\triangle e_ae_{b'}e_c) \leq \area(\triangle e_ae_be_c)$ implies that $\area(\triangle e_ae_{b'}e_{c'}) <\area(\triangle e_ae_be_{c'})$.
	
	2. Assume $e_a,e_{a'},e_b,e_{b'},e_c$ are distinct and lie in clockwise order (see Figure~\ref{fig:Obs-back-forw-F}~(b)).
	Suppose $\area(\triangle e_ae_{b'}e_c)$ and $\area(\triangle e_{a'}e_{b'}e_c)$ are both finite -- this means $e_a\prec e_{b'},e_{b'}\prec e_c, e_c\prec e_{a'}$.
	Then, $\area(\triangle e_ae_{b'}e_c) \leq \area(\triangle e_ae_be_c)$ implies that $\area(\triangle e_{a'}e_{b'}e_c) <\area(\triangle e_{a'}e_be_c)$.
\end{observation}

\begin{proof}[Proof of Observation~\ref{Obs:back-forw-F}] We only prove part~1;
	the proof of part~2 is similar and is omitted.
	
	See Figure~\ref{fig:Obs-back-forw-F}~(a). By the assumption, $\area(\triangle e_ae_{b'}e_{c'})$ is finite.
	Assume $\area(\triangle e_ae_be_{c'})$ is also finite (hence $e_b\prec e_{c'}$), otherwise it is trivial that
	$\area(\triangle e_ae_{b'}e_{c'}) <\area(\triangle e_ae_be_{c'})$.
	
	Assume $\ell_b$ intersects $\ell_a,\ell_c,\ell_{c'}$ at $X,X',X''$ and $\ell_{b'}$ intersects $\ell_a,\ell_c,\ell_{c'}$ at $Y,Y',Y''$, respectively. Let $O=\ell_b\cap \ell_{b'}$.
	Because $\area(\triangle e_ae_{b'}e_c) \leq \area(\triangle e_ae_be_c)$, $\area(\triangle OXY)\leq \area(\triangle OX'Y')$.
	Hence $\area(\triangle OXY)<\area(\triangle OX''Y'')$, i.e. $\area(\triangle e_ae_{b'}e_{c'}) <\area(\triangle e_ae_be_{c'})$.
\end{proof}

\begin{proof}[Proof of Lemma~\ref{lemma:OPTmono}]
	1. First, assume $\D^*_b=\D_c$. In this case, $\OPT_{b,c}$ is the previous edge of $\D_c$ (by definition).
	If $\D_{c+1}=\D_c=\D^*_b$, edges $\OPT_{b,c},\OPT_{b,c+1}$ are both equal to the previous edge of $\D_c$.
	If $\D_{c+1}\neq \D_c$, edge $\OPT_{b,c+1}$ lies in $[\D^*_b\circlearrowright \D_{c+1}]=[\D_c\circlearrowright \D_{c+1}]$
	and hence lies after $\OPT_{b,c}$ (clockwise).
	
	Assume $\D^*_b\neq \D_c$. Let $e_x=\OPT_{b,c}$ and take $e_a$ in $[\D^*_b\circlearrowright v_x]$. See Figure~\ref{fig:OPTmono}~(a).
	As $\OPT_{b,c}=e_x$, $\area(\triangle e_xe_be_c)<\area(\triangle e_ae_be_c)$.
	Applying Observation~\ref{Obs:back-forw-F}, $\area(\triangle e_xe_be_{c+1})<\area(\triangle e_{a}e_be_{c+1})$.
	Therefore, $\OPT_{b,c+1}\neq e_{a}$. This means $\OPT_{b,c+1}$ lies in $[v_x\circlearrowright \D_{c+1}]$ and hence equals or lies after $\OPT_{b,c}$.
	
	\medskip \noindent 2. When $\D^*_b=\D_c$, it is trivial that $\D^*_{b+1}=\D_c$. Therefore, in this case both $\OPT_{b,c}$ and $\OPT_{b+1,c}$ equal to the previous edge of $\D_c$. Next, assume that $\D^*_b\neq \D_c$. Let $e_x=\OPT_{b,c}$.
	Similar as in the above case, we shall prove that $\OPT_{b+1,c}\neq e_a$ for any $e_a$ in $[\D^*_b\circlearrowright v_x]$.
	See Figure~\ref{fig:OPTmono}~(b).
	
	If it holds that $e_a$ is not chasing $e_{b+1}$, we can get $\OPT_{b+1,c}\neq e_a$ immediately
	as $\area(\triangle e_ae_{b+1}e_c)$ is infinite in this case. (To make it rigorous, we need to show that $\OPT_{b+1,c}\neq e_a$ even if
	$\D^*_{b+1}=\D_c$. In that case, $\OPT_{b+1,c}$ is defined as the previous edge of $\D_c$, which is in $[v_x\circlearrowright \D_c]$ and cannot be $e_a$.)
	
	Assume now $e_a\prec e_{b+1}$. Because $\OPT_{b,c}=e_x$, $\area(\triangle e_xe_be_c)<\area(\triangle e_ae_be_c)$.
	Applying Observation~\ref{Obs:back-forw-F}, $\area(\triangle e_xe_{b+1}e_c)<\area(\triangle e_ae_{b+1}e_c)$. It follows that $\OPT_{b+1,c}\neq e_a$.
\end{proof}

\begin{proof}[Proof of Theorem~\ref{thm:F-3-stable}]
	\emph{Correctness.} Recall the proof of correctness in section~\ref{sect:3-stable}.
	That proof can easily be adapted to show the correctness of Algorithm~\ref{alg:RK-F}.
	
	\noindent \emph{Running time.}
	Because $e_b,e_c$ move clockwise during the Rotate-and-Kill process, $\OPT_{b,c}$ moves clockwise by the bi-monotonicity of $\OPT_{b,c}$ (Lemma~\ref{lemma:OPTmono}) and thus can be computed in amortized $O(1)$ time using the unimodality of $\area(\triangle e_ae_be_c)$ (Lemma~\ref{lemma:area-unimodal}).
	Checking whether an all-flush triangle $\triangle e_ae_be_c$ is F-3-stable (not necessarily) reduces to checking if $e_a,e_b,e_c$ are stable in this triangle, which takes $O(1)$ time due to the unimodality in Lemma~\ref{lemma:area-unimodal}.
	Computing $\Kill_F(b,c)$ takes amortized $O(1)$ time according to property \textcircled{2}. Altogether, Algorithm~\ref{alg:RK-F} runs in $O(n)$ time.
\end{proof}

\section{Find one 3-stable and one F-3-stable triangle in $O(n)$ time}\label{sect:one}

\begin{description}
	\item[forw-stable.]
	$v_i$ is \emph{back-stable} in $\triangle v_i v_j v_k$ if $\area(\triangle v_i v_j v_k)\geq \area(\triangle v_{i-1} v_j v_k)$ (or $i-1=k$).
	\item [back-stable.]
	$v_i$ is \emph{forw-stable} in $\triangle v_i v_j v_k$ if $\area(\triangle v_i v_j v_k)\geq \area(\triangle v_{i+1} v_j v_k)$ (or $i+1=j$).\\
\end{description}

We define \emph{back-stable} and \emph{forw-stable} for $v_j$ and $v_k$ similarly (note $v_i,v_j,v_k$ in clockwise order).
Observe that if $v_i$ is back-stable and forw-stable, it is stable in $\triangle v_i v_j v_k$.
(This follows from a unimodality of function $d(X)$ for $X\in [v_j\circlearrowright v_k]$ where $d(X)$ denotes the distance from $X$ to $\overleftrightarrow{v_jv_k}$.)

Our algorithm for finding one 3-stable triangle consists of two steps, as shown in the following.

\noindent \textbf{Step~1.} Find the largest area triangle $\triangle v_rv_sv_t$ with $r=1$. See the pseudo code in Algorithm~\ref{alg:one-3-stable-Step1}.

\begin{algorithm}[h]
	\caption{Step 1 of our algorithm for finding a 3-stable triangle.}\label{alg:one-3-stable-Step1}
	$r=1,s=2,t=3,k=3$\;
	\For{$j$= $2$ to $n-1$}{
		\lIf{$k=j$}{$k\leftarrow k+1$}
		\lWhile{$v_k$ is not forw-stable in $\triangle v_rv_jv_k$}{$k\leftarrow k+1$ \label{code:k++}}
		\lIf{$\area(\triangle v_rv_jv_k) > \area(\triangle v_rv_sv_t)$} {$(s,t)\leftarrow (j,k)$}
	}
\end{algorithm}

After Line~\ref{code:k++}, $v_k$ will be the vertex with the largest distance to $\overleftrightarrow{v_iv_j}$ among all vertices on the right of $\overrightarrow{v_iv_j}$, and thus $\triangle v_rv_jv_k$ will be the triangle with the largest area among all those triangles rooted at $v_r,v_j$.
Further since we try every $j$ at the for-loop, Algorithm~\ref{alg:one-3-stable-Step1} finds the largest area triangle with one corner at $v_1$.
Obviously, this algorithm runs in $O(n)$ time.

\medskip \noindent \textbf{Step~2.}
This step is shown in Algorithm~\ref{alg:one-3-stable-Step2} and its analysis is given by the following lemma.

\begin{algorithm}[h]
	\caption{Step 2 of our algorithm for finding a 3-stable triangle.}\label{alg:one-3-stable-Step2}
	\KwIn{$r,s,t$ output by step~1.}
	\If{$v_r$ is not forw-stable (in $\triangle v_r v_s v_t$; which is omitted in the following)}{
		\While{$v_r$ is not forw-stable\label{code:alg-one-while1}}{
			$r\leftarrow r+1$ \label{code:alg-one-r++}\;
			\Repeat{none of the above two conditions hold}{
				\lWhile{$v_s$ is not forw-stable}{$s\leftarrow s+1$\label{code:alg-one-s++}}
				\lWhile{$v_t$ is not forw-stable}{$t\leftarrow t+1$\label{code:alg-one-t++}}
			}
		}
	}
	\lElse{
		\While{$v_r$ is not back-stable\label{code:alg-one-while2}}{
			$r\leftarrow r-1$\;
			\Repeat{none of the above two conditions hold}{
				\lWhile{$v_s$ is not back-stable}{$s\leftarrow s-1$}
				\lWhile{$v_t$ is not back-stable}{$t\leftarrow t-1$}
			}
		}
	}
\end{algorithm}

\begin{lemma}\label{lemma:one-stable-correctness}
	1. Initially (namely, at the beginning of Algorithm~\ref{alg:one-3-stable-Step2}), $v_s,v_t$ are stable in $\triangle v_rv_sv_t$.
	
	2. If $v_r$ is not forw-stable initially, $v_r,v_s,v_t$ will be stable after the while-loop (beginning at Line~\ref{code:alg-one-while1}).
	
	3. If $v_r$ is not back-stable initially, $v_r,v_s,v_t$ will be stable after the while-loop (beginning at Line~\ref{code:alg-one-while2}).
	
	4. The algorithm terminates in $O(n)$ steps.
\end{lemma}

\begin{proof}
	1. Initially, because $\triangle v_rv_sv_t$ is the largest triangle with $r=1$, $v_s,v_t$ are stable in $\triangle v_rv_sv_t$.
	
	\smallskip \noindent
	2. Obviously, \emph{a corner is forw-stable or back-stable}.
	When $v_r$ is not forw-stable, it must be back-stable.
	Therefore, $v_r,v_s,v_t$ are back-stable before entering the while-loop (beginning at Line~\ref{code:alg-one-while1}).
	
	We then argue that $v_r,v_s,v_t$ are back-stable throughout the loop. This follows from two facts:
	
	(i) \emph{A back-stable corner (e.g. the one at $v_r$) remains back-stable when we change another corner (e.g. the one at $v_s$ or $v_t$) forwardly (e.g. $s\leftarrow s+1$ or $t\leftarrow t+1$).} This is due to Observation~\ref{Obs:back-forw} below.
	
	(ii) Upon the execution of $r\leftarrow r+1$ at Line~\ref{code:alg-one-r++},
	it holds that $v_r$ is not forw-stable, which means $v_{r+1}$ is back-stable in $\triangle v_{r+1} v_s v_t$.
	Therefore, after increasing $r$, the new $v_r$ itself is back-stable.
	Similarly, $v_s$ is back-stable after $s\leftarrow s+1$ (Line~\ref{code:alg-one-s++}) and
	$v_t$ is back-stable after $t\leftarrow t+1$ (Line~\ref{code:alg-one-t++}).
	
	On the other hand, according to the conditions of Line~\ref{code:alg-one-while1} and Line~\ref{code:alg-one-while2},
	it is obvious that $v_r,v_s,v_t$ are forw-stable after the while-loop.
	Therefore, $v_r,v_s,v_t$ are stable after the while-loop (beginning at Line~\ref{code:alg-one-while1}).
	(This applies the trivial fact that \emph{a corner is stable when it is forw-stable and back-stable}.)
	
	\smallskip \noindent
	3. The part is symmetric to part~2; proof omitted.
	(We need the following fact which is symmetric to (i): \emph{A forw-stable corner remains forw-stable when we change another corner backwardly.})
	
	\smallskip \noindent
	4. As the area of $\triangle v_r v_s v_t$ increases at every change of $r,s,t$, the algorithm terminates eventually.
	Because $r,s,t$ can only increase throughout whereas $r$ cannot return to $1$ (otherwise it implies a triangle with a corner at $v_1$
	whose area is larger than the triangle given by Step~1, contradiction),
	the total number of changes of $r,s,t$ is bounded by $O(n)$; namely, the algorithm has $O(n)$ steps.
\end{proof}

The following observation is a counterpart of Observation~\ref{Obs:back-forw-F}. Its trivial proof is omitted.

\begin{observation}\label{Obs:back-forw}
	(a) Assume $v_a,v_b,v_{b'},v_c,v_{c'}$ are distinct and lie in clockwise order.
	Then, $\area(\triangle v_av_{b'}v_c) \geq \area(\triangle v_av_bv_c)$ implies that $\area(\triangle v_av_{b'}v_{c'}) >\area(\triangle v_av_bv_{c'})$.
	
	(b) Assume $v_a,v_{a'},v_b,v_{b'},v_c$ are distinct and lie in clockwise order.
	Then, $\area(\triangle v_av_{b'}v_c) \geq \area(\triangle v_av_bv_c)$ implies that $\area(\triangle v_{a'}v_{b'}v_c) >\area(\triangle v_{a'}v_bv_c)$.
\end{observation}

\paragraph*{Extension.}
Our algorithm for finding a 3-stable triangle easily extends to finding a F-3-stable triangle.
In the first step, we find the smallest area all-flush triangle $\triangle e_r e_s e_t$ with $r=1$.
Thus, we obtain $r,s,t$ where $e_s$ and $e_t$ are stable in $\triangle e_r e_s e_t$.
In the second step, if $e_r$ is not stable, we adjust $r,s,t$ using the strategy given in Algorithm~\ref{alg:one-3-stable-Step2}
(of course, $v_r,v_s,v_t$ should be modified to $e_r,e_s,e_t$).

Here, we define the concepts back-stable and forw-stable on edges of an all-flush triangle in the same manner.
Assume that $\area(\triangle e_i e_j e_k)$ is finite.
We state that $e_i$ is \emph{back-stable} in $\triangle e_ie_je_k$ if $\area(\triangle e_i e_j e_k)\leq \area(\triangle e_{i-1} e_j e_k)$ (or $i-1=k$), and $e_i$ is \emph{forw-stable} in $\triangle e_ie_je_k$ if $\area(\triangle e_i e_j e_k)\leq \area(\triangle e_{i+1} e_j e_k)$ (or $i+1=j$).
Obviously,
(i) \emph{If $e_i$ is back-stable and forw-stable in $\triangle e_ie_je_k$, it is stable.}
(ii) \emph{$e_i$ is back-stable or forw-stable in $\triangle e_ie_je_k$.}
These facts follow from Lemma~\ref{lemma:area-unimodal}.
Moreover,
(iii) \emph{A back-stable edge (e.g. the one at $e_r$) remains back-stable when we change another edge (e.g. the one at $e_s$ or $e_t$) forwardly (e.g. $s\leftarrow s+1$ or $t\leftarrow t+1$).}
This fact is due to Observation~\ref{Obs:back-forw-F}.

The analysis of the algorithm for finding a F-3-stable triangle is almost the same as that given in Lemma~\ref{lemma:one-stable-correctness}. (The three underlying facts (i), (ii), and (iii) are keys to the analysis.)

\begin{remark}
	Our algorithm given in section~\ref{sect:one} (denoted by Alg-One) is different from Alg-DS.
	First, step~1 of Alg-One sets the initial value of $(r,s,t)$ differently from the initial value $(1,2,3)$ used by Alg-DS.
	Second, step~2 of Alg-One has two symmetric subroutines (one beginning at Line~\ref{code:alg-one-while1}, and the other beginning at Line~\ref{code:alg-one-while2}), whereas Alg-DS has only one.
	Moreover, Alg-One finds one 3-stable triangle but Alg-DS does not.
	
	The only thing in common between Alg-One and Alg-DS is that
	they both contain steps 3--7 of Algorithm~\ref{alg:one-3-stable-Step2},
	yet this does not indicate that Alg-One originates from Alg-DS.
\end{remark}

\section{Find all G-3-stable triangles in $O(n)$ time}\label{sect:G-3-stable}

\begin{figure}[h]
	\centering \includegraphics[width=.95\textwidth]{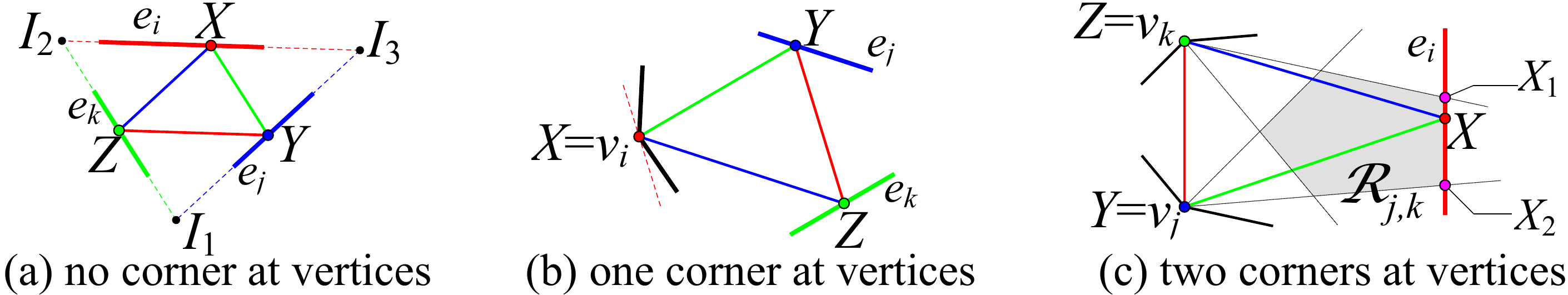}
	\caption{Some G-3-stable triangles that are not 3-stable.}\label{fig:4cases}
\end{figure}

Above all, some G-3-stable triangles that are not 3-stable are shown in Figure~\ref{fig:4cases}.

\begin{note}\label{note:canonical}
	As shown by Figure~\ref{fig:4cases}~(c), the number of G-3-stable triangles could be infinite if an edge and a chord of $P$ are parallel. In this case, the only two extremal triangle $\triangle X_1YZ$ and $\triangle X_2YZ$ are \emph{canonical}. G-3-stable triangles shown in (a) and (b) and 3-stable triangles are also \emph{canonical}.
\end{note}

Given $r,s,t$ so that $\triangle v_rv_sv_t$ is 3-stable, in this section we design an algorithm for finding all (canonical) G-3-stable triangles. This algorithm generalizes the one for finding all 3-stable triangles.


For ease of discussion, call each edge or vertex of $P$ a \emph{unit}.
Denote the $2n$ units $v_1,e_1,\ldots,v_n,e_n$ by $u_1,\ldots,u_{2n}$.
Assume $u_{r'}=v_r,u_{s'}=v_s,u_{t'}=v_t$ (i.e., $r'=2r-1,s'=2s-1,t'=2t-1$).
Denote
$$\U_{r,s,t}=\{(u_b,u_c)\mid b\in \{s',s'+1,\ldots,t'\}, c\in \{t',t'+1,\ldots,r'\},(b,c)\notin \{(t'-1,t'),(t',t')\} \}.$$

A unit pair $(u_b,u_c)$ is \emph{G-dead} if there is no
G-3-stable $\triangle XYZ$ such that $Y\in u_b$ and $Z\in u_c$ (and that $X,Y,Z$ lie in clockwise order).
Be aware that we assume \emph{the edges of $P$ do \textbf{not} contain their endpoints}.
Therefore, when we state $X\in e_i$, we disallow $X$ to be equal to any endpoint of $e_i$.
When we state $Y\in u_i$, we disallow $Y$ to be equal to any endpoint of $u_i$ if $u_i$ is an edge.

Be aware that \textbf{G-dead is different from dead} for $(u_b,u_c)=(v_j,v_k)$. G-dead implies dead but the reverse is not true. For example,
in Figure~\ref{fig:4cases}~(c), $(v_j,v_k)$ is dead but not G-dead.

%
%

\begin{algorithm}[b]
	\caption{Computing the edge pairs and vertex pairs in $\U_{r,s,t}$ that are not G-dead.}\label{alg:RK-G}
	$(b,c)\leftarrow (s',t')$; ~ $a\leftarrow r$;~ $\List\leftarrow \{(u_b,u_c)\}$\;
	\Repeat{$(b,c)=(t',r')$}
	{
		Let $j$ be the index so that $u_b\in \{v_j,e_{j-1}\}$ and $k$ be the index so that $u_c\in \{v_k,e_{k-1}\}$\;
		Compute the vertex $v_a$ on the right of $\overrightarrow{v_jv_k}$ furthest to $\overleftrightarrow{v_jv_k}$\;
		\lIf {$\Kill_G(b,c)$=`b'} {$b\leftarrow b+1$} \lElse {$c\leftarrow c+1$}
		$\List\leftarrow \List\cup\{(u_b,u_c)\}$\;
	}
\end{algorithm}

\paragraph*{Finding all G-3-stable triangles (sketch).}
Using a Rotate-and-Kill process (which is shown in Algorithm~\ref{alg:RK-G}),
we find out \emph{all the edge pairs and vertex pairs in $\U_{r,s,t}$ that are not G-dead}.
(There could be pairs of a vertex and an edge in $\U_{r,s,t}$ that are not G-dead, but we do not need to find them out.)
The function (oracle) for killing is $\Kill_G$, whose description is given right below in \eqref{eqn:def-Kill-G}.

Denote the set of unit pairs output by the above process by $\List_{r,s,t}$, which contains $O(n)$ unit pairs.
Using the same method, we compute two other set of unit pairs $\List_{s,t,r}$ and $\List_{t,r,s}$.
Given $\List_{r,s,t}$, $\List_{s,t,r}$, $\List_{t,r,s}$, we then compute all the G-3-stable triangles.
This last step is not difficult (by observing that if $\triangle X_1X_2X_3$ is G-3-stable,
it has a pair of neighboring pairs $X_i,X_{i+1}$ which lie in two vertices simultaneously or lie in two edges simultaneously,
and the vertex pair or edge pair containing $X_i,X_{i+1}$ belongs to $\List_{r,s,t} \cup \List_{s,t,r} \cup \List_{t,r,s}$).
See the algorithm in subsection~\ref{subsect:list-to-G-3stable}.

\paragraph*{Description of $\Kill_G$.} Recall Definition~\ref{def:R} for $H_{j,k},I_{j,k},J_{j,k},K_{j,k}$.
For each $(u_b,u_c)\in \U_{r,s,t}$, define
\begin{equation}\label{eqn:def-Kill-G}
	\Kill_G(b,c)=\left\{
	\begin{array}{ll}
		\text{`b'}, & \hbox{$(u_b,u_c)=(v_j,v_k)$ and $v_a\leq I_{j,k}$;} \\
		\text{`c'}, & \hbox{$(u_b,u_c)=(v_j,v_k)$ and $v_a>I_{j,k}$;} \\
		\text{`b'}, & \hbox{$(u_b,u_c)=(e_{j-1},e_{k-1})$ and $v_a\leq H_{j,k}$;} \\
		\text{`c'}, & \hbox{$(u_b,u_c)=(e_{j-1},e_{k-1})$ and $v_a>H_{j,k}$;} \\
		\text{`b'}, & \hbox{$(u_b,u_c)=(v_j,e_{k-1})$ and $v_a< J_{j,k}$;} \\
		\text{`c'}, & \hbox{$(u_b,u_c)=(v_j,e_{k-1})$ and $v_a\geq J_{j,k}$;} \\
		\text{`b'}, & \hbox{$(u_b,u_c)=(e_{j-1},v_k)$ and $v_a\leq K_{j,k}$;} \\
		\text{`c'}, & \hbox{$(u_b,u_c)=(e_{j-1},v_k)$ and $v_a> K_{j,k}$,}
	\end{array}
	\right.
\end{equation}
where $v_a$ denotes the vertex on the right of $\overrightarrow{v_jv_k}$ furthest to $\overleftrightarrow{v_jv_k}$,
and where ``$X<Y$'' means that $X$ is closer than $Y$ \emph{to $\overleftrightarrow{v_jv_k}$} (and ``$\leq,>,\geq$'' are defined similarly).
Definition (\ref{eqn:def-Kill-G}) generalizes (\ref{eqn:def-Kill}).

\begin{note}
	In the vertex-edge case ``$(u_b,u_c)=(v_j,e_{k-1})$'' the condition is ``$<$'' or ``$\geq$'', whereas in the other three cases it is
	``$\leq$'' or ``$>$''. This is not a typo although it looks like one for the readers.
\end{note}

An illustration of Algorithm~\ref{alg:RK-G} by a concrete example can be found in appendix~\ref{subsect:example-r-k}.

\subsection*{Analysis of Algorithm~\ref{alg:RK-G}}

The following lemma is crucial to the analysis of Algorithm~\ref{alg:RK-G}. It is proved in the next subsections.

\begin{lemma}\label{lemma:G-correct}
	Assume $(u_b,u_c)\in \U_{r,s,t}$.
	\begin{enumerate}
		\item If $\Kill_G(b,c)=$ `b', all the edges pairs and vertex pairs in $\{(u_b,u_{c+1}),\ldots,(u_b,u_{r'})\}$ are G-dead.
		\item If $\Kill_G(b,c)=$ `c', all the edges pairs and vertex pairs in $\{(u_{b+1},u_c),\ldots,(u_{t'},u_c)\}$ are G-dead.
	\end{enumerate}
\end{lemma}

In fact, the following stronger statements hold (but these statements are neither used nor proved in this manuscript).
1.~If $\Kill_G(b,c)=$ `b', all the unit pairs $(u_b,u_{c+1}),\ldots,(u_b,u_{r'})$ are G-dead.
2.~If $\Kill_G(b,c)=$ `c', all the unit pairs $(u_{b+1},u_c),\ldots,(u_{t'},u_c)$ are G-dead.

\begin{theorem}
	In $O(n)$ time, Algorithm~\ref{alg:RK-G} computes a set of $O(n)$ unit pairs, namely $\List$, which contains all the edge pairs and vertex pairs in $\U_{r,s,t}$ that are not G-dead.
\end{theorem}

\begin{proof}
	Because $\triangle v_rv_sv_t=\triangle u_{r'}u_{s'}u_{t'}$ is 3-stable, $(u_{t'},u_{r'})$ is not G-dead.
	This implies that the algorithm eventually arrives at a state with $(b,c)=(t',r')$, at which it terminates.
	Upon termination, each edge pair and vertex pair in $\U_{r,s,t}$ that is not G-dead is in $\List$, as a corollary of Lemma~\ref{lemma:G-correct}.
	
	(To be rigorous, we also need to prove that $(b,c)\notin \{(t'-1,t'),(t',t')\}$ throughout the process.
	This easily follows from the fact that $\Kill_G(t'-2,t')=$`c', which is implicitly shown in Note~\ref{note:rigorous}.)
	
	Throughout Algorithm~\ref{alg:RK-G}, indices $b,c$ and hence $j,k$ increase monotonously.
	This implies that $v_a$ can be computed in $O(1)$ amortized time.
	Therefore, computing $\Kill_G(b,c)$ only takes $O(1)$ amortized time. So, Algorithm~\ref{alg:RK-G} runs in $O(n)$ time.
\end{proof}


\subsection{Preliminaries for proving Lemma~\ref{lemma:G-correct}}\label{subsect:G-correct-pre}

\begin{observation}\label{obs:EEE-VEE-EVV}
	Assume $\triangle XYZ$ is G-3-stable ($X,Y,Z$ in clockwise order). Recall Figure~\ref{fig:4cases}.
	
	1. If $X,Y,Z$ lie in $e_i,e_j,e_k$ respectively (recall that we disallow $X$ to be equal to any endpoint of $e_i$ when we state that $X_i$ lies in $e_i$),
	then $e_i,e_j,e_k$ are distinct and in clockwise order and $X,Y,Z$ are respectively
	the midpoints of $I_2I_3$, $I_3I_1$, and $I_1I_2$,
	where $I_1,I_2,I_3$ respectively denote the intersecting point between $\ell_j,\ell_k$, and that between $\ell_k,\ell_i$, and that between $\ell_i,\ell_j$.
	
	2. If $X$ lies at a vertex $v_i$ and $Y,Z$ lie in $e_j,e_k$ respectively,
	then $v_i,e_j,e_k$ are in clockwise order, and
	$e_j$ cannot admit $v_i$ as an endpoint (otherwise $Y$ cannot be stable) and
	$e_k$ cannot admit $v_i$ as an endpoint (otherwise $Z$ cannot be stable),
	and $Y$ is the unique point in $e_j$ such that $v_iY \parallel e_k$, and
	$Z$ is the unique point in $e_k$ such that $v_iZ \parallel e_j$.
	
	3. If $X$ lies in $e_i$ and $Y,Z$ lie at vertices $v_j,v_k$ respectively,
	then $e_i,v_j,v_k$ are in clockwise order and $v_jv_k\parallel e_i$ (otherwise $X$ cannot be stable).
	Moreover, $X$ lies in $X_1X_2$ -- the intersecting segment of $e_i$ and $\R_{j,k}$.
	In this case, the area of $\triangle X'v_jv_k$ is a constant for any $X'$ in $X_1X_2$.
\end{observation}

\begin{figure}[b]
	\begin{minipage}[b]{0.55\textwidth}
		\centering \includegraphics[width=.85\textwidth]{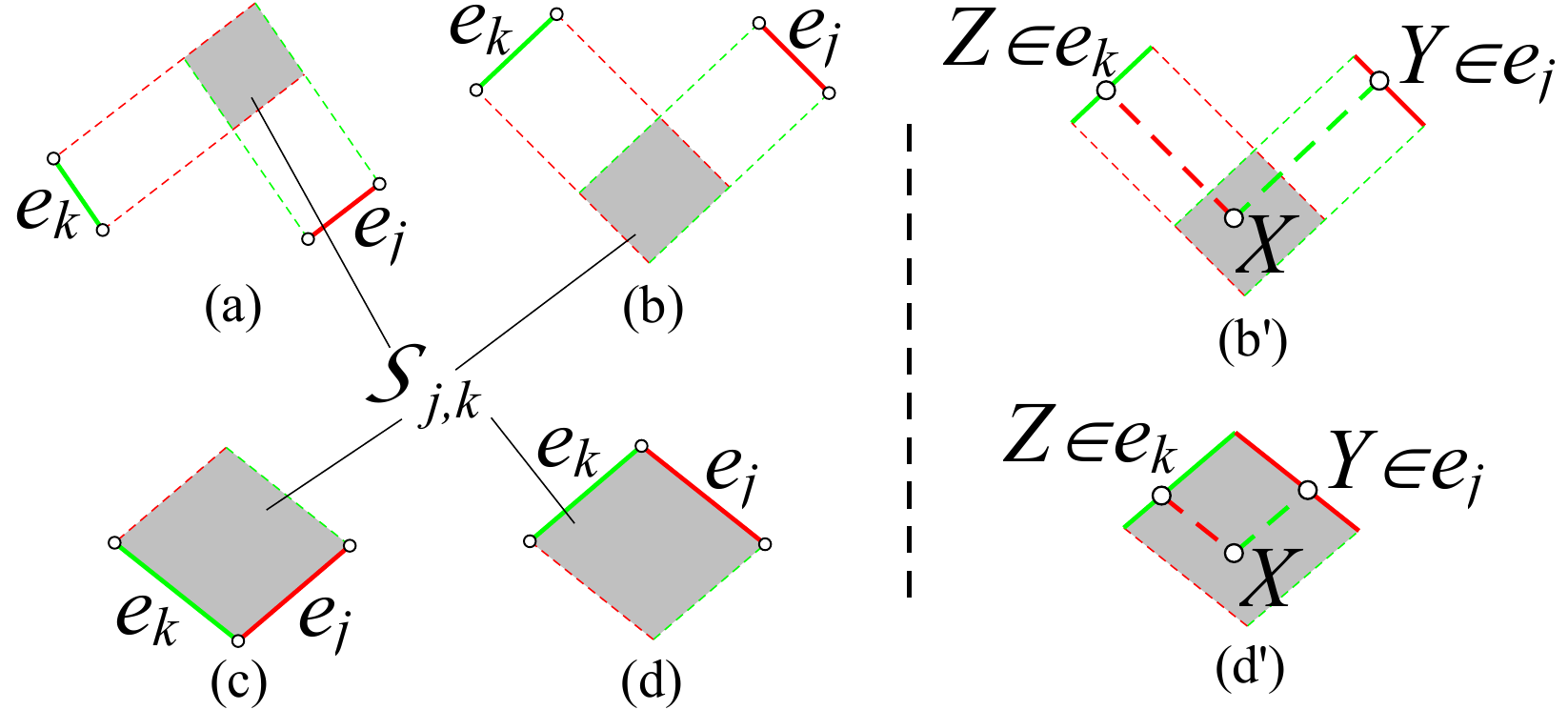}
		\caption{Illustration of the definition of $\SSS_{j,k}$.}\label{fig:QEDef}
	\end{minipage}
	\begin{minipage}[b]{0.45\textwidth}
		\flushright \includegraphics[width=.75\textwidth]{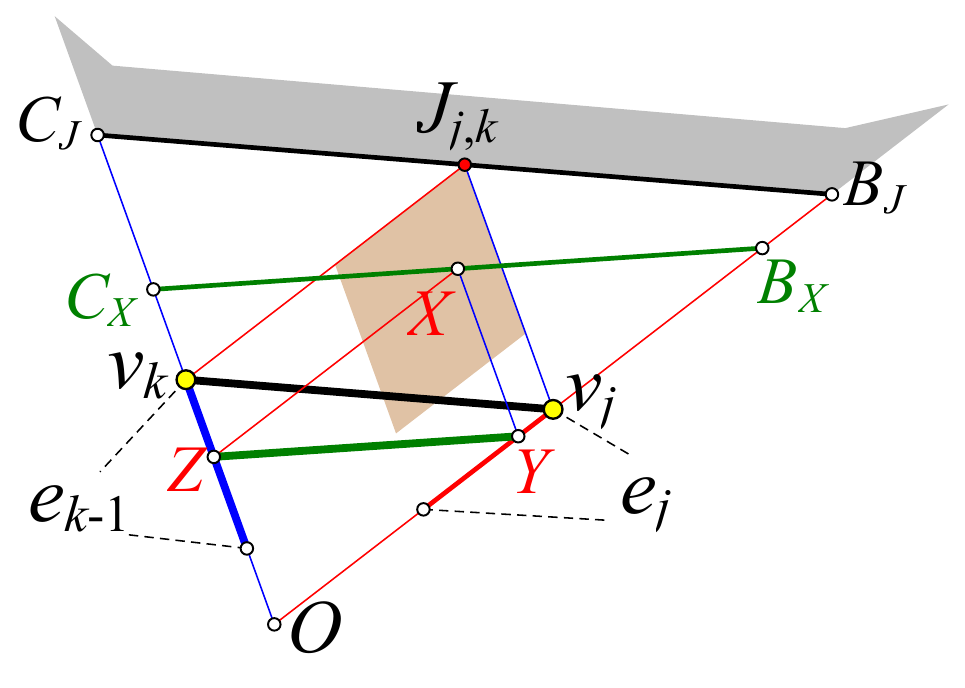}
		\caption{Proof of Observation~\ref{obs:R-E}~part~3}\label{fig:EE-Gdead}
	\end{minipage}
\end{figure}

\begin{proof}
	1. See Figure~\ref{fig:4cases}~(a).
	Since $X,Y,Z$ are stable, $XY\parallel e_k, YZ\parallel e_i$ and $ZX\parallel e_j$.
	So, $|I_2X|: |XI_3| = |I_1Y|:|YI_3| = |I_1Z|: |ZI_2|=|I_3X|: |XI_2|$. Hence $|I_3X|=|XI_2|$, so $X=(I_2+I_3)/2$.
	Similarly, $Y=(I_3+I_1)/2$ and $Z=(I_1+I_2)/2$.
	
	Part~2 is trivial. The proof of part~3 is similar to that of Observation~\ref{obs:R}~part~1.
\end{proof}

\begin{definition}
	For each edge pair $(e_j,e_k)$, define $\SSS_{j,k}$ as the intersecting region of two strips:
	The first strip is bounded by the two lines parallel to $e_k$ at the two endpoints of $e_j$ respectively.
	The second strip is bounded by the two lines parallel to $e_j$ at the two endpoints of $e_k$ respectively.
	See Figure~\ref{fig:QEDef} (a)-(d). Regard that $\SSS_{j,k}$ does \textbf{not} contain its boundaries.
	Note that $\SSS_{j,k}=\varnothing$ if $e_j\parallel e_k$.
\end{definition}

\begin{observation}\label{obs:R-E}
	\begin{enumerate}
		\item Given two distinct edges $e_j,e_k$ and assume that there is a $\triangle XYZ$ with $Y\in e_j, Z\in e_k$, and $X\in \partial P$, where $Y,Z$ are stable, then $X$ must lie in $\SSS_{j,k}$.
		\item A pair $(e_j,e_k)$ is G-dead if (1) $\SSS_{j,k}$ does not intersect $P$ or (2) $e_k\prec e_j$ .
		\item A pair $(e_j,e_{k-1})$ is G-dead if there exists some point $A$ in $P$ which lies on the right of $\overrightarrow{v_jv_k}$ such that $A\geq J_{j,k}$ in the distance to $\overleftrightarrow{v_jv_k}$.
		\item A pair $(v_j,v_k)$ is G-dead if (1) $\R_{j,k}$ does not intersect $P$ or (2) $e_k\prec e_j$ or (3) there exists some point $A$ in $P$ which lies on the right of $\overrightarrow{v_jv_k}$ such that $A>K_{j,k}$ in the distance to $\overleftrightarrow{v_jv_k}$.
	\end{enumerate}
\end{observation}

\begin{proof}
	1. Since $Y\in e_j$ is stable, we get $XZ\parallel e_j$. So $X$ lies in the aforementioned strip parallel to $e_j$.
	Since $Z\in e_k$ is stable, we get $XY\parallel e_k$. So $X$ lies in the aforementioned strip parallel to $e_k$.
	
	\smallskip
	2.  When $\SSS_{j,k}$ does not intersect $P$, there is no G-3-stable $\triangle XYZ$ such that $Y,Z$ are restricted to $e_j,e_k$ respectively according to part~1 of this observation; namely, $(e_j,e_k)$ is G-dead.
	
	Next, suppose that $e_k\prec e_j$ yet $(e_j,e_k)$ is not G-dead. Then, there is a G-3-stable $\triangle XYZ$ such that $Y\in e_j$ and $Z\in e_k$.
	We require that $X,Y,Z$ lie in clockwise order.
	However, $X,Y,Z$ lie in counterclockwise order since (i) $X\in \SSS_{j,k}$ (by part~1) and (ii) $e_k\prec e_j$; see Figure~\ref{fig:QEDef}~(b') and (d').
	
	\smallskip
	3. Suppose $(e_j,e_{k-1})$ is not G-dead.
	It implies a G-3-stable $\triangle XYZ$ with $Y\in e_j$ and $Z\in e_{k-1}$; see Figure~\ref{fig:EE-Gdead}.
	We claim that \emph{$A>X$ in the distance to $\overleftrightarrow{YZ}$} (and hence $X$ is not stable in $\triangle XYZ$).
	
	Make a parallel line of $YZ$ at $X$. Assume that it intersects $\ell_j,\ell_{k-1}$ at $B_{X},C_{X}$, respectively.
	Make a parallel line of $v_jv_k$ at $J_{j,k}$. Assume that it intersects $\ell_j,\ell_{k-1}$ at $B_J,C_J$, respectively.
	Let $\phi$ be the \emph{closed} region on the right of $\overrightarrow{B_JC_J}$, the right of $e_j$, and the right of $e_{k-1}$.
	Let $O= \ell_j \cap \ell_{k-1}$.
	
	Following the assumption that $A\in P$ and $A\geq J_{j,k}$ in the distance to $\overleftrightarrow{v_jv_k}$,
	point $A$ lies in $\phi$.
	Observe that $|OB_{X}|=2|OY|<2|Ov_j|=|OB_J|$ and $|OC_{X}|=2|OZ|<2|Ov_k|=|OC_J|$.
	Therefore, any point in $\phi$ (in particular, point $A$) is further than $X$ in the distance to $\overleftrightarrow{YZ}$.
	So the claim holds.
	
	\smallskip
	4. Observation~\ref{obs:R} states that $(v_j,v_k)$ is dead under cases (1)-(3).
	In fact, using the proof of Observation~\ref{obs:R}, we can obtain the stronger result that states $(v_j,v_k)$ is G-dead under these cases.
\end{proof}

\subsection{Proof of Lemma~\ref{lemma:G-correct}}\label{subsect:G-correct}

According to the definition of $\Kill_G$ in (\ref{eqn:def-Kill-G}), proving Lemma~\ref{lemma:G-correct} reduces to proving the following $4\times 2=8$ arguments.
(Assume $j\in [s,t],k\in[t,r]$, and $j\neq k$ in the following.)
\begin{equation*}
	\begin{aligned}
		\text{VV1:}\quad & \text{if } v_a\leq   I_{j,k}, &&(v_j,v_{k'})         & \text{ is G-dead for }k'\in & [k+1,r].\\
		\text{VV2:}\quad & \text{if } v_a>      I_{j,k}, &&(v_{j'},v_k)         & \text{ is G-dead for }j'\in & [j+1,t].\\
		\text{EE1:}\quad & \text{if } v_a\leq   H_{j,k}, &&(e_{j-1},e_{k'})     & \text{ is G-dead for }k'\in & [k,r-1].\\
		\text{EE2:}\quad & \text{if } v_a>      H_{j,k}, &&(e_{j'},e_{k-1})     & \text{ is G-dead for }j'\in & [j,t-1].\\
		\text{VE1:}\quad & \text{if } v_a<      J_{j,k}, &&(v_j,v_{k'})         & \text{ is G-dead for }k'\in & [k,r].\\
		\text{VE2:}\quad & \text{if } v_a\geq   J_{j,k}, &&(e_{j'},e_{k-1})     & \text{ is G-dead for }j'\in & [j,t-1].\\
		\text{EV1:}\quad & \text{if } v_a\leq   K_{j,k}, &&(e_{j-1},e_{k'})     & \text{ is G-dead for }k'\in & [k,r-1].\\
		\text{EV2:}\quad & \text{if } v_a>      K_{j,k}, &&(v_{j'},v_k)         & \text{ is G-dead for }j'\in & [j,t].\\
	\end{aligned}
\end{equation*}

Note that EE1 and EE2 are respectively contained by EV1 and VE2 (using $K_{j,k}>H_{j,k}>J_{j,k}$).

In the proof of Lemma~\ref{lemma:correct}, some vertex pairs are proved dead under certain cases,
and it is easy to see that these vertex pairs are actually G-dead therein.
Therefore, the proof of Lemma~\ref{lemma:correct} can be borrowed to prove VV1 and VV2 without much change.

Note that VE1 is almost contained by VV1. The only claim in VE1 not covered by VV1 is the following:
(i) \emph{$(v_j,v_k)$ is G-dead when $v_a<J_{j,k}$}.
Note that EV2 is almost contained by VV2. The only claim in EV2 not covered by VV2 is the following:
(ii) \emph{$(v_j,v_k)$ is G-dead when $v_a>K_{j,k}$}.
We leave the proofs of claims (i) and (ii) as exercises for the reader (they should already be obvious), and
we prove EV1 and VE2 in the following to complete the proof of Lemma~\ref{lemma:G-correct}.

\begin{figure}[h]
	\centering \includegraphics[width=.95\textwidth]{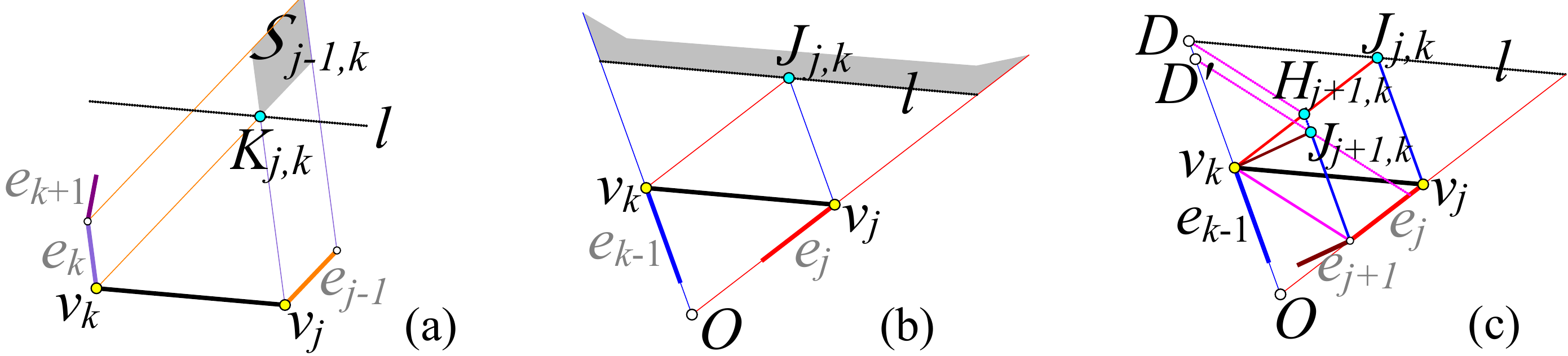}
	\caption{Illustration of the proofs of EV1 and VE2.}\label{fig:march-6}
\end{figure}

\begin{proof}[Proof of EV1.]
	First, we argue that $(e_{j-1},e_k)$ is G-dead.
	Note that $\SSS_{j-1,k}=\varnothing$ if $e_{j-1}\parallel e_k$.
	Therefore, $(e_{j-1},e_k)$ is G-dead when $e_{j-1}\parallel e_k$ or $e_k \prec e_{j-1}$ (applying Observation~\ref{obs:R-E}~part~2).
	Assume $e_{j-1}\prec e_k$ in the following.
	See Figure~\ref{fig:march-6}~(a).
	Let $l$ be the line at $K_{j,k}$ that is parallel to $v_jv_k$, and let $\phi$ denote the (closed) half-planed delimited by $l$ and containing $v_jv_k$.
	Following the assumption that $v_a\leq K_{j,k}$, polygon $P$ is contained in $\phi$.
	Moreover, $\SSS_{j-1,k}$ clearly is disjoint with $\phi$.
	Together, $\SSS_{j-1,k}$ does not intersect $P$, and thus $(e_{j-1},e_k)$ is G-dead (due to Observation~\ref{obs:R-E}~part~2).
	The same argument applies to any pair $(e_{j-1},e_{k'})$ where $k'\in [k,r-1]$.
\end{proof}

\begin{proof}[Proof of VE2.]
	Because $v_a\geq J_{j,k}$, we get $J_{j,k}\neq \infty$ and $e_j\prec e_{k-1}$. See Figure~\ref{fig:march-6}~(b).
	Let $d_{j,k}(X)$ denote the distance from $X$ to $\overleftrightarrow{v_jv_k}$ for any pair $(v_j,v_k)$ such that $j\neq k$.
	Let $l$ be the line at $J_{j,k}$ that is parallel to $v_jv_k$.
	Denote by $\phi$ the closed half-plane delimited by $l$ and not containing $v_b,v_c$.
	
	Since $v_a\geq J_{j,k}$, pair $(e_j,e_{k-1})$ is G-dead (due to Observation~\ref{obs:R-E}~part~3).
	
	See Figure~\ref{fig:march-6}~(c).
	Denote by $D$ the intersecting point between $l$ and $\ell_{k-1}$. Make a parallel line of $v_{j+1}v_k$ at $J_{j+1,k}$,
	and denote its intersecting point with $\ell_{k-1}$ by $D'$.
	Since $v_a\geq J_{j,k}$, point $v_a$ lies in $\phi$.
	Since that $v_a\in P$, this point also lies in $p_{k-1}$.
	Together, we obtain that $d_{j+1,k}(v_a)\geq d_{j+1,k}(D)$.
	Moreover, it is clear that $d_{j+1,k}(D)=d_{j+1,k}(H_{j+1,k})>d_{j+1,k}(J_{j+1,k})$.
	Altogether, we get $d_{j+1,k}(v_a)>d_{j+1,k}(J_{j+1,k})$,
	which implies that $(e_{j+1},e_{k-1})$ is G-dead (due to Observation~\ref{obs:R-E}~part~3).
	By induction, we can see $(e_{j'},e_{k-1})$ is G-dead for $j'\in [j,t-1]$.
\end{proof}

\subsection{Computing all G-3-stable triangles from $\List_{r,s,t},\List_{s,t,r},\List_{t,r,s}$}\label{subsect:list-to-G-3stable}

Suppose we are given $r,s,t$ such that $\triangle v_rv_sv_t$ is 3-stable. Suppose $\List_{r,s,t}$, $\List_{s,t,r}$, $\List_{t,r,s}$ are preprocessed. We show how to find all canonical G-3-stable triangles in the following.

Recall Note~\ref{note:canonical} for canonical G-stable triangles. Henceforth, for convenience, we abuse the notation a little so that
\textbf{a G-3-stable triangle refers to a canonical G-3-stable triangle}.

Denote by $Q_{r,s,t}$ the set of G-3-stable $\triangle XYZ$ where
$Y\in [v_s \circlearrowright v_t]$ and $Z\in [v_t\circlearrowright v_r]$, and where neither $Y$ nor $Z$ or both of them lie at vertices of $P$ (and $X,Y,Z$ lie in clockwise order).

\begin{observation}
	$Q_{r,s,t}\cup Q_{s,t,r}\cup Q_{t,r,s}$ contains all the G-3-stable triangles.
\end{observation}

\begin{proof}
	Take any G-3-stable triangle $\triangle XYZ$.
	It has two corners which both lie at vertices of $P$ or both lie in edges of $P$.
	Without loss of generality, assume that \emph{$Y$ and $Z$} are such a pair of corners.
	By Lemma~\ref{lemma:interleaving}~part~3, $\triangle XYZ$ interleaves $\triangle v_rv_sv_t$.
	Therefore, there are three possibilities:
	1. $Y\in [v_s\circlearrowright v_t]$ and $Z\in[v_t\circlearrowright v_r]$. In this case, $\triangle XYZ \in Q_{r,s,t}$.
	2. $Y\in [v_t\circlearrowright v_r]$ and $Z\in[v_r\circlearrowright v_s]$. In this case, $\triangle XYZ \in Q_{s,t,r}$.
	3. $Y\in [v_r\circlearrowright v_s]$ and $Z\in[v_s\circlearrowright v_t]$. In this case, $\triangle XYZ \in Q_{t,r,s}$.
\end{proof}

Denote $Q=Q_{r,s,t}$ and $\List=\List_{r,s,t}$. We show how to compute $Q$ from $\List$ in the following, and
we can compute $Q_{s,t,r}$ from $\List_{s,t,r}$ and compute $Q_{t,r,s}$ from $\List_{t,r,s}$ using the same method.

Consider the vertex (or two vertices) with the largest distance to $\overleftrightarrow{v_jv_k}$ on the right of $\overrightarrow{v_jv_k}$.
If there is only one such vertex, we define it to be $A_{j,k}$ and $A^*_{j,k}$.
Otherwise, there are two such vertices.
The (clockwise) first one is defined to be $A_{j,k}$ and the last one is defined to be $A^*_{j,k}$.

\begin{observation}\label{obs:range-X}
	Assume that $\triangle XYZ$ is G-3-stable, where $Y\in u_b,Z\in u_c$.
	We claim that $X\in [A_{j,k}\circlearrowright A^*_{j,k}]$ if $(u_b,u_c)=(v_j,v_k)$,
	and that $X\in [A_{j,k}\circlearrowright A^*_{j+1,k+1}]$ if $(u_b,u_c)=(e_j,e_k)$.
\end{observation}

\begin{figure}[b]
	\centering \includegraphics[width=\textwidth]{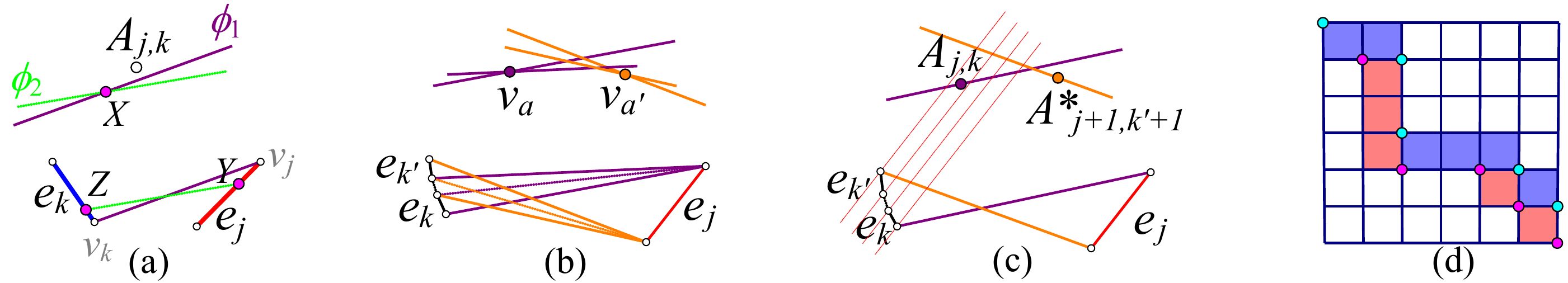}
	\caption{(a) illustrates Observation~\ref{obs:range-X}. (b) illustrates that the running time of Algorithm~\ref{alg:List-to-Q-naive} could be $\Omega(n^2)$; see an explanation right below. (c) and (d) illustrate the batch technique given below.
	}\label{fig:finalstep}
\end{figure}

\begin{proof}
	We only show the proof of the second claim, as the first claim is trivial.
	Suppose $(u_b,u_c)=(e_j,e_k)$ yet $X\notin [A_{j,k}\circlearrowright A^*_{j+1,k+1}]$.
	First, notice that $e_j\prec e_k$. This follows from Observation~\ref{obs:R-E}~part~2 and the fact that $(e_j,e_k)$ is not G-dead.
	Next, there are two symmetric cases: $X\in [v_{k+1}\circlearrowright A_{j,k}]$ or $X\in [A^*_{j+1,k+1}\circlearrowright v_j]$.
	Without loss of generality, assume the first case occurs.
	
	See Figure~\ref{fig:finalstep}~(a). Denote by $\phi_1$ the half-plane at $X$ which is parallel to $v_jv_k$ and does not contain $v_jv_k$.
	Denote by $\phi_2$ the half-plane at $X$ which is parallel to $YZ$ and does not contain $YZ$.
	By the definition of $A_{j,k}$, it lies in $\phi_1$.
	This implies that $A_{j,k}\in \phi_2$ since $YZ$ intersects $v_jv_k$ whereas $X$ lies between $v_{k+1}$ and $A_{j,k}$.
	It follows that $X$ is not stable in $\triangle XYZ$. Contradictory.
\end{proof}

Our first algorithm for computing $Q$ is given in Algorithm~\ref{alg:List-to-Q-naive}
(some details of this algorithm are elaborated below).
The correctness of Algorithm~\ref{alg:List-to-Q-naive} follows from Observation~\ref{obs:range-X}.

\begin{algorithm}[h]
	\caption{Compute $Q$ from $\List$ (the first naive algorithm).}\label{alg:List-to-Q-naive}
	Compute $A_{j,k}$ and $A^*_{j,k}$ for each vertex pair $(v_j,v_k)\in \List$ \label{code:U^*-to-quasi-precomp1}\;
	\lForEach{vertex pair $(v_j,v_k)\in \List$ \emph{and} unit $u_i$ in $[A_{j,k}\circlearrowright A^*_{j,k}]$}
	{
		Output all G-3-stable triangle $\triangle XYZ$ such that $X\in u_i,Y\in v_j$ and $Z\in v_k$ \label{code:U^*-to-quasi-1}
	}
	Compute $A_{j,k}$ and $A^*_{j+1,k+1}$ for each edge pair $(e_j,e_k)\in \List$ \label{code:U^*-to-quasi-precomp2}\;
	\lForEach{edge pair $(e_j,e_k)\in \List$ \emph{and} unit $u_i$ in $[A_{j,k}\circlearrowright A^*_{j+1,k+1}]$ \label{code:u^*-to-quasi-search-edgepair}}
	{
		Output all G-3-stable triangle $\triangle XYZ$ such that $X\in u_i,Y\in e_j$ and $Z\in e_k$ \label{code:U^*-to-quasi-2}
	}
\end{algorithm}

\subparagraph{Running time analysis.}
According to Algorithm~\ref{alg:RK-G}, the pairs $(u_{b_1},u_{c_1})$, $\ldots$, $(u_{b_m},u_{c_m})$ in $\List$ has a monotonicity property that
$b_1,\ldots,b_m$ are increasing (non-strictly) and so as $c_1,\ldots,c_m$.
Therefore, if $(v_j,v_k)$ is taken over all vertex pairs in $\List$ in order,
$A_{j,k}$ and $A^*_{j,k}$ move in clockwise and thus can be computed in amortized $O(1)$ time.
Therefore, Line~\ref{code:U^*-to-quasi-precomp1} costs $O(n)$ time. So as Line~\ref{code:U^*-to-quasi-precomp2}.

Following Observation~\ref{obs:EEE-VEE-EVV}, given any three units $u_i,u_b,u_c$, we can compute in $O(1)$ time all the (canonical) G-3-stable
triangles with corners restricted to these units respectively.
So Lines \ref{code:U^*-to-quasi-1} and \ref{code:U^*-to-quasi-2} both take $O(1)$ time.
Therefore, the total running time of Algorithm~\ref{alg:List-to-Q-naive} is $O(T)$, where
\begin{equation}
	T={\sum}_{(v_j,v_k)\in \List} | A_{j,k}\circlearrowright A^*_{j,k}|+
	{\sum}_{(e_j,e_k)\in \List} | A_{j,k}\circlearrowright A^*_{j+1,k+1}|,
\end{equation}
where $|X \circlearrowright Y|$ denotes the number of vertices in $[X\circlearrowright Y]$.

The first term of $T$ is easily bounded by $O(n)$ using the monotonicity property of $\List$.
However, the second term of $T$ could be $\Omega(n^2)$ in the worst case!
Assume $(e_j,e_k),\ldots,(e_j,e_{k'})$ are $\Omega(n)$ edge pairs in $\List$.
Assume $A_{j,k}=\ldots=A_{j,k'}=v_a$ and $A^*_{j+1,k+1}=\ldots=A^*_{j+1,k'+1}=v_{a'}$ and there are $\Omega(n)$ units in $[v_a\circlearrowright v_{a'}]$.
See Figure~\ref{fig:finalstep}~(b).
In this case, the second term of $T$ would be $\Omega(n^2)$.\bigskip

In the remaining part of this section, we optimize Algorithm~\ref{alg:List-to-Q-naive} by a batch technique.

\begin{lemma}\label{lemma:batch}
	1. For $(e_j,e_k),\ldots,(e_j,e_{k'})\in \List$,
	we can use a \emph{batch} to find all G-3-stable $\triangle XYZ$ where $Y\in e_j$ and $Z$ lies in some edge in $e_k,\ldots,e_{k'}$,
	in $O(k'-k+|A_{j,k} \circlearrowright A^*_{j+1,k'+1}|)$ time.
	
	2. For $(e_j,e_k),\ldots,(e_{j'},e_k)\in \List$,
	we can use a \emph{batch} to find all G-3-stable $\triangle XYZ$ where $Z\in e_k$ and $Y$ lies in some edge in $e_j,\ldots,e_{j'}$,
	in $O(j'-j+|A_{j,k} \circlearrowright A^*_{j'+1,k+1}|)$ time.
\end{lemma}

\newcommand{\Ans}{\mathsf{Ans}}
\begin{proof}
	1. For each $e_h\in \{e_k,\ldots,e_{k'}\}$, let $\Ans_h$ be the set of G-3-stable $\triangle XYZ$ with $Y\in e_j$ and $Z\in e_h$.
	Make two parallel lines of $e_j$ at the two endpoints of $e_h$, as shown in Figure~\ref{fig:finalstep}~(c).
	This defines a stripe $\phi_h$. Assume $\phi_h$ does not contain its boundaries.
	Denote $\rho=[A_{j,k}\circlearrowright A^*_{j+1,k'+1}]$.
	
	For each $\triangle XYZ\in \Ans_h$, we have (i) \emph{$X\in \phi_h$} and (ii) \emph{$X\in \rho$}.
	Fact~(i) is because $XZ\parallel e_j$, whereas fact~(ii) comes from Observation~\ref{obs:range-X}.
	Based on these two facts, computing $\Ans_h$ reduces to
	enumerating all the units in $\rho$ that intersect $\phi_h$.
	Therefore, computing $\Ans_k,\ldots,\Ans_{k'}$ reduces to
	enumerating all the units in $\rho$ that intersect $\phi_k,\ldots,\phi_{k'}$ respectively.
	
	Since the stripes $\phi_k,\ldots, \phi_{k'}$ do not overlap and has a monotonicity property
	(to prove this, we need to apply the fact that $e_j\prec e_k,\ldots,e_j\prec e_{k'}$.
	Note that $\Ans_h$ is empty if $e_j$ is not chasing $e_h$;
	without loss of generality, we can assume that $e_j\prec e_k,\ldots,e_j\prec e_{k'}$),
	enumerating the aforementioned units only take $O(k'-k+ |A_{j,k}\circlearrowright A^*_{j+1,k'+1}|)$ time. Therefore, part~1 holds.
	
	\smallskip The proof of part~2 is symmetric to the proof of part~1 and is thus omitted.
\end{proof}

\paragraph*{Optimize Algorithm~\ref{alg:List-to-Q-naive} to $O(n)$ time (using the batches mentioned in Lemma~\ref{lemma:batch}).}
We change Lines~\ref{code:u^*-to-quasi-search-edgepair} and \ref{code:U^*-to-quasi-2} to the following scheme with multiple rounds.
In round one, take the first edge pair in $\List$
and take as many edge pairs in $\List$ as long as they are still in the same row.
Handle the edge pairs taken in this round (which are in a row) by a batch.
In round two, take the next edge pair in $\List$ and take as many edge pairs in $\List$ as long as they are still in the same column.
Handle the edge pairs taken in this round (which are in a column) by a batch.
Repeat these two rounds alternatively until all edge pairs in $\List$ are handled. See an illustration in Figure~\ref{fig:finalstep}~(d).
Following Lemma~\ref{lemma:batch}, the total running time of the odd rounds can be bounded by $O(n)$, so as the total running time of the even rounds. Therefore, the overall running time is $O(n)$.

As a summary, we can compute all the G-3-stable triangles in $O(n)$ time.

\paragraph*{A connection between the G-3-stable and the minimum enclosing triangle.}
Assume $T=\triangle xyz$ is a locally minimum area triangle among the triangles enclosing $P$.
It is proved in \cite{Tri-Enclose-Area-kleeMid} (and rediscovered in many places) that
\emph{the midpoint of each side of $T$ touches $P$.}
Therefore, $X,Y,Z$ lie in $\partial P$, where
$Z,X,Y$ denote the midpoints of the three sides $xy,yz,zx$ of $T$.

Moreover, $X,Y,Z$ are all stable in $\triangle XYZ$, because $\triangle xyz$ contains $P$, $yz \parallel YZ$, $zx\parallel ZX$, and $xy \parallel XY$.
Therefore, $\triangle XYZ$ is G-3-stable. (Yet it may not be canonical; see Note~\ref{note:canonical}. However, if it is not canonical, we can find a canonical one with the same area).

Further since it is easy to compute $x,y,z$ in $O(1)$ time from $X,Y,Z$,
finding the locally minimum triangles enclosing $P$ reduces to computing the (canonical) G-3-stable triangles.

\clearpage

\section{Find the all-flush triangle with the minimum perimeter}\label{sect:MPFT}

We give some preliminaries first.

\begin{figure}[h]
	\centering
	\includegraphics[width=.95\textwidth]{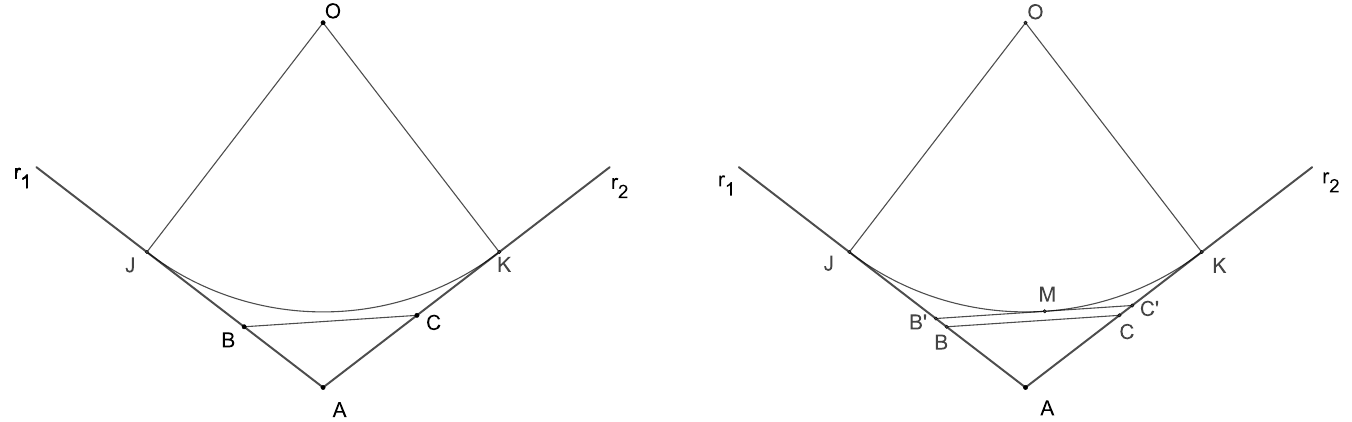}\\
	\caption{Illustration of Observation~\ref{Obs:MFPT-circle-tangent}}\label{fig:MFPT-1}
\end{figure}

\begin{observation}\label{Obs:MFPT-circle-tangent}
	Assume rays $r_1,r_2$ originate from point $A$, and $r_2$ is on the right of $r_1$; see Figure~\ref{fig:MFPT-1}.
	Assume circle $O$ is tangent to $r_1$ and $r_2$ at points $J,K$ respectively.
	Let $\widehat{JK}$ denote the arc starting from $J$ and counterclockwise to $K$.
	Consider any segment $BC$ that connects $r_1$ and $r_2$.
	
	1. If $BC$ is tangent to $\widehat{JK}$, the perimeter of $\triangle ABC$ equals $2|AJ|$.
	
	2. If $BC$ lies below $\widehat{JK}$ as shown in Figure~\ref{fig:MFPT-1}, the perimeter of $\triangle ABC$ is smaller than $2|AJ|$.
	
	3. Otherwise, the perimeter of $\triangle ABC$ is larger than $2|AJ|$.
\end{observation}

\begin{proof}
	We only prove the first claim. The other claims follow from the first one immediately.
	
	Consider any segment $B'C'$ tangent to $\widehat{JK}$ (see the right picture of Figure~\ref{fig:MFPT-1} for example).
	The perimeter of $\triangle AB'C'$ equals $(|AB'|+|B'M|)+(|MC'|+|C'A|)=|AJ|+|AK|=2|AJ|$.
\end{proof}

\begin{figure}[h]
	\centering
	\includegraphics[width=.95\textwidth]{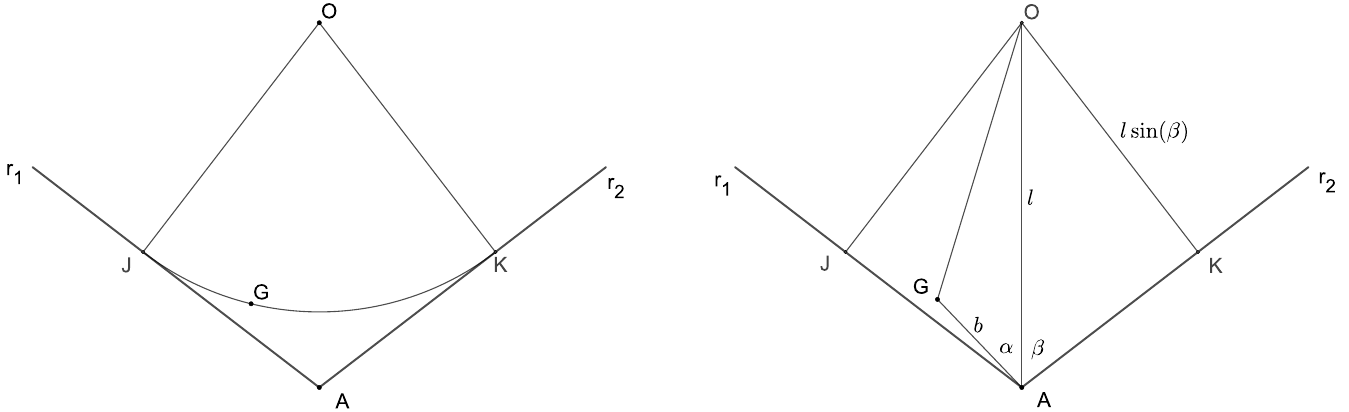}\\
	\caption{Illustration of Observation~\ref{Obs:MFPT-unique-circle-pass-G}}\label{fig:MFPT-2}
\end{figure}

\begin{observation}\label{Obs:MFPT-unique-circle-pass-G}
	Assume rays $r_1,r_2$ originate from point $A$, and $r_2$ is on the right of $r_1$;
	assume moreover that point $G$ is on the right of $r_1$ and on the left of $r_2$; see Figure~\ref{fig:MFPT-2}.
	We claim that there exists a unique circle $O$ tangent to $r_1,r_2$ such that the arc $\widehat{JK}$ passes through $G$,
	where $J$ denotes the tangent point of circle $O$ and $r_1$, and where $K$ denotes the tangent point of circle $O$ and $r_2$.
\end{observation}

\begin{proof}
	First, we argue that there cannot be more than one such circles.
	We prove it by contradiction. Suppose that $O_1,O_2$ are two such circles.
	Let $B_1C_1$ be the segment connecting $r_1,r_2$ which is tangent to $O_1$ at $G$.
	Let $B_2C_2$ be the segment connecting $r_1,r_2$ which is tangent to $O_2$ at $G$.
	
	Be aware that $B_2C_2$ is different from $B_1C_1$ (otherwise $O_1=O_2$).
	Since $B_1C_1$ the tangent line of $O_1$ at $G$, segment $B_2C_2$ cuts the circle $O_1$.
	Applying Observation~\ref{Obs:MFPT-circle-tangent}, $\triangle AB_2C_2$ has a larger perimeter than $\triangle AB_1C_1$.
	Since $B_2C_2$ the tangent line of $O_2$ at $G$, segment $B_1C_1$ cuts the circle $O_2$.
	Applying Observation~\ref{Obs:MFPT-circle-tangent}, $\triangle AB_1C_1$ has a larger perimeter than $\triangle AB_2C_2$.
	Contradiction.
	
	Next, we argue that there exists at least one circle $O$ satisfying the requirement.
	
	See the right picture of Figure~\ref{fig:MFPT-2}.
	Let $\alpha,\beta$ denote the angle as shown in the figure and let $b=|AG|$, which are fixed by the assumption.
	As circle $O$ must be tangent to $r_1,r_2$,
	point $O$ lies on the bisector of $r_1,r_2$, and hence its position can be specified by the length $l=|OA|$.
	It reduces to showing that there exists a number $l>0$ so that $OK=OG$ and $\angle AOG < \angle AOK$.
	
	According to the requirement $OK=OG$, we get $OK^2=OG^2$, namely, $l^2\sin^2\beta= b^2+l^2-2bl \cos \alpha$.
	In other words, $\cos^2\beta \cdot l^2 - 2 b \cos \alpha \cdot l + b^2=0$.
	One solution to this equation is
	\begin{equation}\label{eqn:MFT(P)-l}
		l= b\cdot \bigg(\cos \alpha + \sqrt{\cos^2\alpha-\cos^2\beta}\bigg)/\cos^2\beta.
	\end{equation}
	
	For this particular $l$, we have $OK=OG$ and it remains to be shown that $\angle AOG < \angle AOK$.
	Because $\alpha<\beta$, we have $\frac{\cos \alpha}{\cos \beta} > \frac{\sin \alpha}{\sin \beta}$.
	So, $\frac{l\cos \beta}{b}>\frac{\sin \alpha}{\sin \beta}$ (applying \eqref{eqn:MFT(P)-l}).
	So $\cos \beta>\frac{b\sin \alpha}{l\sin \beta}$.
	Further since $\frac{\sin \angle AOG}{b}= \frac{\sin \alpha}{l\sin \beta}$,
	we get $\sin \angle AOG= \frac{b\sin \alpha}{l\sin \beta}<\cos \beta=\sin \angle AOK$.
	So, $\angle AOG<\angle AOK$.
\end{proof}

\subsection{Unimodality and bi-monotonicity}

Let $M_A(B,C)$ denote the unique point $M$ on $BC$ such that $|AB|+|BM|= |AC|+|CM|$.

\begin{figure}[h]
	\centering
	\includegraphics[width=.95\textwidth]{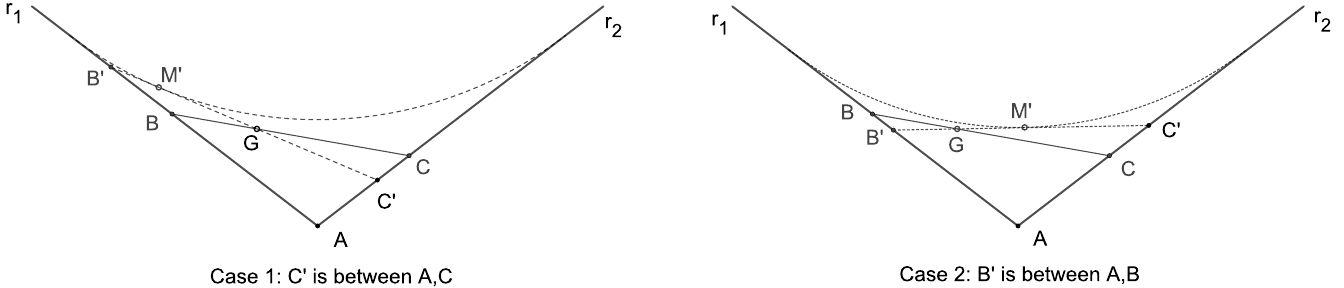}\\
	\caption{Illustration of Observation~\ref{Obs:MFPT-left-right-M}}\label{fig:MFPT-3}
\end{figure}

\begin{observation}\label{Obs:MFPT-left-right-M}
	Let $r_1,r_2,G$ be the same as Observation~\ref{Obs:MFPT-unique-circle-pass-G}.
	Let $BC$ be the segment connecting $r_1,r_2$, passing through $G$, and tangent to the circle $O$ mentioned in Observation~\ref{Obs:MFPT-unique-circle-pass-G};
	see Figure~\ref{fig:MFPT-3}.
	
	Consider another segment $B'C'$ connecting $r_1,r_2$ and passing through $G$. Let $M'=M_A(B',C')$.
	
	If $C'$ is between $A,C$, point $M'$ is to the left of $G$ (namely, $B',M',G,C'$ lie in order).
	
	If $B'$ is between $A,B$, point $M'$ is to the right of $G$ (namely, $B',G,M',C'$ lie in order).
\end{observation}

\begin{proof}
	We only show the proof of the first case. The proof of the other case is symmetric.
	
	Because $BC$ is tangent to the circle $O$ defined in Observation~\ref{Obs:MFPT-left-right-M}, whereas $B'C'$ cuts circle $O$,
	$\triangle ABC$ has a smaller perimeter than $\triangle AB'C'$ (due to Observation~~\ref{Obs:MFPT-circle-tangent}).
	Build the circle $O'$ tangent to $r_1,r_2$ and $B'C'$, a part of which is drawn as the dotted arc
	(see the left part of Figure~\ref{fig:MFPT-3}).
	Since $\triangle ABC$ has a smaller perimeter than $\triangle AB'C'$, segment $BC$ lies below the arc (by Observation~\ref{Obs:MFPT-circle-tangent}). Therefore, $GC'$ is below the arc. It follows that the tangent point $M'$ is between $B'$ and $G$.
\end{proof}

\begin{observation}\label{Obs:MFPT-monotone-perimeter}
	Let $r_1,r_2,G,B,C,B',C'$ be the same as Observation~\ref{Obs:MFPT-left-right-M}. see Figure~\ref{fig:MFPT-3}.
	
	The perimeter of $\triangle AB'C'$ grows if we move $B'$ along $r_1$ away from $A$ starting at $B$
	(meanwhile move $C'$ along the opposite direction of $r_2$ so that $B'C'$ passes through the fixed point $G$).
	
	The perimeter of $\triangle AB'C'$ grows if we move $C'$ along $r_2$ away from $A$ starting at $C$
	(meanwhile move $B'$ along the opposite direction of $r_1$ so that $B'C'$ passes through the fixed point $G$).
\end{observation}

\begin{proof}
	We only show the proof of the claim on $B'$. The proof of the claim on $C'$ is symmetric.
	
	Suppose there is another segment $B''C''$ connecting $r_1r_2$, passing through $G$, with $|AB''|>|AB'|>|AB|$.
	We shall prove the following (x): $\triangle AB''C''$ has a larger perimeter than $\triangle AB'C'$.
	
	Let $O'$ be the circle tangent to $r_1,r_2$ and $B'C'$. By Observation~\ref{Obs:MFPT-left-right-M},
	the tangent point $M'$ of segment $B'C'$ and circle $O'$ is on the left of $G$, as shown in the left picture of Figure~\ref{fig:MFPT-3}.
	It follows that $B''C''$ cuts circle $O'$.
	Further due to Observation~\ref{Obs:MFPT-circle-tangent}, we obtain the argument (x) above.
\end{proof}

\begin{lemma}[Unimodality of $\peri(\triangle e_ie_je_k)$]\label{lemma:MFPT-unimodal}
	Fix two indices $i,j$ such that $e_i\prec e_j$.
	Let $E$ be the set of $e_k$ such that $e_k\prec e_i$ and $e_j\prec e_k$, which is obviously an interval of edges. We claim that
	the perimeter of $\triangle e_ie_je_k$ is unimodal when $e_k$ is taken over the edges in $E$ in clockwise.
	More precisely, for $e_{k-1},e_k,e_{k+1}\in E$, we claim that it cannot occur that
	\begin{equation}\label{eqn:MFT(P)-2}
		\peri(\triangle e_ie_je_k)\geq \peri(\triangle e_ie_je_{k-1})\text{ and }\peri(\triangle e_ie_je_k)\geq \peri(\triangle e_ie_je_{k+1}).
	\end{equation}
\end{lemma}

\begin{figure}[h]
	\centering
	\includegraphics[width=.9\textwidth]{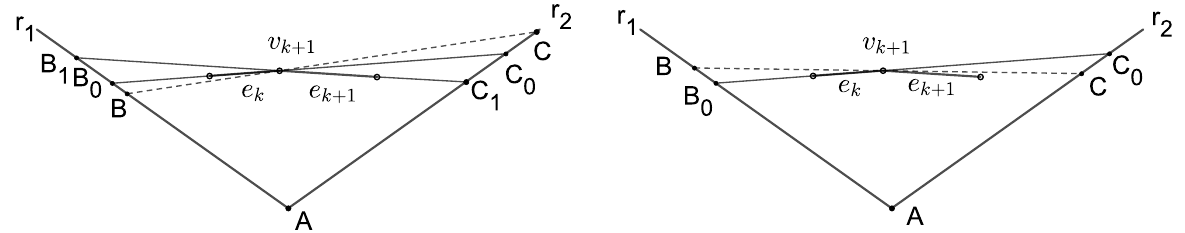}\\
	\caption{Illustration of the proof of Lemma~\ref{lemma:MFPT-unimodal}}\label{fig:MFPT-4}
\end{figure}

\begin{proof}
	Let $r_1,r_2$ be the rays containing $e_j,e_i$ respectively as shown in Figure~\ref{fig:MFPT-4}.
	Assume $\ell_k$ intersect $r_1,r_2$ at $B_0,C_0$ respectively, and $\ell_{k+1}$ intersect $r_1,r_2$ at $B_1,C_1$ respectively.
	
	This lemma follows from the following three facts immediately.
	
	1. If $\peri(\triangle e_ie_je_k)\geq \peri(\triangle e_ie_je_{k+1})$, point $M_A(B_0,C_0)$ is on the right of $v_{k+1}$.
	
	2. If $\peri(\triangle e_ie_je_k)\geq \peri(\triangle e_ie_je_{k-1})$, point $M_A(B_0,C_0)$ is on the left of $v_k$.
	
	3. Point $M_A(B_0,C_0)$ cannot be on the right of $v_{k+1}$ and on the left of $v_k$.
	
	Fact~3 is trivial. Fact~2 is symmetric to Fact~1. We prove Fact~1 in the following.
	
	Choose $G=v_{k+1}$ and let $BC$ be the segment connecting $r_1,r_2$ that is tangent to the circle $O$ given in
	Observation~\ref{Obs:MFPT-unique-circle-pass-G}.
	First, we argue that $|AC|<|AC_0|$. Suppose to the opposite that $|AC|\geq |AC_0|$, as shown in the left picture of Figure~\ref{fig:MFPT-4}.
	We obtain from Observation~\ref{Obs:MFPT-monotone-perimeter} that
	$\triangle AB_1C_1$ has a larger perimeter than $\triangle AB_0C_0$, namely,
	$\peri(\triangle e_ie_je_k)<\peri(\triangle e_ie_je_{k+1})$.
	
	Next, observe the right picture of Figure~\ref{fig:MFPT-4}, where $|AC|<|AC_0|$.
	Since $|AC_0|>|AC|$, we obtain from Observation~\ref{Obs:MFPT-left-right-M} that
	$M_A(B_0,C_0)$ is on the right of $v_{k+1}$.
\end{proof}

\begin{definition}
	For any two indices $i,j$ such that $e_i\prec e_j$,
	denote by $\Opt_{i,j}$ the index $k$ in set $\{k\mid e_j\prec e_k,e_k\prec e_i\}$ which minimizes $\peri(\triangle e_ie_je_k)$.
	Be aware that there could be two consecutive indices which minimize $\peri(\triangle e_ie_je_k)$ according to the unimodality of $\peri(\triangle e_ie_je_k)$ (Lemma~\ref{lemma:MFPT-unimodal}); in this case, we denote by $\Opt_{i,j}$ the first index between the two indices.
\end{definition}

\begin{lemma}[Bi-monotonicity of $\Opt_{i,j}$]\label{lemma:MFPT-bimono}
	It holds that $\Opt_{i+1,j}\geq \Opt_{i,j}$ and $\Opt_{i,j+1}\geq \Opt_{i,j}$.
\end{lemma}

Note: $\Opt_{i+1,j}$ and $\Opt_{i,j}$ are in the interval $[j+1,i]$.
The inequality $\Opt_{i+1,j}\geq \Opt_{i,j}$ means that $\Opt_{i+1,j}=\Opt_{i,j}$
or $\Opt_{i+1,j}$ is after $\Opt_{i,j}$ when we enumerate indices in $[j+1,i]$ from $j+1$ to $i$.
The inequality $\Opt_{i,j+1}\geq \Opt_{i,j}$ has a similar meaning. (Similar to Lemma~\ref{lemma:OPTmono}.)

\begin{figure}[h]
	\centering
	\includegraphics[width=\textwidth]{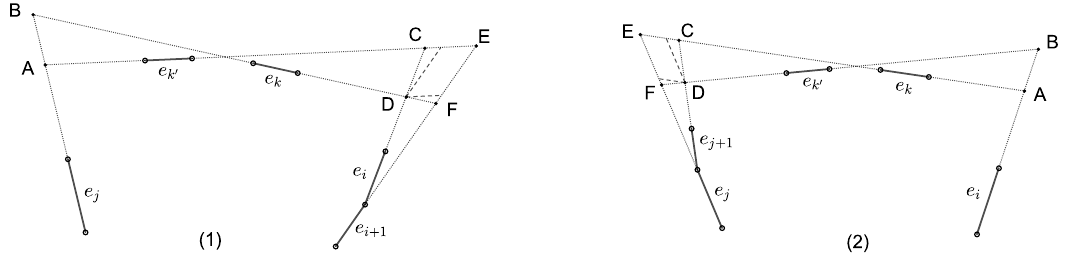}\\
	\caption{Illustration of Lemma~\ref{lemma:MFPT-bimono}}\label{fig:MFPT-5}
\end{figure}

\begin{proof}
	First, we prove $\Opt_{i+1,j}\geq \Opt_{i,j}$.
	Suppose to the opposite that $\Opt_{i+1,j}=k' < k=\Opt_{i,j}$, as shown in Figure~\ref{fig:MFPT-5}~(1).
	
	Because $\Opt_{i,j}=k>k'$, we know $|AB|+|BD|<|AC|+|CD|$.
	
	Observe that $|CD|+|DF|<|CE|+|EF|$ by the triangle inequality (see dashed line).
	
	Summing up, we get $|AB|+|BD|+|DF|<|AC|+|CE|+|EF|$; i.e.,
	$|AB|+|BF| < |AE|+|EF|$; i.e. $\peri(\triangle e_{i+1}e_je_k)<\peri(\triangle e_{i+1}e_je_{k'})$, contradicting with $\Opt_{i+1,j}=k'$.
	
	Next, we prove $\Opt_{i,j+1}\geq \Opt_{i,j}$.
	Suppose to the opposite that $\Opt_{i,j+1}=k' < k=\Opt_{i,j}$, as shown in Figure~\ref{fig:MFPT-5}~(2).
	
	Because $\Opt_{i,j}=k>k'$, we know $|AE|+|EF|<|AB|+|BF|$.
	
	Observe that $|CD|+|DF|<|CE|+|EF|$ by the triangle inequality (see dashed line).
	
	Summing up, $|AE|+|CD|+|DF|<|AB|+|BF|+|CE|$;  i.e.,
	$|AC|+|CD| <|AB|+|BD|$; i.e. $\peri(\triangle e_{i}e_{j+1}e_k)<\peri(\triangle e_{i}e_{j+1}e_{k'})$, contradicting with $\Opt_{i,j+1}=k'$.
\end{proof}

\begin{definition}
	We say $e_i$ is \emph{stable} in $\triangle e_ie_je_k$ if $\peri (\triangle e_ie_je_k)$ cannot be reduced by changing $e_i$.
	We state that $\triangle e_ie_je_k$ is a \emph{3-stable} triangle, if $e_i,e_j,e_k$ are all stable in $\triangle e_ie_je_k$.
\end{definition}

\begin{lemma}\label{lemma:MFPT-interleaving}
	The 3-stable triangles are pairwise interleaving.
\end{lemma}

\begin{figure}[h]
	\centering
	\includegraphics[width=.6\textwidth]{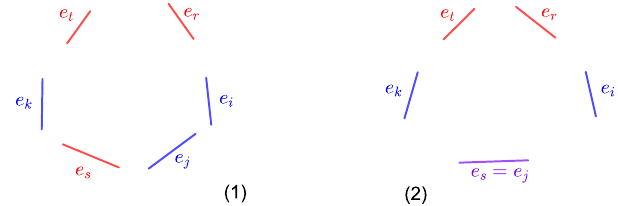}\\
	\caption{Illustration of the proof of Lemma~\ref{lemma:MFPT-interleaving}}\label{fig:MFPT-6}
\end{figure}

\begin{proof}
	Suppose $T_1=\triangle e_ie_je_k$ and $T_2=\triangle e_re_se_t$ are 3-stable triangles that are not interleaving.
	As they are not interleaving, sets $\{e_i,e_j,e_k\}$ and $\{e_r,e_s,e_t\}$ share at most one common element.
	Therefore, there are two essentially different cases (as shown in Figure~\ref{fig:MFPT-6}):
	
	Case~1. $e_t,e_r,e_i,e_j,e_s,e_k$ are distinct and lie in clockwise order.
	
	Case~2. $e_t,e_r,e_i,e_j=e_s,e_k$ are distinct and lie in clockwise order.
	
	Because $e_i$ is stable in $\triangle e_ie_je_k$,  we have $\peri(\triangle e_ie_je_k)\leq \peri(\triangle e_re_je_k)$.
	This further implies that $\peri(\triangle e_ie_se_t)<\peri(\triangle e_re_se_t)$.
	A proof of this implication is basically the same as the proof of the bi-monotonicity of $\Opt_{i,j}$ (Lemma~\ref{lemma:MFPT-bimono}) and is omitted.
	
	As $e_i$ is better than $e_r$ for $(e_s,e_t)$, edge $e_r$ is not stable in $\triangle e_re_se_t$. Contradiction.
\end{proof}

\subsection{Key lemmas for designing the killing function}

\begin{figure}[h]
	\centering
	\includegraphics[width=.85\textwidth]{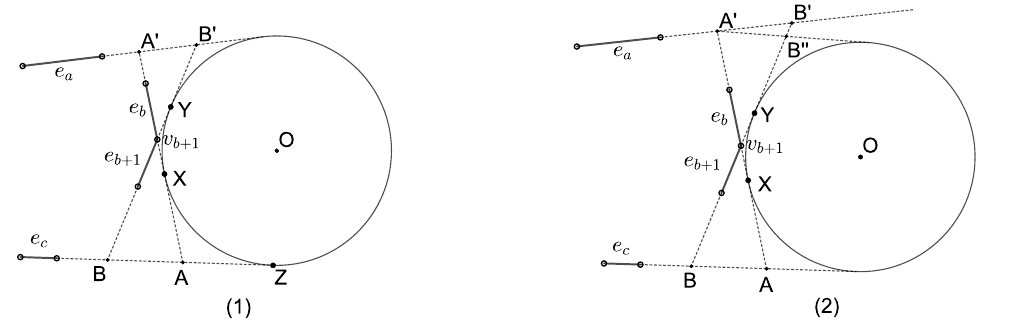}\\
	\caption{Illustration of the proof of Observation~\ref{obs:MFPT-keyObservation}}\label{fig:MFPT-7}
\end{figure}

\begin{observation}\label{obs:MFPT-keyObservation}
	Assume $e_c\prec e_a$. Assume $e_{b},e_{b+1}$ are chased by $e_a$ and are chasing $e_c$.
	Circle $O$ is tangent to $\ell_{b},\ell_{b+1},$ and $\ell_c$ on the right of $e_c$ and left of $e_{b},e_{b+1}$; see Figure~\ref{fig:MFPT-7}.
	
	$\peri(\triangle  e_ae_{b}e_c)=\peri(\triangle e_ae_{b+1}e_c)$ $\Leftrightarrow$ $\ell_a$ is tangent to $O$ .
	
	$\peri(\triangle  e_ae_{b}e_c)<\peri(\triangle e_ae_{b+1}e_c)$ $\Leftrightarrow$ $\ell_a$ is disjoint with $O$ .
	
	$\peri(\triangle  e_ae_{b}e_c)>\peri(\triangle e_ae_{b+1}e_c)$ $\Leftrightarrow$ $\ell_a$ cuts $O$ .
\end{observation}

\begin{proof}
	First, assume that $\ell_a$ is tangent to $O$, as shown in Figure~\ref{fig:MFPT-7}~(1).
	
	Observe that $|BA|+|AX|=|BA|+|AZ|=|BZ|=|BY|=|Bv_{b+1}|+|Xv_{b+1}|$.
	
	Therefore, $|BA|+|Av_{b+1}| = |BA|+|AX|+|Xv_{b+1}|=|Bv_{b+1}|+2|Xv_{b+1}|$.
	
	Therefore, $2|Xv_{b+1}|=|BA|+|Av_{b+1}|-|Bv_{b+1}|$. Similarly, $2|Yv_{b+1}|=|A'B'|+|B'v_{b+1}|-|A'v_{b+1}|$.
	
	Further since $|Xv_{b+1}|=|Yv_{b+1}|$, we get $|BA|+|Av_{b+1}|-|Bv_{b+1}|=|A'B'|+|B'v_{b+1}|-|A'v_{b+1}|$.
	In other words, $|AB|+|AA'|= |A'B'|+|BB'|$, namely, $\peri(\triangle  e_ae_{b}e_c)=\peri(\triangle e_ae_{b+1}e_c)$.
	
	Next, assume that $\ell_a$ is disjoint with $O$, as shown in Figure~\ref{fig:MFPT-7}~(2).
	
	Make a tangent line of $O$ at point $A'$ (other than $\ell_b$) and denote its intersection with $\ell_{b+1}$ by $B''$.
	From the above analysis we have $|A'B''|+|BB''|=|AB|+|AA'|$.
	By the triangle inequality, we have $|A'B'|+|B'B''|>|A'B''|$.
	Therefore, $|A'B'|+|B'B|=|A'B'|+|B'B''|+|BB''| > |A'B''|+ |BB''|=|AB|+|AA'|$.
	It follows that $\peri(\triangle  e_ae_{b}e_c)<\peri(\triangle e_ae_{b+1}e_c)$.
	
	Similarly, we get $\peri(\triangle  e_ae_{b}e_c)>\peri(\triangle e_ae_{b+1}e_c)$ when $\ell_a$ cuts circle $O$.
\end{proof}

\begin{definition}
	For $b\neq c$, denote by $O_{b,c}$ the circle tangent to $e_{b},e_{b+1},e_c$
	on the right of $e_c$ and on the left of $e_{b},e_{b+1}$.
	(It is the circle $O$ introduced in Observation~\ref{obs:MFPT-keyObservation}; see Figure~\ref{fig:MFPT-7}.)
\end{definition}

\begin{figure}[h]
	\begin{minipage}[b]{.65\textwidth}
		\centering \includegraphics[width=\textwidth]{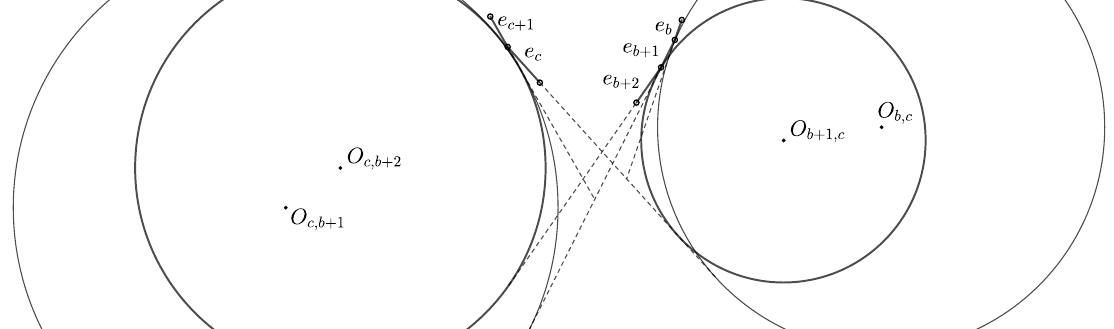}
	\end{minipage}
	\begin{minipage}[b]{.34\textwidth}
		\centering \includegraphics[width=\textwidth]{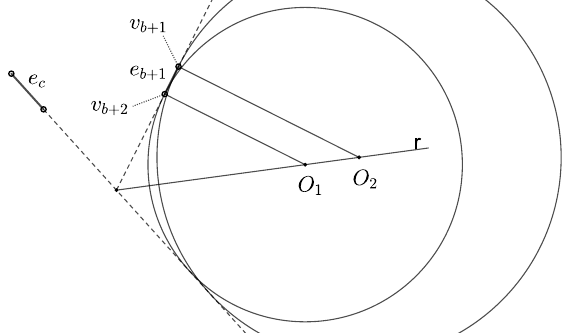}
	\end{minipage}
	\caption{Illustration of the proof of Observation~\ref{observation:MFPT-circle-mono1}}\label{fig:MFPT-8}
\end{figure}

\begin{observation}\label{observation:MFPT-circle-mono1}
	Assume $e_b\prec e_{c+1}$ and assume that $e_b,e_{b+1},\ldots,e_{c+1}$ have at least five edges.
	See Figure~\ref{fig:MFPT-8}. Consider any edge $e_a\in \{e_{c+2},e_{c+3},\ldots,e_{b-1}\}$. We claim that
	
	1. If $\ell_a$ intersects circle $O_{c,b+2}$, it also intersects circle $O_{c,b+1}$.
	
	2. If $\ell_a$ intersects circle $O_{b+1,c}$, it also intersects circle $O_{b,c}$.
	
	To be clear, ``intersects'' means ``cuts or is tangent to''.
\end{observation}

\begin{proof}
	The proof to the first claim is rather trivial, and can be obtained by slightly modifying the proof of Lemma~\ref{lemma:MFPT-bimono} (the bi-monotonicity of $\Opt$); so we omit it.
	
	To prove the second claim, we construct two auxiliary circles $O_1,O_2$ which are both tangent to $\ell_c$ and $\ell_{b+1}$,
	as shown in the right part of Figure~\ref{fig:MFPT-8}. Point $O_1$ is on the perpendicular line of $\ell_{b+1}$ at $v_{b+2}$,
	and point $O_2$ is on the perpendicular line of $\ell_{b+1}$ at $v_{b+1}$. We point out three facts:
	
	a. If $\ell_a$ intersects circle $O_1$, it also intersects circle $O_2$ (trivial proof omitted).
	
	b. If $\ell_a$ intersects circle $O_{b+1,c}$, it also intersects circle $O_1$.
	
	c. If $\ell_a$ intersects circle $O_2$, it also intersects circle $O_{b,c}$.
	
	Combine these three facts, we obtain the second claim of this lemma.
	
	Denote by $r$ the ray originating from the intersection of $\ell_{b+1}$ and $\ell_c$ and is toward $O_1,O_2$.
	Imagine a process where $e_{b+2}$ rotates in counterclockwise around $v_{b+2}$ until it lies on $\ell_{b+1}$.
	Point $O_{b+1,c}$ would move along $r$ (get further from the origin of $r$) and arrive $O_1$, which
	implies fact~b.
	
	The proof of fact~c is similar to that of fact~b (but rotate $e_b$ around $v_{b+1}$ in clockwise).
\end{proof}

%
%
%

\begin{figure}[h]
	\centering \includegraphics[width=.67\textwidth]{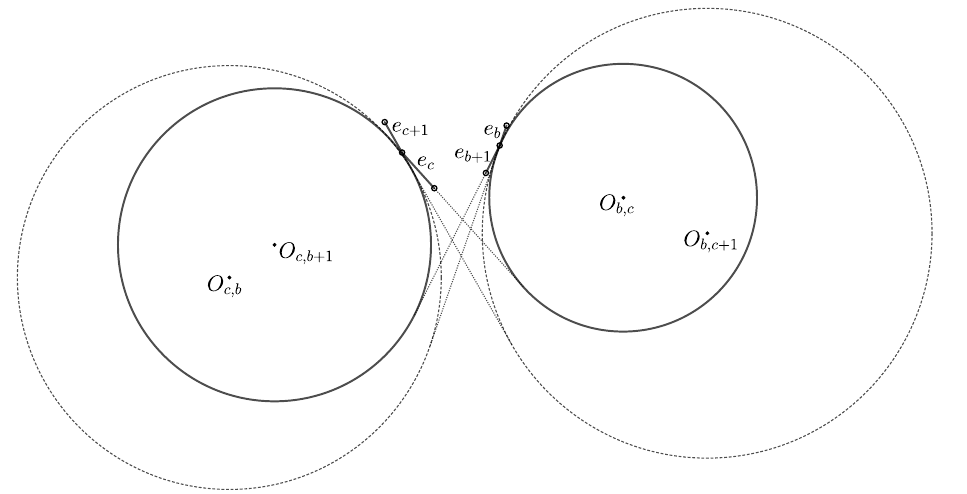}
	\caption{Illustration of two key lemmas: Lemma~\ref{lemma:MFPT-2conditions-prepare} and Lemma~\ref{lemma:MFPT-2conditions-final}}\label{fig:MFPT-9}
\end{figure}

\begin{lemma}\label{lemma:MFPT-2conditions-prepare}
	Assume $e_b\prec e_{c+1}$ and $e_{b+1}\prec e_c$ as shown in Figure~\ref{fig:MFPT-9}. We claim that
	
	1. If $\triangle e_{b'}e_ce_a$ is 3-stable for some $e_{b'}\in\{e_{b+1},\ldots,e_{c-1}\}$ and $e_a\in \{e_{c+2},\ldots,e_{b-1}\}$,
	then $\ell_a$ intersects $O_{c,b+1}$ and $O_{b,c}$ (i.e. the two small circles in the figure).
	
	2. If $\triangle e_be_{c'}e_a$ is 3-stable for some $e_{c'}$ in $\{e_{c+1},\ldots,e_{b-1}\}$  and $e_a\in \{e_{c+2},\ldots,e_{b-1}\}$, then $\ell_a$ avoids $O_{b,c+1}$ and $O_{c,b}$ (i.e. the two big circles in the figure).
	
	To be clear, ``avoid'' means ``is disjoint with or tangent to''.
\end{lemma}

\begin{proof}[Proof of Claim~1.]
	Suppose $\triangle e_{b'}e_ce_a$ is 3-stable.
	
	Because $e_c$ is stable in $\triangle e_{b'}e_ce_a$, $\peri(\triangle e_{b'}e_ce_a)\leq \peri(\triangle e_{b'}e_{c+1}e_a)$.
	Applying Observation~\ref{obs:MFPT-keyObservation}, $\ell_a$ intersects $O_{c,b'}$.
	Applying Observation~\ref{observation:MFPT-circle-mono1}.1, $\ell_a$ also intersects $O_{c,b'-1},\ldots,O_{c,b+1}$.
	
	Because $e_{b'}$ is stable in $\triangle e_{b'}e_ce_a$, $\peri(\triangle e_{b'}e_ce_a)\leq \peri(\triangle e_{b'-1}e_ce_a)$.
	Applying Observation~\ref{obs:MFPT-keyObservation}, $\ell_a$ intersects $O_{b'-1,c}$.
	Applying Observation~\ref{observation:MFPT-circle-mono1}.2, $\ell_a$ also intersects $O_{b'-2,c},\ldots,O_{b,c}$.
\end{proof}

\begin{proof}[Proof of Claim~2.]
	Suppose $\triangle e_be_{c'}e_a$ is 3-stable.
	
	Because $e_b$ is stable in $\triangle e_be_{c'}e_a$, $\peri(\triangle e_be_{c'}e_a)\leq \peri(\triangle e_{b+1}e_{c'}e_a)$.
	Applying Observation~\ref{obs:MFPT-keyObservation}, $\ell_a$ avoids $O_{b,c'}$.
	It follows (from an observation symmetric to Observation~\ref{observation:MFPT-circle-mono1}.1) that $\ell_a$ also avoids $O_{b,c'-1},\ldots,O_{b,c+1}$.
	
	Because $e_{c'}$ is stable in $\triangle e_be_{c'}e_a$, $\peri(\triangle e_be_{c'}e_a)\leq \peri(\triangle e_be_{c'-1}e_a)$.
	Applying Observation~\ref{obs:MFPT-keyObservation}, $\ell_a$ avoids $O_{c'-1,b}$.
	It follows (from an observation symmetric to Observation~\ref{observation:MFPT-circle-mono1}.2) that $\ell_a$ also avoids
	$O_{c'-2,b},\ldots,O_{c,b}$.
\end{proof}

We state that $(e_i,e_j)$ is \emph{dead} if there is no $e_k$ such that $\triangle e_ie_je_k$ is 3-stable.

\begin{lemma}\label{lemma:MFPT-2conditions-final}
	Assume $e_b\prec e_{c+1}$ and $e_{b+1}\prec e_c$ as shown in Figure~\ref{fig:MFPT-9}. We claim that
	
	1. $(e_{b+1},e_c),(e_{b+2},e_c),\ldots$ are dead if it holds that
	\begin{equation}\label{eqn:MFT(P)-condition-1}
		\text{none of $\ell_{c+2},\ldots,\ell_{b-1}$ intersects $O_{c,b+1}$ and $O_{b,c}$},
	\end{equation}
	
	and 2. $(e_b,e_{c+1}),(e_b,e_{c+2}),\ldots$ are dead if it holds that
	\begin{equation}\label{eqn:MFT(P)-condition-2}
		\text{none of $\ell_{c+2},\ldots,\ell_{b-1}$ avoids $O_{b,c+1}$ and $O_{c,b}$}.
	\end{equation}
\end{lemma}

\begin{proof}
	1. Suppose some $(e_{b'},e_c)$ in the list $(e_{b+1},e_c),\ldots$ is not dead.
	Clearly, $e_{b'} \in \{e_{b+1},\ldots,e_{c-1}\}$.
	
	Because $(e_{b'},e_c)$ is not dead, there exists $e_a$ such that $\triangle e_ae_{b'}e_c$ is 3-stable.
	Be aware that $e_c\prec e_a$. Therefore $e_a \in \{e_{c+1},\ldots,e_{b-1}\}$.
	Moreover, $e_a\prec e_{b'}$ and so $e_a\neq e_{c+1}$. Together, $e_a \in \{e_{c+2},\ldots,e_{b-1}\}$.
	Applying Lemma~\ref{lemma:MFPT-2conditions-prepare}, $\ell_a$ intersects $O_{c,b+1}$ and $O_{b,c}$. Contradictory.
	
	2. Suppose some $(e_b,e_{c'})$ in the list $(e_b,e_{c+1}),\ldots$ is not dead.
	Clearly, $e_{c'}$ in $\{e_{c+1},\ldots,e_{b-1}\}$.
	
	Because $(e_b,e_{c'})$ is not dead, there exists $\triangle e_ae_be_{c'}$ which is 3-stable.
	Clearly, $e_a\in$ $\{e_{c'+1},\ldots,e_{b-1}\}$ and hence $e_a\in \{e_{c+2},\ldots,e_{b-1}\}$.
	Applying Lemma~\ref{lemma:MFPT-2conditions-prepare}, $\ell_a$ avoids $O_{b,c+1}$ and $O_{c,b}$. Contradictory.
\end{proof}

\begin{remark}
	To continue, we may try the strategy used in subsection~\ref{subsect:Kill-F-MFTA}:
	Define a common tangent of circles $O_{b,c},O_{c,b}$,
	and a common tangent of circles $O_{b,c+1},O_{c,b+1}$,
	and find a line $l$ whose angle is between these two tangent lines
	and moreover, use the relation between $l$ and $P$ to determine that \eqref{eqn:MFT(P)-condition-1} or \eqref{eqn:MFT(P)-condition-2} holds.
	This is correct but we are not able to determine the relation between $l$ and $P$ in amortized $O(1)$ time.
	To determine the relation in such time, we must prove a monotonicity of the angles resembling the one given in Observation~\ref{obs:d-mono-2}, but such a monotonicity does not hold here.
\end{remark}

\subsection{Designing an efficient killing functions}

We state that $(e_i,e_j)$ is \emph{DEAD}, if there is no $e_k$ such that $\triangle e_ie_je_k$
has the minimum perimeter among all the all-flush triangles.
Note that ``dead'' implies ``DEAD'', but the reverse is false.

\begin{figure}[h]
	\centering \includegraphics[width=.7\textwidth]{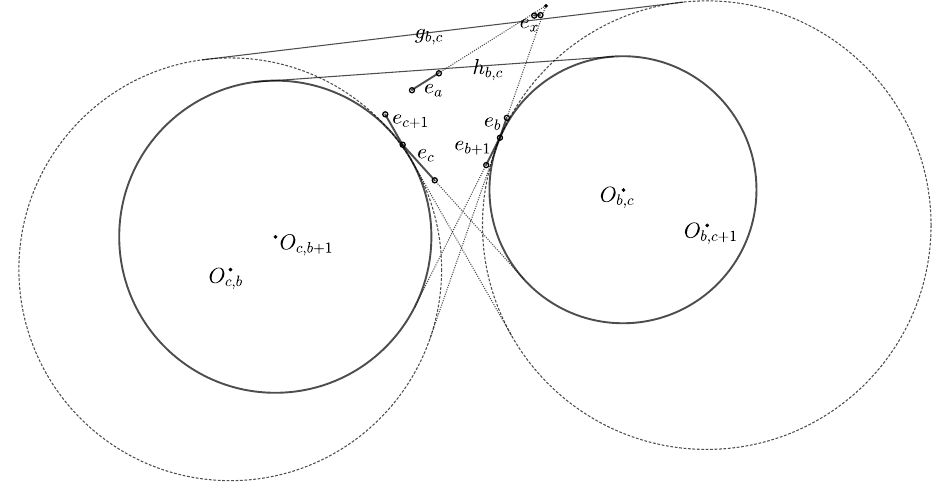}
	\caption{Illustration of our killing function}\label{fig:MFPT-10}
\end{figure}

\begin{lemma}\label{lemma:MFPT-kill-function}
	Assume $e_b\prec e_{c+1}$ and $e_{b+1}\prec e_c$. See Figure~\ref{fig:MFPT-10}.
	Denote by $g_{b,c}$ the common tangent of circles $O_{c,b},O_{b,c+1}$
	(that intersects the ray $\overrightarrow{v_cv_{c+1}}$).
	Denote by $h_{b,c}$ the common tangent of circles $O_{b,c},O_{c,b+1}$
	(that intersects the ray $\overrightarrow{v_cv_{c+1}}$).
	Denote $a=\Opt_{b,c}$. We claim that
	
	1. If $e_a$ is not (entirely) below line $h_{b,c}$, pairs $(e_{b+1},e_c),(e_{b+2},e_c),\ldots$ are dead.
	
	2. If $e_a$ is (entirely) below line $h_{b,c}$, pairs $(e_b,e_{c+1}),(e_b,e_{c+2}),\ldots$ are DEAD.\\
	(In fact, we conjecture that these pairs are also dead. Yet we have no clue how to prove it.)
\end{lemma}

Following this lemma, we obtain the following killing function which is similar to \eqref{eqn:def-Kill}.
\begin{equation}\label{eqn:def-Kill-MFPT}
	\text{$\Kill_p(b,c)=$`b' if $e_a$ is below $h_{b,c}$, and $\Kill_p(b,c)=$`c' otherwise.}
\end{equation}

\begin{proof}[Proof of Claim~1.]
	If $e_a$ is not (entirely) below line $h_{b,c}$,
	none of $\ell_{c+2},\ldots,\ell_{b-1}$ can intersect both $O_{c,b+1}$ and $O_{b,c}$
	(otherwise the convex polygon $P$ is entirely below $h_{b,c}$ and so as $e_a$).
	It follows that $(e_{b+1},e_c)$, $(e_{b+2},e_c)$, $\ldots$ are dead according to Lemma~\ref{lemma:MFPT-2conditions-final} claim~1.
\end{proof}

We now proceed to the proof of Claim~2.

When $e_a$ is below $h_{b,c}$, there could be some $\ell_x$ in $\{\ell_{c+2},\ldots,\ell_{b-1}\}$ which avoids $O_{b,c+1}$ and $O_{c,b}$;
see Figure~\ref{fig:MFPT-10}. So, condition \eqref{eqn:MFT(P)-condition-2} does not hold
and hence we cannot apply Lemma~\ref{lemma:MFPT-2conditions-final} claim~2 directly.
However, we can show that even if there is $\ell_x$ as mentioned above,
$\triangle e_be_{c'}e_x$ cannot be the minimum perimeter all-flush triangle for $e_{c'}\in \{e_{c+1},\ldots,e_{b-1}\}$,
as $\triangle e_be_ce_a$ has a smaller perimeter.

We state three inequalities for $\ell_x\in \{\ell_{c+2},\ldots,\ell_{b-1}\}$ that avoids circles $O_{b,c+1}$ and $O_{c,b}$.
\begin{eqnarray}
	\peri(\triangle e_ae_be_c) &<&\peri(\triangle ge_be_c), \label{eqn:MFT(P)-inequality-1} \\
	\peri(\triangle ge_be_c) &\leq &\peri(\triangle ge_be_{c'}),\label{eqn:MFT(P)-inequality-2} \\
	\peri(\triangle ge_be_{c'}) &\leq &\peri(\triangle e_xe_be_{c'}), \label{eqn:MFT(P)-inequality-3}
\end{eqnarray}
where $g$ is short for $g_{b,c}$. Combining these inequalities, we prove Claim~2 as follows.

\begin{proof}[Proof of Claim~2.]
	Suppose some pair $(e_b,e_{c'})$ in the list $(e_b,e_{c+1}),\ldots$ is not DEAD.
	This means there is $\triangle e_xe_be_{c'}$ which is (one of) the globally minimum solution
	(and hence is 3-stable).
	Clearly, $e_x\in \{e_{c'+1},\ldots,e_{b-1}\}$ and hence $e_x\in \{e_{c+2},\ldots,e_{b-1}\}$, whereas
	$e_{c'}$ in $\{e_{c+1},\ldots,e_{b-1}\}$.
	Applying Lemma~\ref{lemma:MFPT-2conditions-prepare}, $\ell_x$ avoids $O_{b,c+1}$ and $O_{c,b}$.
	Then, applying the inequalities \eqref{eqn:MFT(P)-inequality-1}-\eqref{eqn:MFT(P)-inequality-3} above,
	$\peri(\triangle e_ae_be_c)< \peri(\triangle e_xe_be_{c'})$. So $\triangle e_xe_be_{c'}$ is not the minimum. Contradiction.
\end{proof}

\begin{proof}[Proof of \eqref{eqn:MFT(P)-inequality-2}.]
	Since $g$ is tangent to circle $O_{c,b}$, $\peri(\triangle ge_be_c)=\peri(\triangle ge_be_{c+1})$
	due to Observation~\ref{obs:MFPT-keyObservation}.
	By the unimodality of the perimeters (Lemma~\ref{lemma:MFPT-unimodal}),
	$\peri(\triangle ge_be_c)=\peri(\triangle ge_be_{c+1})<\peri(\triangle ge_be_{c+2})<\cdots,$
	which implies \eqref{eqn:MFT(P)-inequality-2}.
\end{proof}

\begin{figure}[h]
	\begin{minipage}{.56\textwidth}
		\flushleft \includegraphics[width=\textwidth]{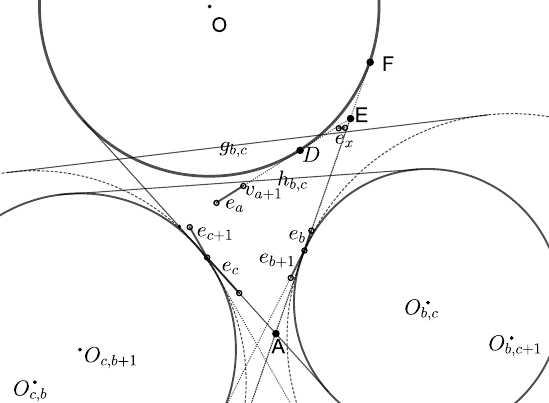}
	\end{minipage}
	\begin{minipage}{.44\textwidth}
		\flushright \includegraphics[width=.85\textwidth]{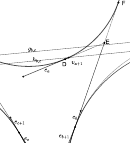}
	\end{minipage}
	\caption{Illustration of the proof of \eqref{eqn:MFT(P)-inequality-1}}\label{fig:MFPT-10p}
\end{figure}

\begin{proof}[Proof of \eqref{eqn:MFT(P)-inequality-1}.]
	Construct a circle $O$ tangent to $\ell_b,\ell_c$, and $\ell_a$, on the right of $e_b,e_c$ and the left of $e_a$; see Figure~\ref{fig:MFPT-10p}. Applying Observation~\ref{Obs:MFPT-circle-tangent},
	it reduces to proving that line $g$ cuts circle $O$.
	This further reduces to proving that $O$ contains a point above $g$ and a point below $g$.
	
	Denote by $D$ the tangent point of $\ell_a$ and circle $O$.
	Denote by $F$ the tangent point of $\ell_b$ and circle $O$.
	As $O$ contains $D$ and $F$, it suffices to show that (i) $F$ is above $g$ and (ii) $D$ is below $g$.
	
	\emph{Proof of (i)}:  Because $\ell_x$ avoids circles $O_{b,c+1}$ and $O_{c,b}$,
	edge $e_x$ is above $g$. This means that the intersection of $\ell_a$ and $\ell_b$, denoted by $E$, must be above $g$.
	Further since $E$ lies on $\ell_a$ whereas circle $O$ (including point $F$) is on the left of $e_a$,
	we see that $F$ must also be above $g$.
	
	\emph{Proof of (ii)}: Dividing $\ell_a$ at $v_{a+1}$ returns two rays, one of which intersects $\ell_b$; denote it by $r$.
	We first show that $D$ does not lie in $r$.
	Suppose to the opposite that $D$ lies in $r$, as shown in the left part of Figure~\ref{fig:MFPT-10p}.
	Obviously, $\ell_{a+1}$ must be disjoint with circle $O$, which implies that
	$\peri(\triangle e_be_ce_{a+1})<\peri(\triangle e_be_ce_a)$ due to Observation~\ref{Obs:MFPT-circle-tangent},
	contradicting with $a=\Opt_{b,c}$.
	Therefore, $D$ must lie in the ray at $v_{a+1}$ toward $v_a$, as shown in the right part of Figure~\ref{fig:MFPT-10p}.
	Applying the assumption that $e_a$ lies below $h_{b,c}$ whereas the intersection $E$ is above $g$ (proved above),
	the ray at $v_{a+1}$ toward $v_a$, which contains point $D$, is below $h_{b,c}$. It follows that $D$ is below $g$.
\end{proof}

\begin{proof}[Proof of \eqref{eqn:MFT(P)-inequality-3}.]
	If $e_x$ is parallel to $g$, the inequality is trivial. Otherwise, there are two cases:
	
	\emph{Case 1}: $v_x$ is closer to $g$ than $v_{x+1}$, as shown in the left part of Figure~\ref{fig:MFPT-10p-more}.
	Denote by $O'$ the circle tangent to $g,\ell_{c'}$, and $\ell_b$ (above $g$ and on the right of $e_b,e_{c'}$).
	Build two parallel lines $t,t'$ of $e_x$ tangent to $O',O_{b,c+1}$ respectively.
	Since (1) $v_x$ is closer to $g$ than $v_{x+1}$ and (2) $e_x \prec e_b$,
	extreme positions for $t$ are $\ell_b$ and $g$. It follows that $t$ intersects $O_{c,b}$.
	Similarly, $t'$ also intersects $O_{c,b}$.
	So, all lines parallel to $e_x$ between $t$ and $t'$ intersects $O_{c,b}$.
	Further since $\ell_x$ avoids $O_{c,b}$ and $O_{b,c+1}$, $\ell_x$ is above $t$,
	which implies that $\peri(\triangle e_be_{c'}g)<\peri(\triangle e_be_{c'}e_x)$ due to Observation~\ref{Obs:MFPT-circle-tangent}.
	
	\begin{figure}[h]
		\begin{minipage}{.5\textwidth}
			\centering \includegraphics[width=.95\textwidth]{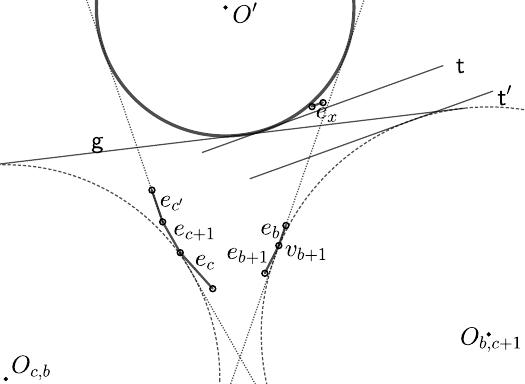}
		\end{minipage}
		\begin{minipage}{.5\textwidth}
			\centering \includegraphics[width=.95\textwidth]{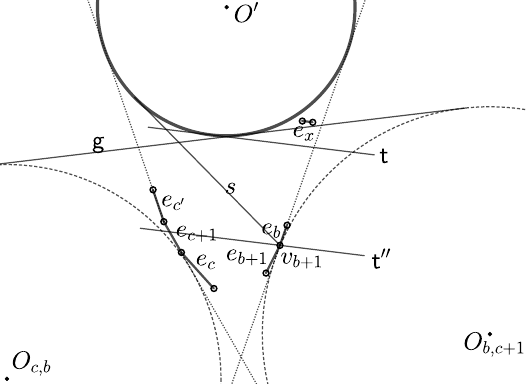}
		\end{minipage}
		\caption{Illustration of the proof of \eqref{eqn:MFT(P)-inequality-3}}\label{fig:MFPT-10p-more}
	\end{figure}
	
	\emph{Case 2}: $v_{x+1}$ is closer to $g$ than $v_x$, as shown in the right part of Figure~\ref{fig:MFPT-10p-more}.
	Define $O'$ and $t$ the same as above. Denote by $t''$ the line at $v_{b+1}$ which is parallel to $e_x$.
	Denote by $s$ the tangent line of $O'$ at $v_{b+1}$ other than $\ell_b$.
	Observe that $s$ cuts circle $O_{b,c+1}$ (see Figure~\ref{fig:MFPT-10p-obs} for an easy illustration).
	
	Imagine a line $l$ rotates in clockwise around $v_{b+1}$, starting at the direction parallel to $g$.
	If $l$ meets $t''$ later than $s$, line $t''$ has to cut $O'$. So, the parallel line $\ell_x$ of $t''$ (above $t''$) cuts or is above $O'$.
	If on the contrary $l$ meets $t''$ earlier than $s$ (as shown in the figure), $t''$ cuts $O_{b,c+1}$, and so does $t$,
	hence $\ell_x$ (which avoids $O_{b,c+1}$) can only be above $t$. In both subcases, it holds that $\ell_x$ cuts or is above $O'$.
	Applying Observation~\ref{Obs:MFPT-circle-tangent}, this means $\peri(\triangle e_be_{c'}g)<\peri(\triangle e_be_{c'}e_x)$.
\end{proof}

\begin{figure}[h]
	\centering \includegraphics[width=.6\textwidth]{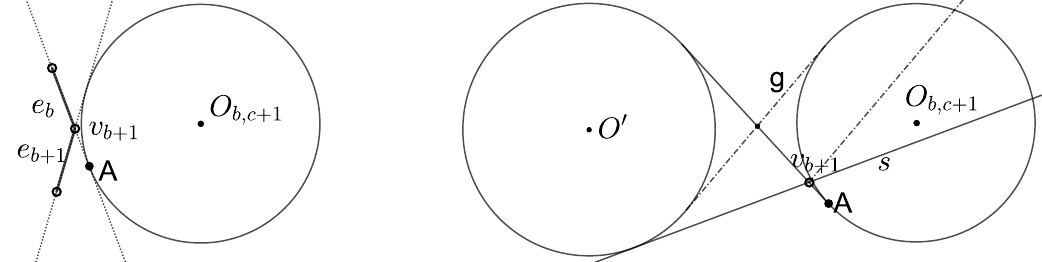}
	\caption{Illustration of an observation used in the proof of \eqref{eqn:MFT(P)-inequality-3} (Case~2)}\label{fig:MFPT-10p-obs}
\end{figure}

\begin{remark}
	The proof of Lemma~\ref{lemma:MFPT-kill-function} is very similar to the proof of Lemma~\ref{lemma:correct}.
	Basically, for claim~1 it applies only the assumption that $e_b,e_c$ (or, $v_b,v_c$ in Lemma~\ref{lemma:correct}) are stable;
	but for claim~2 it applies the other assumption that $e_a$ (or, $v_a$ in Lemma~\ref{lemma:correct}) is stable as well.
\end{remark}

\section{Comparison of Alg-A to the previous Alg-CM and Alg-K}\label{sect:compare}

\paragraph*{Outline of Alg-K \cite{Kallus17b}.}
Take any $\alpha\in [0,2\pi)$.
Kallus \cite{Kallus17b} proved that
among those $\triangle ABC$ in which $A,B,C$ lie in $P$'s boundary and $\theta[\overrightarrow{BC}]=\alpha$,
there is a unique one with the maximum area, denoted by $T_\alpha=\triangle A_\alpha B_\alpha C_\alpha$.
It is also proved that, when $\alpha$ increases, $A_\alpha,B_\alpha,C_\alpha$ will all move in clockwise around $P$'s boundary (non-strictly).
Moreover, this rotating process of $A_\alpha,B_\alpha,C_\alpha$ can be simulated discretely;
in $O(1)$ time we can jump from one ``critical point'' of $(\alpha,\triangle A_\alpha B_\alpha C_\alpha)$ to the next.
Finally, the maximum area triangle can be found among $\{\triangle A_\alpha B_\alpha C_\alpha \mid \alpha \hbox{ is a critical point}\}$.

\begin{figure}[h]
	\centering\includegraphics[width=.95\textwidth]{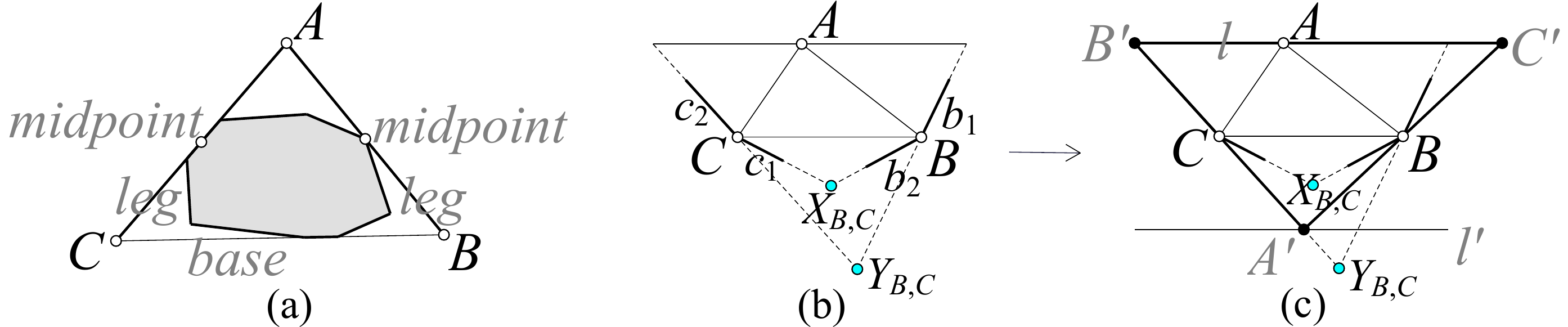}\\
	\caption{Equivalence between Alg-K and Alg-CM.}\label{fig:equivalence}
\end{figure}

\paragraph*{Outline of Alg-CM \cite{linear-correct}.}
Chandran and Mount define the \emph{$P$-stable} triangles (outside $P$) as follows.
All sides of a $P$-stable triangle are touched by $P$.
In particular, two of them (called \emph{legs}) have their midpoints touched by $P$, whereas the remaining one is called the \emph{base}.
Moreover, one of the following holds: (1) The base is flushed with (i.e. contains an edge of) $P$.
(2) One of the legs is flushed with an edge of $P$ and has as its midpoint a vertex of this edge.
See Figure~\ref{fig:equivalence}~(a).

If such a triangle satisfies (1), it is called \emph{$P$-anchored}.
The $P$-anchored triangles were introduced earlier in \cite{Tri-Enclose-Area-nlogn2},
where it is proved that \emph{they contain all the local minimums of enclosing triangles of $P$}.
It is proved in \cite{Tri-Enclose-Area} that the $P$-anchored triangles are ``interspersing'' (similar to interleaving).
Using the Rotating-Caliper technique \cite{rotatingcaliper} with some clever algorithmic tricks,
\cite{Tri-Enclose-Area} computes the $P$-anchored triangles in linear time and thus find the minimum enclosing triangles.

Using some ``more involved'' (as said in \cite{linear-correct}) observations,
Chandran and Mount \cite{linear-correct} compute all the $P$-stable triangles in linear time by the Rotating-Caliper technique,
and show that \emph{all the local maximums of triangles in $P$ can be computed easily from $P$-stable triangles}.
(It follows that Alg-CM computes all the maximum area triangles in $P$ and all the minimum area triangles enclosing $P$.)

\paragraph*{An equivalence between the triangles explored by the two algorithms.}
We find that the triangles explored by Alg-K are the same as the $P$-stable triangles explored by Alg-CM.
Assume $T_\alpha=\triangle ABC$. See Figure~\ref{fig:equivalence}~(b).
Let $b_1,b_2$ be the two neighboring edges of $B$ and $c_1,c_2$ be the two neighboring edges of $C$ as shown in the figure.
Assume the extended lines of $b_2,c_1$ intersect at point $X_{B,C}$ whereas the extended lines of $b_1,c_2$ intersect at point $Y_{B,C}$.
Let $d(X)$ denote the distance from point $X$ to $\overleftrightarrow{BC}$.
It is easy to prove that the following are equivalent:
(i) $\triangle ABC$ is the maximum area triangle with $\theta[\overrightarrow{BC}]=\alpha$;
and (ii) $d(A)\in [d(X_{B,C}),d(Y_{B,C})]$ and A has the largest distance to $\overleftrightarrow{BC}$ among all vertices on the right of $\overrightarrow{BC}$.
As fact~(i) holds, (ii) also holds.
We now construct a triangle $\triangle A'B'C'$ outside $P$ (see Figure~\ref{fig:equivalence}~(c)).
Let $l$ be the line at $A$ parallel to $\overleftrightarrow{BC}$ and $l'$ be its symmetric line with respect to $\overleftrightarrow{BC}$.
Assume the extended line of $c_2$ intersects $l,l'$ at points $B',A'$, respectively, and
let $C'$ be the symmetric point of $A'$ with respect to $B$.
Applying fact (ii), $P$ must be contained in $A'B'C'$, which implies that $A'B'C'$ is $P$-stable.
Similarly, from a $P$-stable triangle $A'B'C'$, we can also construct $\triangle ABC$ satisfying (ii) and hence satisfying (i).

\subsection{Experimental results}\label{subsect:experiment}

We implement Alg-A, Alg-K, and Alg-CM by \emph{c++ programs}.
The running times of these programs on seven randomly generated test cases are shown in Table~\ref{table:experiment},
in which we also record the number of iterations required by each program.
Download our implementations at GitHub \cite{implementation}.
\textbf{Our implementations of Alg-K and Alg-CM are based on comprehensive understanding of their algorithms,
	and the code optimizations in three programs are at the same level.}

\begin{table}[h]
	\begin{footnotesize}
		\centering
		\begin{spacing}{1.2}
			\begin{tabular}{|c||c||c|c|c||c|c|c|}
				\hline
				& N             & T(A)      & T(K)      & T(CM)  & \#(A)        & \#(K)     & \#(CM)\\ \hline
				1     & 100035        & 0.0027       & 0.0157       & 0.0228       & 99878         & 455361        & 455178 \\
				2     & 1000000       & 0.0235       & 0.2001       & 0.2605       & 998511        & 4659599       & 4657908 \\
				3     & 1000000       & 0.0207       & 0.1673       & 0.2355       & 997681        & 4501969       & 4499593 \\
				4     &  999900       & 0.0241       & 0.2109       & 0.2950       & 998933        & 4658202       & 4657109 \\
				5     &  999900       & 0.0209       & 0.2110       & 0.2513       & 998203        & 5001147       & 4999500 \\
				6     & 1000000       & 0.0263       & 0.2319       & 0.2797       & 999673        & 4580226       & 4579936 \\
				7     & 1000000       & 0.0329       & 0.2283       & 0.2789       & 999968        & 4581191       & 4581150 \\ \hline
				Sum   & \bf 6099835   & \bf 0.1511   & \bf 1.2652   & \bf 1.6237   & \bf 6092847   & \bf 28437695  & \bf 28430374 \\
				\hline
			\end{tabular}
		\end{spacing}
		\caption{A test on our implementations of Alg-A, Alg-K, and Alg-CM.
			In the table, symbol ``T'' denotes the running time, ``\#'' denotes the number of iterations taken by the algorithm,
			and ``N'' denotes the total number of vertices in the test case (each test case consists of several polygons).}\label{table:experiment}
	\end{footnotesize}
\end{table}

\begin{description}
	\item [Running environment and experiment result.] We run the experiment on Lenovo Notepad X1 Carbon (i7-8550U CPU, 8G RAM) with Win10 (64bit). We use VC++ compiler on release mode. (We also did an experiment using GCC compiler \cite{implementation}. The result is similar.)
	
	Our experiment shows that the running time of Alg-A is roughly \textbf{one eighth} of the running time of Alg-K, or \textbf{one tenth} of the running time of Alg-CM. (Moreover, the number of iterations required by Alg-CM and Alg-K is roughly \textbf{4.67} times that of Alg-A.)
	
	\item [On convex polygon generation.] We generate all test cases in random.
	Specifically, three methods are applied:
	(1) Generate $N$ random points in a disk and compute their convex hull. (Only test~1 is generated using this method)
	(2) Take $n$ random points in an ellipse.
	(3) A clever and efficient method for generating convex polygons given in \cite{gen-convex-polygon} (which is based on \cite{random-convex-polygon}).
	
	Our seven test cases and our \emph{C++ program} for generating them can be downloaded at \cite{implementation}.
	Test~2 and Test~3 consist of 5000 polygons, each of which has 200 vertices;
	Test~4 and Test~5 consist of 3333 polygons, each of which has 300 vertices;
	Test~6 (Test~7, respectively) consists of 1000 (100, respectively) polygons, each of which has 1000 (10000, respectively) vertices.
	
	As Alg-K and Alg-CM suffers from float issue, we give up polygons with $>10000$ vertices.
	
	\item [Further notes.] 1. We only calculate the time for computing the maximum area triangle
	and does not calculate the time for inputting the data. See \emph{experiment.cpp} in \cite{implementation}.
	2. Theoretically, \#(Alg-K) equals \#(Alg-CM) (see last page).
	However, these two statistics are slightly different in our experiment; see Table~\ref{table:experiment}.
	The difference is mainly due to the degenerate case (where a chord of $P$ is parallel to an edge of $P$) and the float issue of both programs.
	Our implementations of Alg-K and Alg-CM have logical difference in handling degenerate cases.
\end{description}

\subsection{A comparison of Alg-A, Alg-CM, and Alg-K}\label{subsect:advantages}

Alg-A has a smaller constant behind the asymptotic complexity $O(n)$ than others as mentioned in subsection~\ref{subsect:compareKKK}.
Moreover, Alg-A is more stable than the alternatives.
During the iterations of Alg-CM, the coordinates of three corners and two midpoints of a P-stable triangle (see Figure 37) are maintained. These coordinates are computed somehow and their true values can differ from their values stored in the computer. Alg-CM uses an involved subroutine (far more complicated than ours given in Algorithm~1) to update the coordinates in each iteration, which accumulates the inaccuracy of coordinates. Even worse, this subroutine computes three angles and selects the smallest to decide how to proceed each time, and due to float issue it is possible to select a wrong angle when angles are close, which causes the subroutine performs incorrectly.
The following comparison might be subjective, though we tried our best to be unbiased.
(We refer the readers to a recent class note of Rote \cite{classnote-rote} for an excellent comparison from third-party.)

\begin{description}
	\item[On difficulty of analysis.] Comparing the description of the main part of Alg-A (the 7 lines in Algorithm~\ref{alg:all-r-k}) with that of Alg-CM (pages 9--10 of \cite{linear-correct}),
	Alg-A is conceptually simpler. Alg-CM is claimed ``involved'' by its authors as it contains complicated subroutines for handling many subcases.
	Our proof is based on few prerequisite results -- only the trivial pairwise interleaving property,
	whereas Alg-CM is built upon nontrivial results of \cite{Tri-Enclose-Area-kleeMid,Tri-Enclose-Area-nlogn2,Tri-Enclose-Area}.
	
	\item[On difficulty of implementation.]
	We implement all the three known algorithms \footnote{As far as we know, Alg-CM has not been implemented in history.
		An implementation of Alg-K was given by Kallus. However, to compare Alg-K with the other two alternatives, it is better to
		implement it by us instead of using the implementation of Kallus, so that
		the algorithms are implemented using similar programming techniques.}
	(see \cite{implementation}).
	Alg-CM and Alg-K are respectively \textbf{3} times and \textbf{2} times longer in code length comparing to Alg-A.
	Implementing Alg-CM correctly is a real challenge and we spent roughly 4 to 5 days on doing this.
	The difficulties lies in: (1) degenerated cases are not discussed in \cite{linear-correct} 
	and (2) we have to put patches to the program multiple times to avoid all kinds of float issues.
\end{description}

\clearpage

\appendix

%
%

\section{Appendix}\label{sect:misc}

Note that in the following proof of Lemma~\ref{lemma:interleaving}, we apply Observation~\ref{Obs:back-forw-F} stated in subsection~\ref{subsect:F-omitted}.

\begin{proof}[Proof of the interleaving property stated in Lemma~\ref{lemma:interleaving}]
	The first claim is implied by the third, as 3-stable implies G-3-stable.
	So we only show the proof of claim~2 and claim~3.
	
	2. Suppose $T_1=\triangle e_ie_je_k$ and $T_2=\triangle e_re_se_t$ are F-3-stable triangles that are not interleaving.
	As they are not interleaving, sets $\{e_i,e_j,e_k\}$ and $\{e_r,e_s,e_t\}$ share at most one common element.
	First, consider the case where the two sets share one common edge, e.g., $e_j=e_s$.
	Without loss of generality, assume that $e_k,e_t,e_r,e_i,e_j=e_s$ are in clockwise order; see Figure~\ref{fig:F-interleaving}~(a)
	(otherwise $e_t,e_k,e_i,e_r,e_j=e_s$ are in clockwise order and it is symmetric).
	
	Because $\triangle e_ie_je_k$ is F-3-stable, $\area(\triangle e_ie_je_k)\leq \area(\triangle e_re_je_k)$.
	This implies that $\area(\triangle e_ie_je_t)<\area(\triangle e_re_je_t)$ (due to Observation~\ref{Obs:back-forw-F});
	i.e., $\area(\triangle e_ie_se_t)<\area(\triangle e_re_se_t)$. It follows that $e_r$ is not stable in $\triangle e_re_se_t$,
	which contradicts the assumption that $\triangle e_re_se_t$ is F-3-stable.
	
	Second, consider the case where the above mentioned two sets share no common edges.
	In this case, since $T_1$ and $T_2$ are not interleaving,
	among $[v_{j+1}\circlearrowright v_k],[v_{k+1}\circlearrowright v_i]$ and $[v_{i+1}\circlearrowright v_j]$,
	there must be one boundary portion that contains no edge from $\{e_r,e_s,e_t\}$,
	whereas the others respectively contain one and two.
	(Note that $e_r,e_s,e_t$ cannot be contained in the same portion. Otherwise $\triangle e_re_se_t$ is unbounded and hence not F-3-stable.)
	Without loss of generality, assume that $[v_{j+1}\circlearrowright v_k]$ and $[v_{k+1}\circlearrowright v_i]$ respectively contain $\{e_s\}$ and $\{e_t,e_r\}$, as shown in Figure~\ref{fig:F-interleaving}~(b).
	
	Because $\triangle e_ie_je_k$ is F-3-stable, $\area(\triangle e_ie_je_k)\leq \area(\triangle e_re_je_k)$.
	This implies that $\area(\triangle e_ie_se_k)<\area(\triangle e_re_se_k)$ (due to Observation~\ref{Obs:back-forw-F}).
	To be rigorous, check that $\area(\triangle e_ie_se_k)$ is finite. This holds because $e_i\prec e_s$ (since $e_r\prec e_s$)
	and $e_s\prec e_k$ (since $e_s\prec e_t$) and $e_k\prec e_i$.
	Furthermore, it follows that $\area(\triangle e_ie_se_t)<\area(\triangle e_re_se_t)$ (due to Observation~\ref{Obs:back-forw-F} again).
	To be rigorous, check that $\area(\triangle e_ie_se_t)$ is finite. This holds because $e_i\prec e_s$ (since $e_r\prec e_s$)
	and $e_s\prec e_t$ and $e_t\prec e_i$ (since $e_k\prec e_i$).
	It follows that $e_r$ is not stable in $\triangle e_re_se_t$, which contradicts the F-3-stable assumption.
	
	\begin{figure}[h]
		\begin{minipage}[b]{.4\textwidth}
			\flushleft \includegraphics[width=.85\textwidth]{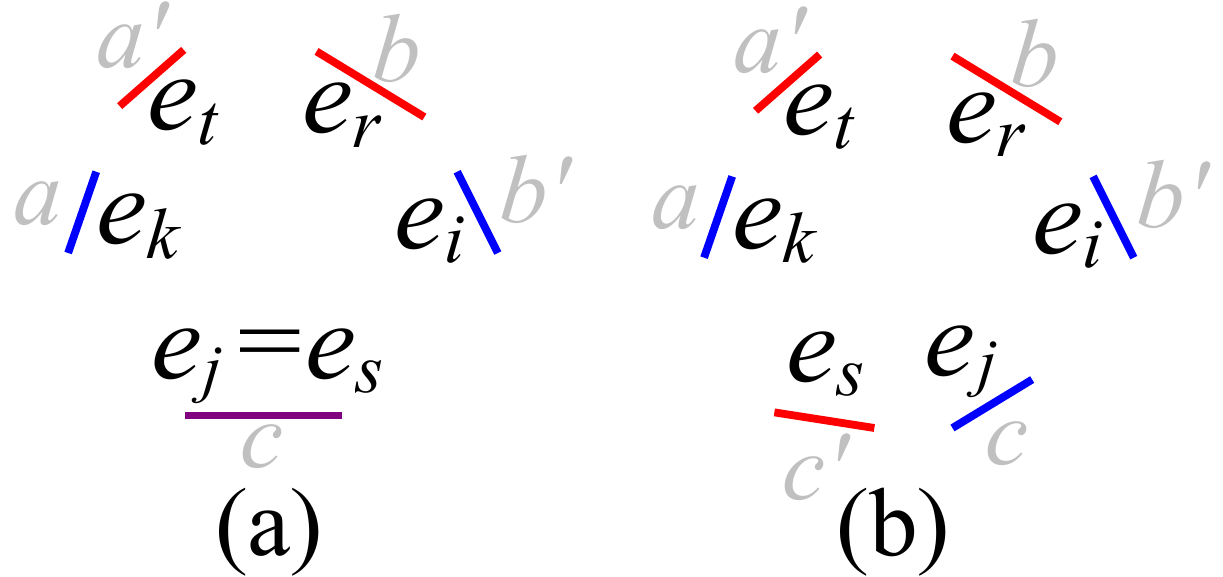}
			\caption{Proof of Lemma~\ref{lemma:interleaving}~part~2.}\label{fig:F-interleaving}
		\end{minipage}
		\begin{minipage}[b]{.6\textwidth}
			\flushright \includegraphics[width=.9\textwidth]{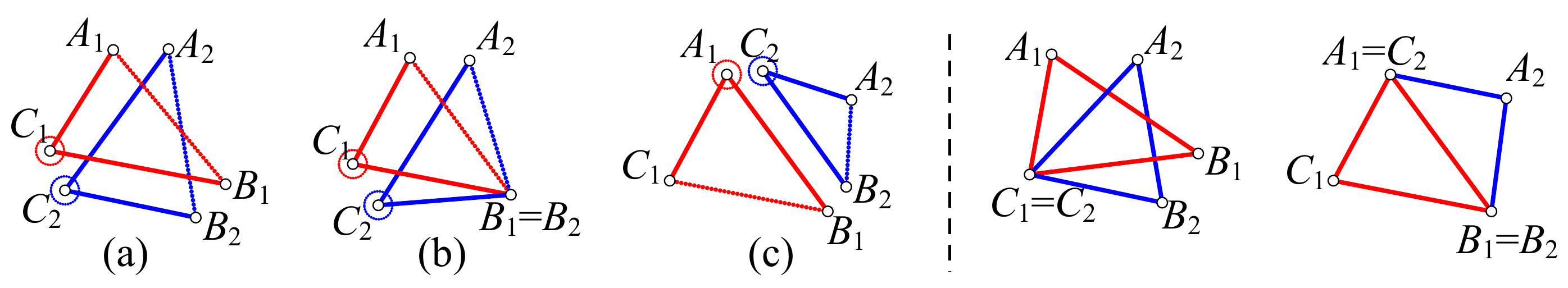}
			\caption{Proof of Lemma~\ref{lemma:interleaving}~part~3.}\label{fig:G-interleaving}
		\end{minipage}
	\end{figure}
	
	3. It reduces to showing that for any two triangles $\triangle A_1B_1C_1,\triangle A_2B_2C_2$ (with corners lying in $\partial P$) that do not interleave,
	at most one of them is G-3-stable.
	When the two triangles do not interleave, there are only three essentially different cases
	as shown in Figure~\ref{fig:G-interleaving}:
	
	\noindent \emph{Case 1.} \emph{$A_1,A_2,B_1,B_2,C_2,C_1$ are different and lie in this order around $\partial P$}.
	See Figure~\ref{fig:G-interleaving}~(a). As $A_1B_1$ intersects $A_2B_2$, at most one of the following holds:
	(i) $C_1$ is stable in $\triangle A_1B_1C_1$; (ii) $C_2$ is stable in $\triangle A_2B_2C_2$.
	This means at most one of the two triangles is G-3-stable.
	
	\noindent \emph{Case 2.} \emph{$A_1,A_2,B_1,B_2,C_2,C_1$ lie in this order around $\partial P$ and are distinct except $B_1=B_2$}.
	See Figure~\ref{fig:G-interleaving}~(b). The proof of this case is the same as above.
	
	\noindent \emph{Case 3.} \emph{$C_1,A_1,C_2,A_2,B_2,B_1$ lie in this order around $\partial P$ and are distinct or are distinct except that $B_1=B_2$.}
	See Figure~\ref{fig:G-interleaving}~(c). At most one of the following holds:
	(i) $A_1$ is stable in $\triangle A_1B_1C_1$; (ii) $C_2$ is stable in $\triangle A_2B_2C_2$.
	This means at most one of the two triangles is G-3-stable.
\end{proof}

\begin{proof}[Proof of Corollary~\ref{corol:number_n}]
	Assume the number of F-3-stable triangles is $m$.
	By Lemma~\ref{lemma:interleaving}, the F-3-stable triangles can be represented as
	$\triangle e_{a_1}e_{b_1}e_{c_1},\ldots,\triangle e_{a_m}e_{b_m}e_{c_m}$
	where $e_{a_1},\ldots,e_{a_m}$, $e_{b_1},\ldots,e_{b_m}$, $e_{c_1},\ldots,e_{c_m}$ lie in clockwise order (in the non-strictly manner as in defining interleaving).
	
	For each $i~(1\leq i\leq m)$, denote $\delta_i$ to be
	$$|\{a_1,a_1+1,\ldots,a_i\}| + |\{b_1,b_1+1,\ldots,b_i\}| + |\{c_1,c_1+1,\ldots,c_i\}|,$$
	where $|\cdot|$ indicates the size of the set here.
	We know $3\leq \delta_1 < \delta_2 < \ldots < \delta_m < n+3$ and this means that $m\leq n$.
	Similarly, the number of 3-stable triangles is at most $n$.
\end{proof}

\subsection{An example of the Rotate-and-Kill process}\label{subsect:example-r-k}

\begin{figure}[h]
	\centering \includegraphics[width=.9\textwidth]{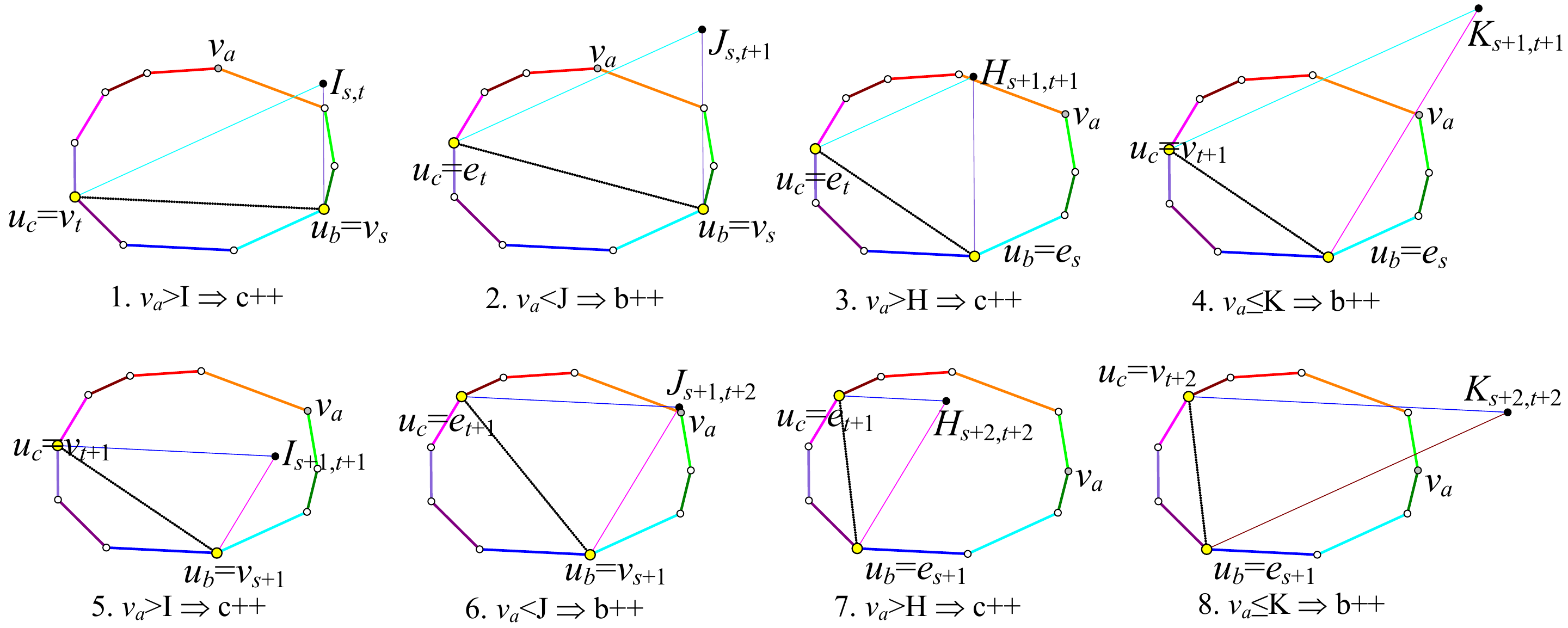}\\ \bigskip
	\centering \includegraphics[width=.9\textwidth]{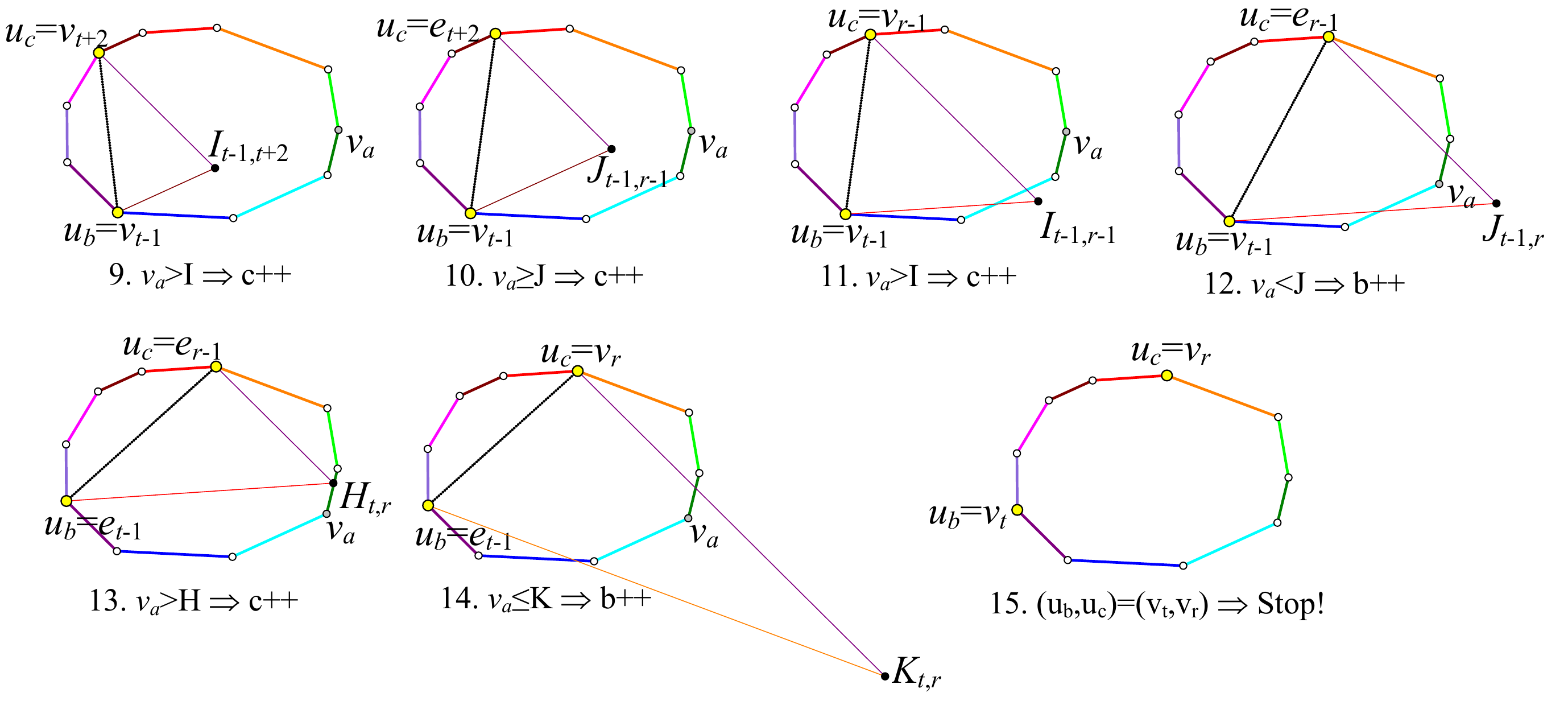}\\
	\caption{A demonstration of the Rotate-and-Kill process given in Algorithm~\ref{alg:RK-G}.}\label{fig:Gexample}
\end{figure}

Figure~\ref{fig:Gexample} demonstrates the Rotate-and-Kill process given in Algorithm~\ref{alg:RK-G}.

%
%

\subsection{Alg-DS may fail to find any 3-stable triangle}\label{subsect:diff-Alg-DS}

\begin{figure}[h]
	\centering \includegraphics[width=.85\textwidth]{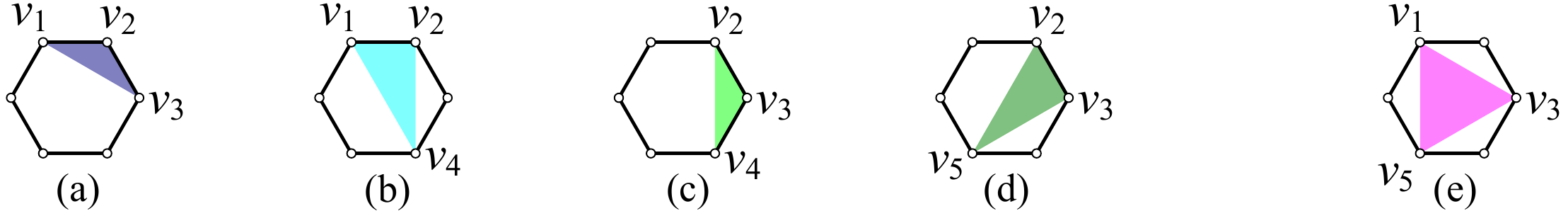}\\
	\caption{Algorithm of Dobkin and Snyder fails to find one 3-stable triangle.}\label{fig:DS-fail}
\end{figure}

Suppose $P$ is a regular hexagon, see Figure~\ref{fig:DS-fail}.
Initially, Alg-DS visits $\triangle v_av_bv_c=\triangle v_1v_2v_3$. See part (a) in the figure.
Next, it increases $c$ by 1 and visits $\triangle v_1v_2v_4$, because $v_4$ is better than $v_3$ when the other two corners are at $v_1,v_2$. See (b).
Now $c$ can no longer be improved, so as $b$, and thus Alg-DS increases $a$ by 1 and visits $\triangle v_2v_3v_4$,
and $\triangle v_2v_3v_5$ subsequently. See (c) and (d).
So on and so forth, it visits twelve triangles: $\triangle v_iv_{i+1}v_{i+2}$ and $\triangle v_iv_{i+1}v_{i+3}$ for each $i\in \{1,\ldots,
6\}$. However, none of them are 3-stable.
Instead, $\triangle v_iv_{i+2}v_{i+4}$ for $i\in \{1,\ldots,6\}$ are 3-stable. See (e).

\subparagraph{Acknowledgement.}
We are sincerely grateful for Professor Zhiyi Huang, who took part in fruitful discussions.

\bibliographystyle{plainurl}
\bibliography{MAT}

\end{document}